\documentclass[aps,prb,twocolumn,nofootinbib,citeautoscript,10pt]{revtex4-2}  
\usepackage{amsmath,amssymb,bm, makecell} 
\usepackage{graphicx}

\usepackage[tight]{subfigure} 

\usepackage{color} 
\usepackage[papersize={8.5in,11in}]{geometry}
\usepackage[colorlinks=true]{hyperref}
\hypersetup{        
    unicode=false,          
    pdftoolbar=true,        
    pdfmenubar=true,        
    pdffitwindow=false,     
    pdfstartview={FitH},    
    pdfkeywords={keyword1} {key2} {key3}, 
    pdfnewwindow=true,      
    colorlinks=true,       
    linkcolor=magenta, 
    citecolor=blue,        
    filecolor=magenta,      
    urlcolor=blue           
} 

\usepackage{graphicx}
\usepackage{dcolumn}
\usepackage{color}
\usepackage{amssymb,amsmath}
\usepackage{tabularx,graphicx}
\usepackage{epstopdf}
\usepackage{latexsym}
\usepackage{colortbl}
\usepackage{psfrag}
\usepackage{bbm,bm,array,physics}
\usepackage{dsfont}

\def \nn{\nonumber \\}
 
 \def\*#1{\mathbf{#1}} 

\newcommand{\bk}{\mathbf{k}}

\newcommand{\bA}{\mathbf{A}}
\newcommand{\be}{\mathbf{e}}

\def\be{\begin{eqnarray}}
\def\ee{\end{eqnarray}}
\def \be{\begin{align}}
\def \ee{\end{align}}
\def \bea{\begin{eqnarray}}
\def \eea{\end{eqnarray}}
\def \nn{\nonumber \\}

\begin{document}

\title{Transport properties in non-Fermi liquid phases of nodal-point semimetals}

\author{Ipsita Mandal}
\email{ipsita.mandal@snu.edu.in}
\affiliation{Department of Physics, Shiv Nadar Institution of Eminence (SNIoE), Gautam Buddha Nagar, Uttar Pradesh 201314, India
\\and
\\ Freiburg Institute for Advanced Studies (FRIAS), University of Freiburg, D-79104 Freiburg, Germany
}

\author{Hermann Freire}
\affiliation{Instituto de F{\'i}sica, Universidade Federal de Goi{\'a}s, 74.001-970,
Goi{\^a}nia-GO, Brazil}


\begin{abstract}
In this review, we survey the current progress in computing transport properties in semimetals which harbour non-Fermi liquid phases. We first discuss the widely-used Kubo formalism, which can be applied to the effective theory describing the stable non-Fermi liquid phase obtained via a renormalization group procedure and, hence, is  applicable for temperatures close to zero (e.g., optical conductivity). For finite-temperature regimes, which apply to the computations of the generalized DC conductivity tensors, we elucidate the memory matrix approach. This approach is based on an effective hydrodynamic description of the system, and is especially suited for tackling transport calculations in strongly-interacting quantum field theories, because it does not rely on the existence of long-lived quasiparticles. As a concrete example, we apply these two approaches to find the response of the so-called \textit{Luttinger-Abrikosov-Benelavskii phase} of isotropic three-dimensional Luttinger semimetals, which arises under the effects of long-ranged (unscreened) Coulomb interactions, with the chemical potential fine-tuned to cut exactly the nodal point. In particular, we focus on the electric conductivity tensors, thermal and thermoelectric response, Raman response, free energy, entropy density, and shear viscosity.
\end{abstract}

\maketitle
\tableofcontents

\section{Introduction}

The Landau's Fermi liquid theory has been incredibly successful in describing most metallic phases, serving as a paradigm to treat condensed matter systems in terms of quasiparticles and their effective interactions. However, we are 
aware of metallic states with unconventional/strange properties, for which the paradigmatic description in terms of Fermi liquids fails. Although of widely-different origins, these systems are commonly known as non-Fermi liquids (NFLs), in which the quasiparticles get destroyed, often brought about by the strong interactions of the soft fluctuations at a Fermi surface/Fermi point with some massless bosonic degrees of freedom. There have been intensive efforts in formulating controlled approximations to describe such NFL phases involving a well-defined Fermi surface \cite{nayak,nayak1, lawler1,  mross,Jiang,  metlsach1, metlsach, ips2, ips3,Shouvik1, Lee-Dalid,  shouvik2, ips-uv-ir1,  ips-uv-ir2, Freire_Pepin_1, Freire_Pepin_2, ips-subir, ips-sc, ips-c2, Lee_2018, ips-fflo, ips-nfl-u1, ips_2kf, Lee_2018, Chowdhury_2022}.
Such metallic states are often dubbed as \textit{strange metals} because of the observation of strange transport properties in many strongly-correlated quantum materials, such as cuprate superconductors \cite{Ando_2007,Legros_2018,Ayres2021}, iron-based superconducting compounds \cite{Hayes2016}, some heavy fermion materials \cite{Nakajima_2006}, and magic-angle twisted bilayer graphene \cite{Cao_2020, ips-rafael}. Instead of a conventional quadratic dependence on the temperature $T$, the normal phases of these materials/heterostructures often show a linear-in-$T$ resistivity within a large temperature window, increasing without any saturation and violating the Mott-Ioffe-Regel limit. Other manifestations of an NFL behaviour involve novel scaling-dependence of optical conductivity \cite{subir-aavishkar, ips-subir, patel2, ips-c2} and enhanced susceptibility towards superconducting instability \cite{ips2, ips3, max_cooper, ips-sc, ips-c2}. In these scenarios, which involve a finite and sharply-defined Fermi surface (although no Landau quasiparticles), the NFL character emerges due to finite-density fermions interacting with a massless boson arising at a quantum critical point \cite{metlsach1, metlsach, Lee-Dalid, ips-uv-ir1, ips-uv-ir2, ips-fflo, ips-rafael, chubukov3, rech, delanna, Chubukov2020, ips_2kf},
or with massless gauge field(s) \cite{olav,ips2, ips3, ips-nfl-u1}, leading to the alternate name of \textit{critical Fermi surface states} \cite{senthil_fs}.
However, additionally, there have been investigations of NFL phases appearing at a Fermi point, i.e., when the chemical potential cuts a band-crossing point in a semimetal \cite{Abrikosov, Abrikosov_Beneslavskii, moon-xu, rahul-sid, ips-rahul, ips-qbt-sc, malcolm-bitan, ips-biref}, which is the focus of this review article.
For these cases, when the Fermi level is tuned to cut at the band-touching point of the node of a semimetal, in the presence of long-ranged Coulomb interactions \cite{Abrikosov, Abrikosov_Beneslavskii, moon-xu, malcolm-bitan, ips-biref}, a stable non-Fermi liquid fixed point emerges.

Transport coefficients, such as the electrical and thermal conductivity and the viscosity tensors, play a central role in describing condensed matter systems. They are experimentally measurable and contain signatures which characterize the different phases of matter. Computing such transport properties, using the effective theories of NFLs states, is challenging because we cannot use the conventional frameworks available, as described below.
Determining transport properties implies that we want to theoretically compute the response of the system to an external perturbation. If the perturbation is small, the response is expected to be a linear function
of the perturbation. This is the basic assumption of the \textit{linear response theory}. In the framework of many-particle
physics, the linear response theory results in the Kubo formula (detailed discussion can be found in Sec.~\ref{seckubo}), which
refers to relations that express the linear response functions in terms of the equilibrium eigenstates of the system, and were first obtained by Kubo \cite{kubo1, Kubo_1957}. However, let us first discuss a different route, mainly to explain why it does not work for the NFL systems we want to analyze. If the external perturbation changes slowly in time and space on atomic scales, we can use a semiclassical description known as the Boltzmann formalism.
This formalism originates from the Boltzmann equation (BE), introduced by Boltzmann in 1872, to study irreversibility in dilute (i.e., low-density) gases from a statistical mechanics point of view. He obtained the BE by visualizing the dynamics of the constituent gas-molecules as free motions, occasionally interrupted by mutual collisions. In the modern literature, the term BE is used in a very wide context (or a more generic sense), referring to any kinetic equation that describes the change of a macroscopic quantity  (such as density of energy, charge, or particle number) in a thermodynamic system. In particular, the transport properties in condensed matter systems are often related to the well-defined quasiparticle-excitations, which, at a phenomenological level, resemble a dilute gas of molecules for which a BE (or a closely related kinetic equation) can indeed be formulated. The quasiparticle concept is crucial to applying the Boltzmann transport concept in condensed matter. Depending on the physical system under consideration, various types of quasiparticles emerge quantum-mechanically (for example, dressed electrons in metals), whose kinetics can then be modelled by an appropriate semiclassical approach of the BE. The idea is to consider the phase space distribution function $f(\mathbf r,\mathbf  k, t)$ of the quasiparticles, and determine the evolution of $f(\mathbf r,\mathbf  k, t)$ in the phase space by using the Liouville equation (arising from the Liouville theorem), modified by adding the correction term arising due to quasiparticle-collisions as a perturbation \cite{ips-kush-review}.
However, the bottomline is, since the BE crucially depends on the existence of well-defined and long-lived fermionic quasiparticles, the absence of quasiparticles in NFLs is a major impediment to using it for NFLs.
Nevertheless, we must add that the methodology of invoking a generalized quasiparticle distribution, within the nonequilibrium Keldysh formalism, has been employed in extracting the behaviour of collective modes of critical Fermi surfaces \cite{prange, kim_qbe, ips-zero-mode, ips-kazi}.

Another limitation of the BE approach stems from the fact that we use a semiclassical
distribution function for computing the transport properties. The Kubo formalism \cite{kubo1, Kubo_1957} remedies this issue by replacing the semiclassical
statistics with quantum statistics. The move from classical statistical mechanics to quantum statistical mechanics is implemented by calculating the average of a macroscopic observable by its trace weighted with the density matrix of the system. Via Kubo’s linear response formalism, we compute the linear-order response coefficient as a correlation function of the operator of interest and the operator which couples to the applied external probe field. In fact, the Kubo formula gives the exact linear-order response functions, as no approximations are made in the derivation. Therefore, it captures the quantum-coherent effects that cannot be captured by Boltzmann transport theory. On the other hand, the time evolution in the Kubo formula is evaluated by tracking the full microscopic dynamics, through the time evolution operator constructed out of the equilibrium Hamiltonian of the system. Therefore, it can work only if we have a controlled perturbative description of the corresponding quantum field theory, where the free part of the quantum effective action can be used to find the exact
eigenstates and energies of the system, which feed into the expressions for the correlation functions. For NFLs, we can thus use the Kubo formula only if we succeed in achieving a controlled perturbative expansion and, using the theory in the vicinity of a stable NFL fixed point, we can then compute the required correlators using the order-by-order perturbative corrections.

We have now understood that, although the Kubo formalism is an improvement upon the Boltzmann transport theory by replacing classical statistical mechanics with quantum statistical mechanics, it does not include the coarse-graining of the microscopic dynamics which renders the Boltzmann transport theory tractable in the thermodynamic limit. The memory matrix formalism \cite{mori0, Mori, zwanzig0, Zwanzig, Woefle, forster-book, andy_review, green_thesis} addresses this problem by devising a method of coarse-graining that is compatible with quantum statistical mechanics. Because the Kubo formalism is an exact quantum statistical mechanics description, it is a very useful starting point. The memory matrix approach is not so much different, as it is a reformulation of the time evolution in the Kubo formalism by making the coarse-graining approximations natural. The general philosophy of a coarse-graining procedure is to (1) find the most relevant collective variables to describe a macroscopic system, (2) separate out the contributions arising from the relevant variables and rewrite the equations of motion for the relevant variables, and (3) approximate the effects of the irrelevant variables on the evolution of the relevant operators. This leads to a macroscopic equation of motion, governing the evolution of the relevant observables, that reproduces the most interesting and leading-order manifestation of the underlying microscopic physics. A simple example is obtained from the Navier–Stokes equation in hydrodynamics, which describes the motion of viscous fluids in terms of the collective variables of pressure, density, and flow, rather than by tracking the positions and velocities
of individual molecules. Such a description is derived from the microscopic physics by applying universal constraints like conservation of particle number and  conservation of momentum, which brings into consideration the behaviour of the average density and the average flow, ignoring the microscopic fluctuations. In a similar spirit, the memory matrix procedure starts with applying the Mori projection on the equations of motion for the operators, obtained using the same time evolution as used in the Kubo formalism. After splitting the dynamics into the subspaces of the relevant and irrelevant operators, the so-called Mori’s projected equations of motion consist of the generalized Langevin equations for the relevant dynamical variables, with the effects of the irrelevant operators appearing in the equations  as correlation functions with the relevant operators.
In recent times, this formalism has attracted a significant amount of interest, especially because it has successfully reproduced the signature transport properties of many strongly-coupled systems, which were earlier computed from nonperturbative approaches (e.g., those obtained by exploiting the gauge-gravity duality between the strongly-coupled quantum field theory and the corresponding classical gravity theory  in one higher spatial
dimension \cite{ sachdev_ads, hartnoll_lectures}).

In view of the discussion presented above, both the Kubo formula and the memory matrix formalism have emerged as central tools to compute the transport coefficients of various NFL phases, exhibited by correlated quantum materials. In this review, we will demonstrate how these two approaches can be employed to compute the scalings of various response coefficients of a nodal-point NFL.

In the recent years, modelling and understanding NFL behaviour have witnessed enormous progress in connection with the various aspects of the transport properties of strange-metallic phases. Such descriptions include strongly-correlated systems like the Sachdev-Ye-Kitaev (SYK) models \cite{SY_1993, Kitaev_SYK, Esterlis, Wang_Yukawa, Patel_SYK, ips_syk, Aldape, Chowdhury_2022, Patel_Science}, strongly-coupled field theories that are holographically dual to black hole physics \cite{sachdev_ads, hartnoll_lectures, Zaanen-CUP, Sachdev-MIT, Phillips_review}, and experimentally-motivated phenomenological models describing high-T$_c$ materials~\cite{Else_2021, Varma_1989, Hartnoll-PRB_2013, Patel-PRB, ips-subir, patel2, ips-c2, Freire-AP_2020, Anurag_2021, PEPIN2023, Pangburn_2023, Berg_2019, ZYMeng, Teixeira_2023}. However, these systems involve a finite Fermi surface (or quantum dots), while we will mostly be interested in NFL phases appearing at Fermi points. Important examples of nodal-point NFLs are described below:
\begin{enumerate}

\item

From the analysis of the electronic bandstructures of compounds like pyrochlore iridates \cite{Kondo_2015}, half-Heusler compounds \cite{Paglione,Taillefer_2013}, grey-tin \cite{Groves}, and rhodium oxides \cite{Chamorro_2024}, a minimal effective low-energy model to describe such systems turns out to be the well-known three-dimensional (3d) Luttinger semimetals \cite{Abrikosov, Abrikosov_Beneslavskii, moon-xu, rahul-sid, ips-rahul, Janssen_Herbut_1, Janssen_Herbut_2, ips-qbt-sc, ips_qbt_plasmons, ips_qbt_tunnel, ips-sandip, krempa, ips-jing}. The resulting dispersion has a single doubly-degenerate quadratic band crossing at the center of the Brillouin zone (i.e., at the $\Gamma$-point). This novel class of materials feature a strong spin-orbit coupling (SOC) and, together with a strong electron-electron interaction caused by unscreened Coulomb interactions, demonstrate Mott correlation physics and NFL behaviour. The itinerant NFL physics was predicted theoretically back in 1974 by Abrikosov \cite{Abrikosov}, who demonstrated, using the quantum field theoretical framework, that the Coulomb interactions drive the system into a stable NFL fixed point, which is now widely known as the \textit{Luttinger-Abrikosov-Beneslavskii (LAB) phase} \cite{moon-xu}. Moon \emph{et al.} \cite{moon-xu} revisited this problem and characterized the NFL fixed point using a dimensional regularization scheme, which enabled them to calculate the universal power-law exponents describing various physical observables like conductivity, susceptibility, specific heat, and the magnetic Gruneisen number. The data from experiments \cite{Kondo_2015} on the pyrochlore iridate (Pr$_2$Ir$_2$O$_7$) show evidence for the LAB phase.

\item 
Quasiparticles with pseudospin-3/2 and having a birefringent linear spectrum with two distinct Fermi velocities, can be realized from simple tight-binding models 
in both two-dimensional (2d) and three-dimensional (3d) systems.
Examples in 2d include decorated $\pi$-flux square lattice \cite{malcolm, prb.85.235119,prb.90.045131}, honeycomb lattices \cite{prb84.195422, watanabe_2011}, and shaken optical lattices \cite{prb.84.165115, prl.107.253001}. The 3d counterparts are captured in various systems having strong spin-orbit coupling
\cite{bradlyn, prb.94.195205}, such as the antiperovskite family \cite{PhysRevB.90.081112} (with the chemical formula A$_3$BX) and the CaAgBi-family materials with a stuffed Wurtzite structure \cite{PhysRevMaterials.1.044201}.
The low-energy effective Hamiltonian of such semimetals shows that Coulomb interactions drive a clean system (i.e. without disorder) into a marginal Fermi liquid phase \cite{malcolm-bitan, ips-biref} in both 2d and 3d.

\end{enumerate}
For additional examples of NFLs in other 2d and 3d nodal-point semimetals, which in turn emerge as a result of topological quantum phase transitions, see Refs.~\cite{Isobe_2016, cho2016novel,Jing-Rong,SangEun,Shi-Xin}.

Using the dimensional regularization scheme (applied to the Kubo formula) and the memory matrix formalism, we computed various transport properties of the LAB phase, which we will summarize in this review \cite{ips_hermann1, ips-hermann2, ips-hermann3, dumitrescu_thesis}. The review is organized as follows. In Sec. \ref{secoptical}, we review the Kubo formalism and the memory matrix technique to set the stage for deriving the transport theories of NFL states. Sec~\ref{secmodel} is devoted to the application of both the methodologies to the specific example of the LAB phase. Finally, in Sec. \ref{final}, we end with a summary and some discussions, providing an outlook concerning the open problems in this field of research.

\section{Transport methodologies beyond the Fermi liquid paradigm}
\label{secoptical}

In this section, we will discuss the basic framework of applying the Kubo formalism and the memory matrix technique to a system which can be described by a field theory action. The need for these two kinds of approaches can be motivated as follows. For NFLs at finite temperatures, there remain great difficulties in calculating transport coefficients analytically due to the lack of a quasiparticle description. This holds true even when the nature of the underlying critical point is well-understood via a controlled approximation (e.g., dimensional regularization and $\varepsilon$-expansion), and the scaling of the transport coefficients are determined by the universal physics at the renormalization group (RG) flow fixed points \cite{dumitrescu_thesis}. As a simple example, let us consider the electrical conductivity $\sigma $ at finite frequencies ($\omega $) and temperatures ($T$), which is obtained from the current-current correlation function via the Kubo formalism. Suppose the expected scaling behaviour is captured by the relation $\sigma (\omega, T) \sim T^{\frac{d-2} {\mathfrak{z}}} \, \mathcal{F}(\omega/T) $, where $d$ is the dimensionality of the system in space, $ \mathfrak{z} $ is the dynamical critical exponent, and $  \mathcal{F} (u) $ is a universal scaling function. It is well-known that the small and the large limits of $\omega$ and $T$ do not commute, leading to very different physical consequences in the two regimes of $\omega \gg T $ and $\omega \ll T $ \cite{subir_book}. 
The $\omega \gg T $ limit essentially corresponds to that of the $T = 0$ stable NFL fixed point of a controlled perturbative approximation of the field theory, thus enabling the current-current correlator to be computed using the 
$T=0$ action. In contrast, the opposite limit of $ T \gg \omega $ falls into the hydrodynamic regime of thermally activated scale-invariant excitations, moving through the sample and interacting amongst themselves over long distance-scales. Calculating the form of the correlators in this limit is significantly more complex, even in a controlled perturbative expansion. As a result, while we can apply the Kubo formalism for computing the optical conductivity, valid in the regime $\omega \gg T $, the $T\gg \omega $ regime's DC conductivity has to be computed by some other method, which we choose here to be the memory matrix technique.

\subsection{Kubo formula}
\label{seckubo}

In this subsection, we review the Kubo formula to calculate the transport properties of a correlated model, following the treatment outlined in Refs.~\cite{luttinger_kubo, jorg_lectures}. Let the quantum system in question, in its thermodynamic equilibrium state, be described by the Hamiltonian $ H^{(0)} $. The expectation value of a physical observable $\mathcal O$ is given by
\begin{align}
\left \langle \mathcal O
\right \rangle \equiv \text{Tr} [ \rho^{(0)} \, \mathcal O]\,,
\end{align}
where 
\begin{align}
\rho^{(0)} = \frac{ e^{-\beta \, H^{(0)}}} {\mathcal Z(\beta)}
\end{align}
is the density operator (or the density matrix), $\beta = 1/\left( k_B \, T \right) $ (with $k_B$ set to unity, when we switch to the natural units), $\mathcal Z (\beta) = \text{Tr} [  e^{-\beta \, H^{(0)}}] $ is the 
partition function, and $T$ is the equilibrium temperature.
We are interested in the measurement of an observable that follows some external perturbation, which helps to shed light
into the inner workings of complex systems and extract dynamical information. 
To this end, we consider the system to be coupled to an external field, that is characterized by a time-dependent interaction part $V(t)$- added to the Hamiltonian, leading to
the total Hamiltonian
\begin{align}
H^{\rm tot}(t)  = H^{(0)} + V(t)\,,
\end{align}
with the new density matrix $\rho(t)$. To emphasize once more, while $H^{(0)}$ describes the quantum system in isolation, an external time-dependent $V(t) $ perturbation is applied. We will consider the scenario where the system is not affected by the perturbation in the infinite past, i.e.,  we are dealing with the case when $\lim_{t \rightarrow  -\infty } V(t) = 0 $. For example, when the external perturbation is a spatially uniform time-dependent electric field, a convenient way to realize this is via
\begin{align}
\label{eqelec}
\mathbf E (t) = \lim \limits_{\delta \rightarrow 0^+} 
\mathbf E^{(0)}\, e^{-i \,\omega \,t + \delta \, t}\,,
\end{align}
where we have included an infinitesimal positive imaginary part to the frequency $\omega$ of a sinusoidal time-dependence.
For the cases of more generic time-dependence of $V(t)$, we would use the form
\begin{align}
V(t)  = \lim \limits_{\delta \rightarrow 0^+} \int_{-\infty}^\infty
\frac{d\omega} {2\pi} \,
 \tilde V(\omega)\, e^{-i \,\omega \,t + \delta \, t}\,,
\end{align}
where $\tilde V(\omega)$ denotes the Fourier transform of $V(t)$ in the temporal space.

\subsubsection{Expectation value in the linear response regime}

As a consequence of the applied external perturbation, the observable $\mathcal O$ becomes time-dependent as well, and its time-evolution is obtained from
\begin{align}
\label{eqexpO}
\left \langle \mathcal O \right \rangle_t = 
 \text{Tr} [ \rho(t) \, \mathcal O]\,,
\end{align} 
where the density matrix itself obeys the von Neumann equation
\begin{align}
\label{eqvon}
i\,\hbar\, \partial_t \rho(t) = \left [ H^{\rm tot}(t), \, \rho(t) \right ].
\end{align}
We would like to emphasize that for $\rho(t)$ and $ V(t)$, we are analyzing the time-dependence of operators that are in the Schrödinger picture, since the von Neumann equation follows for an arbitrary density matrix defined with the help
of many-body wavefunctions obeying the Schrödinger equation with the Hamiltonian $ H^{\rm tot} (t)$.

Since the perturbation is switched on in the infinite past, when the system is assumed to be in equilibrium, we have
\begin{align}
\lim \limits_{t \rightarrow  -\infty } \rho(t) 
= \rho^{(0)}\,.
\end{align}
If we restrict ourselves to small/weak external perturbations, we can focus on the so-called \textit{linear response regime}, which refers to retaining changes that are proportional to linear order in $V (t)$.
In order to proceed further, we need to adopt the interaction picture. In this representation, the density matrix $ \rho^{(I)}(t) $ [of the interaction picture Hamiltonian $H^{\rm tot}(t)$] obeys the equation
\begin{align}
\label{eqrho}
\rho(t) = e^{ -\frac{i \, H^{(0)}  t} {\hbar}} 
\, \rho^{(I)}(t) \,e^{ \frac{i \, H^{(0)}  t } {\hbar} } \,.
\end{align} 
Note that, if considered with respect to $H^{(0)}$, $ \rho^{(I)}(t) $ must be interpreted as
corresponding to the Heisenberg picture. This is the reason why, henceforth, we will state that the operators are taken
in the Heisenberg picture. Eq.~\eqref{eqrho} gives us the time evolution of $ \rho^{(I)}(t) $ as
\begin{align}
& i\,\hbar\, \partial_t \rho (t) = \left [ H^{(0)}(t), \, \rho(t) \right ]
+ i\,\hbar\, e^{ -\frac{i \, H^{(0)}  t} {\hbar}} 
\, \partial_t \rho^{(I)} (t) \,e^{ \frac{i \, H^{(0)}  t } {\hbar} } \nn
& \Rightarrow
e^{ \frac{i \, H^{(0)}  t } {\hbar} } \,
\left [ V(t), \, \rho(t) \right ] \, e^{ -\frac{i \, H^{(0)}  t} {\hbar}} 
= i\,\hbar\, \partial_t \rho^{(I)} (t) \nn
& \Rightarrow i\,\hbar\, \partial_t \rho^{(I)} (t)  
= \left [ V^{(I)} (t), \, \rho^{(I)}(t) \right ] \,,
\end{align}
where we have used Eq.~\eqref{eqvon}. The formal solution of the above differential equation is given by
\begin{align}
\rho^{(I)} (t) = \rho^{(0)} - \frac{i}{\hbar} \int_{-\infty}^t
dt' \left [ V^{(I)} (t'), \, \rho^{(I)}(t') \right ].
\end{align}

Going back to the Schrödinger picture, the solution above leads to
\begin{align}
\rho (t) = \rho^{(0)} - \frac{i}{\hbar} \int_{-\infty}^t
dt' \, e^{ \frac{ -i \, H^{(0)}  (t-t')  } {\hbar} }
\left [ V (t'), \, \rho (t') \right ]
e^{ \frac{ i \, H^{(0)}  (t-t')  } {\hbar} }\,,
\end{align}
which can be solved recursively order-by-order, by generating a systematic expansion with respect to $ V (t) $.
At the zeroth order, we have $\rho (t) = \rho^{(0)} + \order{V (t)}$. At the first order, we insert the zeroth-order result and obtain
\begin{align}
\rho (t) &= \rho^{(0)} - \frac{i}{\hbar} \int_{-\infty}^t
dt' \, e^{ \frac{ -i \, H^{(0)}  (t-t')  } {\hbar} }
\left [ V (t'), \, \rho^{(0)} \right ]
e^{ \frac{ i \, H^{(0)}  (t-t')  } {\hbar} }\nn
& \quad + \, \order{V^2 (t)} \nn
& = \rho^{(0)} - \frac{i}{\hbar} \int_{-\infty}^t
dt' \, 
\left [ V^{(I)} (t'), \, \rho^{(0)} \right ] 
 + \, \order{V^2 (t)}
\,,
\end{align}
which is the regime of linear response.
Hence, it follows from Eq.~\eqref{eqexpO} that
\begin{align}
\label{eqexpO2}
\left \langle \mathcal O \right \rangle_t & = 
 \text{Tr} [ \rho^{(0)} \, \mathcal O]
 - \frac{i}{\hbar} \int_{-\infty}^t
dt' \, \text{Tr} \big[ 
\, [ V (t'), \, \rho^{(0)}  ] \, \mathcal O^{(I)}(t) \big] \nn
& = 
 \left \langle \mathcal O \right \rangle
 - \frac{i}{\hbar} \int_{-\infty}^t
dt' \, \text{Tr} \big[ 
\,  [ \mathcal  O^{(I)}(t), \, V (t') ] \, \rho^{(0)}  \big]
\nn
& = 
 \left \langle \mathcal O \right \rangle
 - \frac{i}{\hbar} \int_{-\infty}^t
dt' \, \left \langle  [ \mathcal  O^{(I)}(t), 
\, V (t') ] \right \rangle ,
\end{align} 
where the symbol $ \left \langle   \cdots \right \rangle $ stands for an equilibrium expectation value with respect to the Hamiltonian $H^{(0)}$.

Using the definition of the retarded correlator for two operators $\mathcal A_1 $
and $\mathcal A_2 $ as
\begin{align}
\left \langle \left \langle  \mathcal A_1^{(I)}(t) ; \,
\mathcal A_2^{(I)} (t') \right \rangle \right \rangle
=   \frac{i} {\hbar} \,\Theta(t-t') 
 \left \langle  [ \mathcal  A_1^{(I)}(t), 
 \, \mathcal  A_2^{(I)} (t') ] \right \rangle,
\end{align}
Eq.~\eqref{eqexpO2} can finally be expressed as
\begin{align}
\label{eqexpO3}
\left \langle \mathcal O \right \rangle_t & = 
 \left \langle \mathcal O \right \rangle
\,-  \int_{-\infty}^\infty dt' 
 \left \langle \left \langle  \mathcal O^{(I)}(t) ; \,
V^{(I)} (t') \right \rangle \right \rangle ,
\end{align}
which demonstrates the core fact that the linear response of a quantum system is characterized by retarded
Green’s functions. This is indeed an extremely useful result, because the inherently nonequilibrium
observable $\left \langle \mathcal O \right \rangle_t$ has been expressed as a correlation function of the system in equilibrium. The physical reason for this remarkable result is that the effects of the interactions between the excitations created in the nonequilibrium state show up at second order in the perturbation and, hence, are not included in linear response.

\subsubsection{The Kubo identity and the retarded Green's function in the frequency domain}

We consider an external and time-dependent perturbation of the form
\begin{align}
V(t) =  -\,  \mathcal B \, F (t) \,,
\end{align}
where $ \mathcal B $ is an observable of the theory and $ F  (t) $ is a classical time-dependent function.
For an observable $ \mathcal A $, Eq.~\eqref{eqexpO3} yields the linear response relation
\begin{align}
\label{eqexpA}
\left \langle \mathcal A \right \rangle_t & = 
 \left \langle \mathcal A\right \rangle
 + \int_{-\infty}^\infty dt' \,  G^R_{ \mathcal A \mathcal B} (t- t')\,
 F  (t') \,,
\end{align}
where
\begin{align}
\label{eqgreen}
G^R_{\mathcal A \mathcal B} (t-t')
&  \equiv  \left \langle \left \langle  \mathcal A (t) ; \,
\mathcal B  (t') \right \rangle \right \rangle
 \nn & 
=   \frac{i} {\hbar} \,\Theta(t) 
 \left \langle  [ \mathcal  A (t), \, 
 \mathcal  B  (t') ] \right \rangle 
\end{align}
is the retarded Green's function\footnote{
An alternate convention, differing by an overall sign, appears in the literature (e.g., in Ref.~\cite{jorg_lectures}) for the definition
of the retarded Green’s function, and $\Theta(t)$ is Heaviside step function. However, the convention adopted here is much more
standard and is the common choice in condensed-matter physics \cite{andy_review}.}
for the operators $ \mathcal A (t)$ and $ \mathcal B (t') $. Here, the time-dependent observables have been expressed in the Heisenberg representation of $H^{(0)}$ (as explained earlier).

We will need to use the Kubo identity
\begin{align}
\label{eqkuboid}
\frac{i} {\hbar}\, [\mathcal O (t'), \, \rho^{(0)} ] = \rho^{(0)} 
\int_0^\beta d\tau \,\dot{\mathcal O} (t) \vert_{t = t'-i\,\hbar\,\tau} \,,
\end{align}
where an overhead dot indicates total derivative with respect to $t$ and,
for a general operator $\mathcal O$ in the Schrödinger picture, the Heisenberg picture version $\mathcal O (t) $ is defined by
\begin{align}
\mathcal O (t) = e^{ \frac{i\, H^{(0)} t} {\hbar } }
\, \mathcal O \, e^{- \frac{i\, H^{(0)} t } {\hbar } }\,.
\end{align}
The Kubo identity can be proved easily via the following steps, starting from the right-hand side of Eq.~\eqref{eqkuboid}:
\begin{align}
& \rho^{(0)} 
\int_0^\beta d\tau \,\frac{ d {\mathcal O} (t) } {dt}
\bigg \vert_{t = t'-i\,\hbar\, \tau}
= \frac{i\, \rho^{(0)} } {\hbar} \,
\int_0^\beta d\tau \,\frac{ d  } {d \tau} \,
{\mathcal O} (t'-i\,\hbar\, \tau)
\nn 
& = \frac{i\,\rho^{(0)}} {\hbar}
 \left[ {\mathcal O} (t'-i\,\hbar\, \beta)
-  {\mathcal O} (t') \right ]\nn
& = \frac{i} {\hbar} \left [
\frac{e^{-\beta \, H^{(0)}}}   {\mathcal{Z} (\beta)} \,\,
e^{ H^{(0)} \,\beta }\, \mathcal O (t') \, e^{-  H^{(0)}\, \beta}
- \rho^{(0)} \,  {\mathcal O} (t')  \right ]
\nn
& = \frac{i} {\hbar} \left [ \mathcal O (t') \, \rho^{(0)} 
- \rho^{(0)} \,  {\mathcal O} (t')  \right ]
= \frac{i} {\hbar}\, [\mathcal O (t'), \, \rho^{(0)} ] \,.
\end{align}

Using Eq.~\eqref{eqkuboid} in Eq.~\eqref{eqgreen}, we get
\begin{align}
\label{eqgreen2}
& G^R_{ \mathcal  A \mathcal  B} (t-t')
 \equiv 
  \frac{i} {\hbar} \,\Theta(t) \,
 \text{Tr} \big[ \rho^{(0)} \,[ \mathcal  A (t), \, 
 \mathcal  B (t') ] \big]  \nn
& = - \frac{i} {\hbar} \,\Theta(t-t') \,
 \text{Tr} \big[ 
   \rho^{(0)}\, \mathcal  B (t') \,  \mathcal  A (t)  
 - \mathcal  B  (t') \,
  \rho^{(0)} \, \mathcal  A (t)   \big]  \nn
& =  - \Theta(t-t')  \,
 \text{Tr} \big[ 
\, \frac{i} {\hbar}\,[   \rho^{(0)} ,\, \mathcal  B  (t') ]\, 
 \mathcal  A (t)    \big]  \nn
& =  \Theta(t-t')  \,
 \text{Tr} \big[ 
\, \rho^{(0)} 
\int_0^\beta d\tau \,{\dot{\mathcal B}} (t'-i\,\hbar\,\tau)
\, \, \mathcal  A (t)    \big] \nn
& =  \Theta(t-t')  \int_0^\beta d\tau 
 \left \langle 
{\dot{\mathcal B}} (t'-i\, \hbar\, \tau)
\, \, \mathcal  A (t)  \right \rangle .
\end{align}
Further simplifying the expressions, we can write
\begin{align}
\label{eqgreen3}
  G^R_{  \mathcal A \mathcal B} (t)
&=  \Theta(t)  \int_0^\beta d\tau 
 \left \langle 
{\dot{\mathcal B}} (-i\, \hbar\, \tau)
\, \, \mathcal  A^{(I)}(t)  \right \rangle\nn
&= \Theta(t)  \int_0^\beta d\tau 
 \left \langle 
{\dot{\mathcal B}} (-t -i\, \hbar\, \tau)
\, \, \mathcal  A  \right \rangle .
\end{align}
We now go to the Fourier space, defining the Fourier transform $\tilde G^R_{  \mathcal A \mathcal B} (\omega ) 
\equiv \int_{-\infty}^\infty dt\, 
e^{i \, \omega_+\, t} \,
G^R_{  \mathcal A \mathcal B} (t)$, where $\omega_+ =  \omega + i\, 0^+$. The part $0^+$ denotes an infinitesimally small positive number, which is incorporated in order to ensure the correct physical result, namely the retarded correlator decays at large times.
The Fourier transform evaluates to
\begin{align}
\label{eqgreen4}
 &  \tilde G^R_{  \mathcal A \mathcal B} (\omega ) 
 =  \int_{0}^\infty dt \int_0^\beta d\tau \, 
e^{i \, \omega_+ \, t} 
 \left \langle 
{\dot{\mathcal B}} (-t -i\, \hbar\, \tau)
\, \, \mathcal  A  \right \rangle
\nn & \hspace{ 1.3  cm} = g_1 + g_2 \,,\nn
& g_1 =   \frac{1} { i\,\omega_+} 
\int_0^\beta d\tau \, 
e^{i \, \omega_+ \, t} 
 \left \langle 
{\dot{\mathcal B}} (-t -i\, \hbar\, \tau)
\, \, \mathcal  A   \right \rangle 
\bigg \vert_{t= 0}^{t=\infty} \,,\nn
& g_2  =   -  \frac{1} { i\,\omega_+} 
\int_{0}^\infty dt \int_0^\beta d\tau \, 
e^{i \, \omega_+ \, t} 
\,\frac{d}{dt} 
 \left \langle 
{\dot{\mathcal B}} (-t-i\, \hbar\, \tau)
\, \, \mathcal  A  \right \rangle .
\end{align}

Let us first analyze the second term, which simplifies to
\begin{align}
& - { i\, \omega_+\, g_2}
\nn & = 
\int_{0}^{-\infty} dt \,e^{ -i \, \omega_+ \, t}  
\,\text{Tr} \Bigg [\,
\left \lbrace \rho^{(0)} \int_0^\beta d\tau 
\,\frac{d}{d t} {\dot{\mathcal B}} ( t-i\, \hbar\, \tau)
\right \rbrace    \mathcal{A}   \Bigg ]
\nn & = \frac{i}  {\hbar}
\int_{0}^{-\infty} dt \,e^{ -i \, \omega_+ \, t}  
\,\text{Tr} \big[
[ {\dot{\mathcal B}} ( t), \,\rho^{(0)}]
\, \,   \mathcal{A}   \big]
\nn & = \frac{i}  {\hbar}
 \int_{0}^{\infty} dt \,e^{ i \, \omega_+ \, t}  
\,\text{Tr} \big[
[ \rho^{(0)}, \,{\dot{\mathcal B}}  ]
\, \,   \mathcal{A}  (t) \big]
\nn & = \frac{i}  {\hbar}
 \int_{0}^{\infty} dt \,e^{ i \, \omega_+ \, t}  
\,\text{Tr} \big[
 \rho^{(0)} \,{\dot{\mathcal B}}  
\,    \mathcal{A}  (t) 
-   \rho^{(0)} \,   \mathcal{A}  (t)
\, {\dot{\mathcal B}} \big]
\nn & =   \int_{-\infty}^{\infty} dt \,e^{ i \, \omega_+ \, t} 
\,\frac{(-i)}  {\hbar} \,\Theta(t) 
\left \langle [   \mathcal{A}  (t),\,
  {\dot{\mathcal B}}   ] \right \rangle
\nn & = 
- \, \tilde   G^R_{  \mathcal A {\dot{\mathcal B}} } (\omega )\,,
\end{align}
where 
\begin{align}
G^R_{  \mathcal A \mathcal {\dot{\mathcal B}} } (t-t')
& \equiv   \frac{i}  {\hbar} \, \Theta(t) 
\left \langle [   \mathcal{A}  (t),\,
  { \dot{\mathcal B}} (t')  ] \right \rangle,
\nn \tilde   G^R_{  \mathcal A {\dot{\mathcal B}} } (\omega ) 
& =  \int_{-\infty}^\infty dt\, 
e^{i \, \omega_+\, t} \,
G^R_{ \mathcal{A} {\dot{\mathcal B}} } (t)\,.
\end{align}
As for the first term $g_1$, the part evaluated at $t = \infty $ must go to zero, as that is the boundary condition, which we have ensured by including convergence factor ($=- 0^+\, t$) in the argument of the exponential.
Thus, $g_1 =   \frac{1} { i\, \omega_+} \chi_{\ell_1 \ell_2} $, where 
\begin{align}
\label{eqsus}
 \, \chi_{ \mathcal{A} {\dot{\mathcal B}} } 
 =
\int_0^\beta d\tau  \left \langle 
{\dot{\mathcal B}} ( -i\, \hbar\, \tau)
\, \, \mathcal  A   \right \rangle 
= \lim_{\omega \rightarrow 0} 
\tilde   G^R_{  \mathcal A {\dot{\mathcal B}} } (\omega ) \,,
\end{align}
which is called the static susceptibility.
Hence, the final expression for the retarded Green's function in the Fourier space reduces to
\begin{align}
\label{eqgreenfin}
 \tilde G^R_{  \mathcal A \mathcal B} (\omega ) 
& = \frac{1} { i \,\omega_+}
\left[ \tilde   G^R_{  \mathcal A {\dot{\mathcal B}}} (\omega )
- \chi_{\mathcal{A} {\dot{\mathcal B}} }  \right].
\end{align}

\subsubsection{Conductivity tensor}

Let us consider a system of charged particles (which are electrons in usual condensed matter systems) subjected to an external spatially homogeneous time-dependent electric field $\mathbf E(t) $ [cf. Eq.~\eqref{eqelec}]. The electric field induces a current, and the conductivity tensor is the linear-response coefficient. For definiteness, let us consider the electrons to be the sole charge carriers, each of which has a charge $\mathfrak q$.

Let us first define the electric current density operator. In the position space representation for the equilibrium Hamiltonian of the system (which is expressed in terms of the fermionic quantum fields), we make the replacement $ -i \, \hbar \,\partial_{r^p} \rightarrow -i \, \hbar \,\partial_{r^p} -\frac{\mathfrak q}{c}  A^p $, which we denote by $\tilde H^{(0)} [\mathbf A]$. The components of $\mathbf{J}$ is then derived as the functional derivatives
\begin{align}
\label{eqcur1}
J_p = -\frac{\delta \tilde H^{(0)} [\mathbf A]}{\delta A^p}\,.
\end{align}
However, for a parabolic spectrum, in addition to the $\mathbf A$-independent term, this generates a linear-in-$\mathbf A$ term. This necessitates gauge-fixing while finding the final expressions. A simpler alternate way is to define the current operator using the relation $\mathbf J = \dot {\boldsymbol{\mathcal P}}  = \frac{i}{\hbar } \,[H^{(0)}, \, {\boldsymbol{\mathcal P}}  ]$, where $\boldsymbol{\mathcal P} $ is the electrical dipole moment operator, which couples to the external electric field. We can express $ \boldsymbol{\mathcal P} $ as $\int d^d\mathbf r \, \mathbf r \, n_c  $, where $  {n}_c $ is the electric charge density operator.
Using this line of argument, we need to set
\begin{align}
\label{eqpol}
V(t) =  -\, \boldsymbol {\mathcal P} \cdot \mathbf E(t)\,.
\end{align}
We can now use Eqs.~\eqref{eqexpA} and \eqref{eqgreen} by setting $ \mathcal{A} = J_p $, $  \mathcal{B} =  \mathcal P_q
 $, and $F(t) = E_q(t) = E^{(0)}_q e^{- i\,\omega_+ \, t }$.
The identity ${\dot{\mathcal P}}_q = J_q $ leads to
\begin{align}
\chi_{J_p  {\dot {\mathcal{ P}}_q } } (t) 
\equiv \chi_{J_p  {J}_q } (t) 
 =   \frac{i}  {\hbar} \, \Theta(t) 
\left \langle [ J_p (t), {J}_q  ] \right \rangle  .
\end{align}
Since the conductivity in the temporal space is defined as
\begin{align}
\left \langle J_p \right \rangle_t & = \int_{- \infty}^\infty dt' \,
\sigma^{\rm temp}_{p q}(t-t') \, E_q(t') \,,
\end{align}
and
Eq.~\eqref{eqexpA} results in (since the average current in the equilibrium distribution is zero)
\begin{align}
\label{eqexpJ}
\left \langle J_p \right \rangle_t & =  
 \int_{-\infty}^\infty dt' \,  G^R_{ J_p {\mathcal P }_q} (t- t')\,
 E_q (t') \,,
\end{align}
we conclude that
\begin{align}
\sigma^{\rm temp}_{p q}(t-t') =    G^R_{ J_p {\mathcal P }_q} (t- t')\,.
\end{align}
Fourier-transforming to the frequency space and using \eqref{eqgreenfin}, we finally obtain
the Fourier-space conductivity as
\begin{align}
\label{eqkubosig}
\sigma_{p q} (\omega) & \equiv 
 \tilde  G^R_{ J_p {\mathcal P }_q} (\omega)
= \frac{ - \,i} {\omega_+} 
\left[   \tilde   G^R_{ J_p  {J}_q} (\omega ) 
- \chi_{ J_p  {J}_q}  \right] .
\end{align}



We would like point out that yet another widely-used procedure \cite{bruus, luttinger_kubo} is to express $\mathbf E $ in terms of the scalar and vector potentials and, then, use
the continuity equation
\begin{align}
\partial_t {n}_c (\mathbf r, t )= - \nabla\cdot \mathbf{J}(\mathbf r, t) \,.
\end{align}

\subsection{Memory matrix formalism}
\label{secmem}

The memory matrix formalism is the second formalism that we are going to focus on in this review, which is a powerful technique for describing transport in strongly correlated systems without quasiparticles. In this subsection, we discuss the procedure to compute the response at finite temperatures, mainly following the treatment in Refs.~\cite{forster-book, andy_review}.

The simplest framework of describing transport without reference to quasiparticles is, in fact, hydrodynamics, which has been known for a very long time. Although hydrodynamics was synonymous with fluid dynamics before the twentieth century, the modern understanding of hydrodynamics is that it describes the long-wavelength and long-time-scale dynamics of an interacting classical or quantum system close to thermal equilibrium, when there is a small number of conserved quantities \cite{kadanoff_hydro}. In other words, the key assumption of hydrodynamics is that the field theory has locally reached a thermal equilibrium. A more generic treatment allows for some of these conserved quantities to decay on long time-scales, while still retaining the nomenclature of ``hydrodynamics'' \cite{andy_review}. The only requirement for applying this generalized treatment is that the list of such conserved quantities must be finite. This requirement cannot be fulfilled in Fermi liquids, because we have occupation numbers of the long-lived quantities at every single wavevector (which form an infinite set). However, for NFLs, it is believed that generic higher-dimensional theories do not admit infinite families of nearly-conserved quantities at strong coupling and, therefore, the requirement is easily satisfied.

Despite the fact that the effective theory of hydrodynamics provides the
relaxation of an interacting classical or quantum system
towards thermal equilibrium, without any reference to the existence of quasiparticles, it is important to note that hydrodynamics is an incomplete description. This implies that although hydrodynamics provides a universal
framework, via a set of constraints that any reasonable (and, at least, approximately translation-invariant) quantum field theory at
finite density and temperature must obey, it does not give us any specific values or temperature-dependence for the microscopic coefficients.
The memory matrix approach provides a way of obtaining the microscopic coefficients. In particular, one does not need
to add a phenomenological momentum-relaxation time, as required in hydrodynamics, as this coefficient can be computed separately using the memory matrix formalism. Although, in principle, the memory matrix technique is an exact microscopic calculation, its practical usefulness stems from the fact that it can be efficiently approximated in a hydrodynamic regime, where there is only a small handful of quantities which do not quickly relax to thermal equilibrium.

As discussed in the introduction, the memory matrix technique is essentially employing the coarse-graining procedure to obtain the time-evolution of the slow/long-lived (hydrodynamic) modes by ``integrating out'' the fast (microscopic) modes. 
This approach was introduced by Mori \cite{mori0, Mori}, Zwanzig \cite{zwanzig0, Zwanzig}, and developed further by G\"{o}tze and W\"{o}lfle \cite{Woefle, forster-book} more than half-a-century ago. Zwanzig separated
the ensemble density into relevant and irrelevant parts by means of a projection, solved the latter part formally in terms of the former one, and substituted the solution back into the equation for the relevant part \cite{mori0}. This exact transformation is particularly suitable for integrating out the fast modes (i.e., the modes having fast variations in time) \cite{Zwanzig}.
Following the earlier arguments, this method works best when quasiparticles are not long-lived, and the only conserved (or approximately-conserved) quantities are charge, energy, and momentum. In the last two decades, this computational tool has become a method of choice for studying transport in NFL phases, arising in one-dimensional \cite{Rosch-PRL,Shimshoni} and higher-dimensional systems \cite{Hartnoll-PRB_2013, Freire-AP_2014, Patel-PRB, Hartnoll_PRB_2014, Zaanen-CUP, andy_review, Freire-AP_2017, Freire-EPL, Sachdev-MIT, Freire-EPL_2018, Berg-PRB, Freire-AP_2020, wang2020low}, as well as strongly-interacting quantum field theories in the context of using gauge-gravity duality \cite{DTSon-PRL_2005,hartnoll_herzog,Blaise,Blake,Blake2}.

Let us now briefly outline the notations required in applying the memory matrix framework. In the following, we will set $\hbar $ and $k_B $ to unity, agreeing to use natural units.
Let us consider the set of linear operators $\lbrace \mathcal A, \, \mathcal B, \, \mathcal C,\, \cdots \rbrace $ in a time-translation-invariant theory. The space of linear operators acting on a Hilbert space is called the Liouville space. 
When dealing with a complicated quantum system, we are usually neither interested in nor capable of describing the time evolution of all its microscopic properties. Rather, most of the times, we want to determine the linear response, which requires the knowledge of the dynamics of only a small set of selected (“relevant”) observables. These relevant observables, together with the identity operator, span a (relatively small) subspace of the Liouville space, which is known as the \textit{level of description} \cite{rau}. Consequently, the dynamics is projected onto the level of description, which yields closed equations of motion for the relevant observables only (although it is, in general, no longer Markovian).

The detection and systematic exploitation of a separation of time scales is the basic
practical merit of the projection-operator method \cite{rau}. A simple example to illustrate this is the Brownian motion of a massive particle within a fluid of small molecules. This process consists of damping on a macroscopic scale
due to the viscosity of the fluid, and fast vibrations due to the stochastic residual forces.
Both the processes are caused by collisions between the particle and the fluid molecules, but they
take place on different time scales --- the damping process is much more relevant for the observable ``position of
the particle'', compared to the vibrations, since the latter vanish when averaged over
time. However, only if the \textit{relevant processes} are filtered out, we can neglect the fast modes.

\subsubsection{Memory function formalism}

A necessary condition for the definition of a projector is the existence of a scalar
product within the Liouville space. There are multiple possibilities to do so. Here, we use the definition put forward by Mori \cite{mori0}, using the inner product
\begin{align} 
\Upsilon_{\mathcal A \mathcal B }(t) & \equiv
 ( \mathcal A(t)|\mathcal B(0))  \,, \nn
 (\mathcal A(t)| \mathcal B(0))  & \equiv
 \int\limits_0^{\beta}
 \mathrm{d}\lambda\; \left\langle 
 \mathcal A^\dagger(t) \, \mathcal B(i \, \lambda)\right \rangle ,
 \label{innprod}
\end{align}
with averages over thermal and quantum fluctuations denoted by the symbol $\langle \cdots\rangle$.
Here, $L$ is the Liouvillian (or Liouville operator), defined by
\begin{align}
\label{eqli}
L \, \circ = [ H^{\rm tot}(t)  , \,\circ]
= [  H^{(0)} , \,\circ ] + [ V(t) , \, \circ ]\,,
\end{align}
and $ | \mathcal A(t)) \equiv e^{i \, L \, t} \,| \mathcal A (0) )$.
Note that the usual convention is to denote $\mathcal A (0)$ simply by $ \mathcal A$ --- hence, we will use these two alternate notations interchangeably. 
The expression follows from the fact that, when written in terms of the Liouvillian, the Heisenberg equation of motion for $\mathcal A$
reads
\begin{align}
\frac{d \mathcal A(t) } {dt} = i\, L\, \mathcal A
\Rightarrow
 \mathcal A(t) =   e^{i \, L \, t} \,A (0)\,.
\end{align}

Rather than working in the time domain, we will take Laplace transforms
to work in the frequency domain.
The Laplace transform of a function $F(t) $ is defined by
\begin{align}
\tilde F ( \omega ) = \int_0^\infty dt\, e^{i\, \omega \, t} \, F(t)\,.
\end{align} 
Furthermore, we need the counterpropagator
\begin{align}
\mathcal R (\omega) 
= 
\int_0^\infty dt\, e^{i\, \omega \, t} \,e^{i\, L\, t}
= \frac{i} { \omega-   L} \,,
\end{align}
where ${\rm Im} ( \omega ) > 0$. We denote the Laplace-transformed correlation function in Eq.~\eqref{innprod} by
the symbol $ \tilde {\Upsilon}_{\mathcal A \mathcal B}(\omega) $, such that
\begin{align}
\tilde {\Upsilon}_{\mathcal A \mathcal B}( \omega ) 
 & \equiv 
\int_0^\infty dt \, e^{i\, \omega\, t} \, {\Upsilon}_{\mathcal A \mathcal B} (t)
\nn & = \int_0^\infty dt \, e^{i\, \omega\, t} \,
(A (0) \, e^{i\, L \, t} |B(0)) 
\nn & = \int_0^\infty dt \, e^{i\, \omega\, t} \,
(A(0) | \, e^{ - i\, L \, t} \,B(0)) 
\nn & =  (A(0) | \, \mathcal R(\omega) \,B (0) ) \,.  
\label{cabz0}
\end{align}
By using $\mathcal R (\omega) = \left [L \,\mathcal R(\omega) 
- L \, \mathcal R(0) \right ] /  \omega$,
we obtain
\begin{align}
\tilde {\Upsilon}_{\mathcal A \mathcal B}(\omega) 
=
\frac{1} {i \, \omega }
\left[ \tilde G^{\mathrm{R}}_{\mathcal A \mathcal B }(\omega) 
- \tilde G^{\mathrm{R}}_{\mathcal A \mathcal B }(i \, 0)\right],  
\label{cabz}
\end{align}
where $ \tilde G^{\mathrm{R}}_{\mathcal A \mathcal B }(\omega) $ is the Laplace transform of the retarded Green's function (in real space and time), defined as
\begin{align}
G^{\mathrm{R}}_{\mathcal A \mathcal B}( t, \mathbf{r}) \equiv 
i \, \Theta(t) \left \langle  [ \mathcal A( t, \mathbf{r}), \, 
\mathcal B( 0, \mathbf{0})] \right \rangle.
\end{align}  
This is the same as Eq.~\eqref{eqgreen}, except that the position-dependence has been suppressed in the former equation, since there we were dealing with a spatially-uniform perturbation.
Henceforth, we will suppress the position-/momentum-dependence of the retarded Green's functions [as already done in
Eq.~\eqref{cabz}] since, in this entire review, we are only interested in the response functions evaluated at zero momentum. Here, we will only be computing thermoelectric transport coefficients, for which $ \tilde G^{\mathrm{R}}_{\mathcal A \mathcal B }(\omega \rightarrow 0) 
\sim i \, \omega \, {\mathfrak s}_{\mathcal A \mathcal B}$, where $ {\mathfrak s}_{\mathcal A \mathcal B }$ is the generalized conductivity between the operators $ \mathcal A$ and $\mathcal B$.
Since $ {\mathfrak s}_{\mathcal A \mathcal B }$ is strictly finite, we have $ \tilde G^{\mathrm{R}}_{\mathcal A \mathcal B }(i \, 0) = 0$ and, hence, the latter can be omitted in the rest of the discussions. Therefore, in this context, $ \tilde \Upsilon_{\mathcal A \mathcal B }(\omega)$ turns out to be equal to $  {\mathfrak s}_{\mathcal A \mathcal B }$. 

Some formal manipulations on the Hilbert space of operators lead to
\begin{align}
{\mathfrak s}_{\mathcal A \mathcal B }(\omega) & \equiv 
 \tilde \Upsilon_{\mathcal A \mathcal B }(\omega) 
\nn
& =  \chi_{\mathcal A \mathcal C}
\left [ \frac{1}
 { M(\omega) + N - i \, \omega \, \chi }
  \right ]_{\mathcal C \mathcal D}
\, \chi_{\mathcal D \mathcal B} \,,  
\label{mmf}
\end{align}
where $ \chi $ is the matrix with elements
\begin{align}
\chi_{\mathcal A \mathcal B } 
 =   ( \mathcal A (0) | \, \mathcal B (0) ) \,.
 \label{chiab}
\end{align}
$\chi_{\mathcal A \mathcal B } $ is known as the static susceptibility between the operators $\mathcal A$ and $\mathcal B$ [same as Eq.~\eqref{eqsus}].
The symbol $M$ stands for the so-called memory matrix, whose components are defined as 
\begin{align}
M_{ \mathcal A \mathcal B }(\omega) = i 
\left(\dot{ \mathcal A}
\left| \, \mathcal{Q}
\,\, \frac{1}
{ \omega- \mathcal{Q} \, L\, \mathcal{Q} }
\, \, \mathcal{Q}
\, \right |\dot{ {\mathcal B } } \right) , 
\label{memmatrix}
\end{align}
where
\begin{align}
\label{eqpq}
\mathcal P = 
\sum_{\mathcal A \mathcal B } 
|\mathcal A (0) ) \,  \chi^{-1} _{\mathcal A \mathcal B} \,
(\mathcal B (0) |
\,, \quad
\mathcal{Q} = \mathbb{I}- \mathcal P\,.
\end{align}
The matrix $N$ has the components
\begin{equation}
N_{\mathcal A \mathcal B } \equiv 
\chi_{ \mathcal A \dot{\mathcal B}} 
= - \chi_{\dot{\mathcal A} \mathcal B} \,.
\end{equation}
Clearly, $N_{\mathcal A \mathcal B }$ is antisymmetric and vanishes identically in a time-reversal-invariant system, if the operators $\mathcal A$ and $ \mathcal B $ transform identically under time-reversal.

We implement the above formula by projecting onto a basis of nearly-conserved operators, which we denote by $\{\xi_i\}$, with a long relaxation time compared to microscopic timescales. These conservation laws are related to symmetries in the model that protect these operators from decaying --- a \textit{small amount of symmetry-breaking} causes these operators to become nearly-conserved (rather than completely-conserved). With this in mind, we define the projection operator $\mathcal P $ [defined above in Eq.~\eqref{eqpq}] to project onto the slow-mode basis as
\cite{forster-book}
\begin{align}
\mathcal{P} =  \sum_{ij}
\left| \xi_i (0) \right)
\,\chi^{-1} _{ \xi_i \xi_j} 
\, \left(\xi_j (0) \right| .
\end{align}
Therefore, the complement operator $\mathcal{Q} $ projects out of the space spanned by these nearly-conserved observables. 
From the Liouville operator, we want to extract that part which changes a given $ \xi_{i_0} $ (for $i = i_0 $) only in the subspace spanned by $\{\xi_i\}$.
Since $\mathcal P +\mathcal Q = \mathbb{I}$, we use the form $ L = L\, \mathcal P + L \, \mathcal  Q $.
By using the operator identity \cite{forster-book}
\begin{align}
\label{eqidentity}
\frac{1} { \mathcal A + \mathcal B }
= \frac{1} { \mathcal A}
- \frac{1}{ \mathcal A } \, \,
\mathcal  B \, \,
\frac{1}{ \mathcal A + \mathcal B } \,,
\end{align}
and using the fact that $\mathcal P$ acts as an identity operator in the space $ \{\xi_i\}$,
we get
\begin{align}
\label{eqmem0}
\tilde \Upsilon_{ \xi_i  \xi_j  }(\omega) 
 & \equiv
(  \xi_i (0)  | \,\frac{i} {\omega-L} \,  |\xi_j(0) )
\nn & = 
(  \xi_i(0) | \,\frac{i} {\omega - L \, \mathcal  Q  - L\, \mathcal P} 
\,  |\xi_j (0) ) \nn
& = 
(  \xi_i (0) | \left[
\frac{i} {\omega - L \, \mathcal  Q  } 
+ \frac{i} {\omega - L \, \mathcal  Q  } \, L \,\mathcal P
\, \frac{1} {\omega - L }
\right ]  |\xi_j (0))\,.
\end{align}
For the first term, we first observe that
\begin{align}
\label{eqseries}
\frac{i} {\omega- L \,\mathcal{ Q } }
=\frac{i} {\omega} \left [ 1
+ \frac{1}{\omega}  L \, \mathcal Q 
+  \frac{1}{ \omega^2}  L \, \mathcal Q \, L \, \mathcal Q + \ldots
\right ],
\end{align}
which shows that, except the first term, each term ends with the operator $\mathcal Q $, and $\mathcal Q \,  |\xi_i ) = 0 $.
As for the second term in Eq.~\eqref{eqmem0}, we observe that
\begin{align}
& \mathcal P \, \frac{i} {\omega- L } \, |\xi_j (0))
 = 
\left| \xi_{i'}(0) \right)
\chi^{-1} _{ \xi_{i'}  \xi_{ j'} } 
 \left(\xi_{j'} (0) \right|
 \frac{i} {\omega- L } \, |\xi_j (0) )
\nn & = 
\left| \xi_{i'}(0) \right)
\chi^{-1} _{ \xi_{i'}  \xi_{ j'} } 
 \tilde \Upsilon_{ \xi_{j'}  \xi_{j}  }(z) \,.
\end{align}
Putting these results together, Eq.~\eqref{eqmem0} evaluates to
\begin{align}
\label{eqmem1}
& \tilde \Upsilon_{ \xi_i  \xi_j  } (\omega) 
 - \frac{i} {\omega} \, \chi_{ \xi_{i}  \xi_{ j} } 
\nn &
= 
(  \xi_i (0) |  \frac{1} {\omega - L \, \mathcal  Q  } \, L 
 \left| \xi_{i'}(0) \right)
\chi^{-1} _{ \xi_{i'}  \xi_{ j'} } 
 \tilde \Upsilon_{ \xi_{j'}  \xi_{j}  }(z)
 \nn & =
(  \xi_i  | 
\left [ \frac{1} {\omega}
+ 
\frac{1}{\omega}
\,  L \, \mathcal  Q 
\, \frac{1} {\omega - L \, \mathcal  Q  } 
\right ] L 
 \left| \xi_{i'} \right)
\chi^{-1} _{ \xi_{i'}  \xi_{ j'} } 
 \tilde \Upsilon_{ \xi_{j'}  \xi_{j}  }(\omega)
 \nn & = \frac{1}{\omega} \left[
(  \xi_i  | \,  L \,| \xi_{i'}  )
+  (  \xi_i  | \,L \, \mathcal  Q 
\, \frac{1} {\omega - \mathcal  Q \,L \, \mathcal  Q  }  
\, \mathcal  Q \, L \, | \xi_{i'} )
\right ] 
 \nn & \quad \times \chi^{-1} _{ \xi_{i'}  \xi_{ j'} } 
 \tilde \Upsilon_{ \xi_{j'}  \xi_{j}  }(\omega)\,.
\end{align}
The last step can be proven with the help of a geometric series [similar to Eq.~\eqref{eqseries}], and using the identity $\mathcal Q^2 = \mathcal Q $.

\subsubsection{Generalized conductivity tensors in the absence of magnetic fields}

\label{secDC}

 Our goal is to compute the conductivity tensors at zero momentum, but finite frequency in general. To start with, we want to consider the case with zero magnetic field, which means that the time-reversal symmetry-breaking matrix $N$ will vanish.

In NFLs, the kinematics of the almost-conserved quantities is entirely different from that in the Fermi liquid theory, and the Wiedemann-Franz law is not expected to hold, even approximately \cite{Hartnoll-PRB_2013}. Since an infinite collection of
conserved densities does not exist in the effective low-energy theory, it follows that, generically, the total electrical and heat currents are affected in distinct ways (unlike in a Fermi liquid). The only conserved quantities are the momenta, up to the effect of irrelevant or weak marginal operators. The conservation of momentum, up to effects that are small at low energies, is a key assumption that allows the memory matrix method to work.
While there are strong interactions in an NFL, it is still possible that there emerges a decoupling of the excitations into patches in the momentum space, for example, on a critical Fermi suface in $(2+1)$-dimensions \cite{Lee-Dalid, ips-uv-ir1, ips-nfl-u1, ips-rafael}. For such cases, there exists a family of conserved momentum vectors. However, here we are interested in NFLs arising at nodal points (with no patch structure), which implies that we will focus on scenarios when there is only one almost-conserved vector operator in the effective low-energy theory.

In the systems we are interested in, the total momentum vector operator $\mathbf P $ is relaxed on a much longer timescale than all other quantities, including the currents \cite{Hartnoll-PRB_2013}. Nevertheless, our discussion can easily be adapted to cases in which the electric current is proportional to the momentum. The only general requirement for the results to hold is that there exists only one almost-conserved vector operator. For simplicity, we will assume that the system is spatially isotropic, such that all the diagonal components of $M$ are equal and all its transverse components vanish (i.e., $M_{P_p P_q} \propto \delta_{p q} $). 

The DC conductivity determines the dissipation due to an arbitrarily low-frequency current being driven through the system. 
In a translationally-invariant field theory at finite density, such that the only conserved vector quantity is the
total momentum, the DC conductivity tensors diverge \cite{hartnoll_241601, Davison_2015}. Intuitively, the conductivity diverges because the current operator has some overlap with the momentum operator, which is conserved. To remedy the infinite-DC-conductivity result, we perturb this theory by an irrelevant operator that breaks translational invariance, so that the infrared is still described by the original fixed point. This renders the total momentum a nearly-conserved operator.

Some standard mechanisms that can cause momentum relaxation in a transport theory are Umklapp processes, coupling to phonons, and impurity scattering. Umklapp scattering is usually exponentially suppressed at low temperatures and, since it depends on details associated with the shape of the underlying Fermi surface of a quantum system, its contribution to transport is expected to be non-universal.
On the other hand, the universal contributions for transport are expected to originate from either disorder effects or coupling to a phonon bath. Moreover, conventional wisdom suggests that, for many systems, scattering of charge carriers by phonons
turns out to be most effective as a relaxation mechanism only at high temperatures, since they tend to be generally suppressed at low-enough temperatures. Because of this, we will choose to concentrate mainly on
impurity scattering as the main mechanism that causes momentum relaxation in the transport theory. 

To implement momentum relaxation induced by impurity scatterings, we will add a weak random impurity potential, that couples to the fermionic density as a quenched disorder (i.e., with no time evolution). If the fermionic field is denoted by $\psi( t, \mathbf{r})$, we need to add the impurity action \cite{hartnoll_144502}
\begin{align}
\label{eqdisorder}
S_{\textrm{imp}}= 
\int d t\, d^d \mathbf{r}\, W(\mathbf{r})
\, \psi^{\dagger}( t, \mathbf{r})\,\psi( t, \mathbf{r})\,.
\end{align}
It is common to take $W(\mathbf{r})$ to be
a zero-mean Gaussian random function, with the mean and the variance obeying \cite{hartnoll_herzog}
\begin{align}
 \overline{   W(\mathbf{r}) }=0 \,,
 \quad \overline {  W(\mathbf{r})\,W(\mathbf{r'}) } =
W_0^2\, \delta^d(\mathbf{r}-\mathbf{r'}) \,,
\end{align}
where the overline denotes disorder-averaging, and $W_0^2$ represents the average magnitude-square of the random potential (experienced by the fermionic fields). We will limit ourselves up to order $W_0^2$.


To apply the memory matrix method explained in the previous subsection, we need to calculate the time dependence of the slowly-varying operator ${P}_p $ using the total Hamiltonian $H^{ (0)} +  H_{\textrm{imp}}$, where
\begin{align}
\label{eqimpO}
 H_{\textrm{imp}} & =
 - \int d^d \mathbf{r}\, W(\mathbf{r})
\, \mathcal{O} ( t, \mathbf{r}) \,,\nn
\text{with } & \mathcal{O} (  t, \mathbf{r})   \equiv 
  \psi ^{\dagger}(  t, \mathbf{r})\,\psi (  t,\mathbf{r})\,,
\end{align} 
is the part of the total Hamiltonian representing the contribution from disorder.
The time evolution of $P_p $ is given by $ {\dot{P}}_p   = 
i \left [
H^{ (0)} +  H_{\textrm{imp}} ,  \,P_p \right ] $, which, in the position-space representation, reduces to
\begin{align}
\dot{P}_p ( t,\mathbf r)
& = i \int d^d\mathbf r' \,  W(\mathbf{r}')
\left [ {P}_p (\mathbf r), \,  \mathcal{O} ( t, \mathbf{r}') \right ]
\nn
& =  \int d^d\mathbf r' \,\delta^d(\mathbf r -\mathbf r')  
\, W(\mathbf{r}') \,
\partial_{r^p}\mathcal{O} ( t, \mathbf{r}')\nn
& = W(\mathbf{r}) \,
\partial_{r^p}\mathcal{O} ( t, \mathbf{r}) \,.
\end{align} 
For convenience, we define
\begin{align}
& F_{ \dot{P}_p \dot{P}_q } ( t, \mathbf{r}, \mathbf{\tilde r} )  
\nn & =
\overline{W(\mathbf{r}) \, W(\mathbf{ \tilde r})} \,
\left \langle  
\left [ \partial_{r^p}\mathcal{O} ( t, \mathbf{r}), \, 
\partial_{{\tilde r}^q}\mathcal{O} ( 0, \mathbf{\tilde r})
\right ] \right \rangle \nn
& = - \, W_0^2  \,
\delta^d(\mathbf{r}-\mathbf{\tilde r})
\nn & \quad  \times
\int \frac{d^d \mathbf k \, d^d \mathbf k'} 
{(2\pi)^{2d}} 
\, \, k_p\, k'_q
\left [ \mathcal{O} ( t, \mathbf{k}), \, \mathcal{O} ( 0, \mathbf{k'})
\right ]
e^{i \,(\mathbf k + \mathbf {k'}) \cdot \mathbf r} \,.
\end{align}
Taking a Fourier transform and evaluating the zero-momentum parts, we get
\begin{align}
& \tilde F_{ \dot{P}_p \dot{P}_q } ( t, \mathbf{\tilde k}, \mathbf{{\tilde k}'} ) 
\big \vert_{ \mathbf{\tilde k} =  \mathbf{{\tilde k}'} = \mathbf 0 }  
\nn & =  - \, W_0^2
\int d^d \mathbf r \, d^d \mathbf {\tilde r}\,
 \delta^d(\mathbf{r}-\mathbf{\tilde r})
\nn & \quad \times
\int \frac{d^d \mathbf k \, d^d \mathbf k'} 
{(2\pi)^{2d}} \, k_p\, k'_q
\left [ \mathcal{O} ( t, \mathbf{k}), \, \mathcal{O} ( 0, \mathbf{k'})
\right ]
e^{i \,(\mathbf k + \mathbf {k'}) \cdot \mathbf r}
\nn & = - \, W_0^2
\int d^d \mathbf r 
\int \frac{d^d \mathbf k \, d^d \mathbf k'} 
{(2\pi)^{2d}} \, k_p\, k'_q
\left [ \mathcal{O} ( t, \mathbf{k}), \, \mathcal{O} ( 0, \mathbf{k'})
\right ]
e^{i \,(\mathbf k + \mathbf {k'}) \cdot \mathbf r}
\nn & = - \, W_0^2
\int \frac{d^d \mathbf k \, d^d \mathbf k'} 
{(2\pi)^{d}} \, \delta(\mathbf k + \mathbf k')\,k_p\, k'_q
\left [ \mathcal{O} ( t, \mathbf{k}), \, \mathcal{O} ( 0, \mathbf{k'})
\right ]
\nn & =  W_0^2
\int \frac{d^d \mathbf k } 
{(2\pi)^{d}} \, k_p\, k_q
\left [ \mathcal{O} ( t, \mathbf{k}), \, \mathcal{O} ( 0, -\mathbf{k})
\right ] \,.
\end{align}
This gives us the retarded Green's function
\begin{align}
 & \tilde G^R_{\dot{P}_p \dot{P}_q }(\omega, \mathbf 0)
 \equiv i\, \int_0^\infty dt\, e^{i \,\omega_+ t}
\tilde F_{ \dot{P}_p \dot{P}_q } ( t, \mathbf 0, \mathbf 0 ) 
\nn &
=  W_0^2
\int \frac{d^d \mathbf k } 
{(2\pi)^{d}} \, k_p\, k_q
\tilde G^R_{\mathcal O \mathcal O} (\omega, \mathbf k) \,,
\end{align}
which we will simply write as $\tilde G^R_{\dot{P}_p \dot{P}_q }(\omega)$, in order
to avoid cluttering of notations. Note that we have defined our retarded Green's function in the presence of disorder
as the one obtained after disorder-averaging.

Due to the presence of the projection operator $\mathcal{Q}$ in Eq.~\eqref{memmatrix}, it is very hard to calculate the memory matrix $M$ exactly. For this reason, we will resort to a useful approximation, which also shows the effectiveness of the approach.
Since $\dot{P}_p $ is linear in the disorder strength $W_0$, the leading contribution to $M_{P_p P_q } $ is of order $W_0^2$. Here, we wish to keep the leading contribution only and,
therefore, we approximate the full Liouville operator (after adding weak disorder) by simply the one without disorder, i.e., we use $L =[H^{(0)},\,\circ ]$. Additionally, we will calculate the ensemble-averages by using only $H^{(0)}$, instead of $ H^{(0)} + H_{\textrm{imp}}$. Since $\mathbf P$ is completely conserved for a clean system, $L \, P_p =0$ and, consequently, $L\, \mathcal{Q} = L$. Hence, the memory matrix [cf. Eq.~\eqref{memmatrix}] can be approximated by
\begin{align}
M_{P_p P_q }(\omega)  
& \approx i  \left( {P}_p
\left| \, L\,\mathcal{Q}
\,\, \frac{1}
{ \omega - \mathcal{Q} \, L\, \mathcal{Q} }
\, \, \mathcal{Q}\,L
\, \right | {P}_q \right)
\nn & = i  \left( {P}_p 
\left| \, L\,\mathcal{Q}
\,\, \frac{1}
{ \omega - L\, \mathcal{Q} }
\, \, L \, \right | {P}_q \right)
\nn & = i  \left( {P}_p
\left| \, L \,\, \frac{1}
{ \omega - L\, \mathcal{Q} }
\, \, L \, \right | {P}_q \right) 
\nn & = i  \left( \dot {P}_p
\left| \, \frac{1} { \omega - L }
\,  L \, \right | \dot{P}_q \right)  
= \tilde {\Upsilon}_{ \dot {P}_p \dot {P}_q }(z) ,
\label{memory22}
\end{align}
using Eq.~\eqref{cabz}.
In the limit of $ z $ going to zero, we define
\begin{align}
\label{memory4}
M_{P_p P_q}(0) & \equiv  \lim \limits_{\omega \rightarrow 0}
\frac{1} {i \, \omega }
\left[ \tilde G^{\mathrm{R}}_{ \dot{P}_p \dot{P}_q } (\omega) 
- \tilde G^{\mathrm{R}}_{ \dot{P}_p \dot{P}_q }( 0)\right]
\nn  & =
 \frac{1} {i}\,\partial_\omega 
 \tilde G^{\mathrm{R}}_{ \dot{P}_p \dot{P}_q }(\omega) \Big \vert_{ \omega=0}
 =
\lim_{ \omega \rightarrow 0}
\frac{\text{Im}\,
\tilde G^{R}_{\dot{P}_{p} \dot{P}_{q}}(\omega)}
{\omega } \,.
\end{align}
The last equality follows from the fact that $\text{Re}\,
\tilde G^{R}_{\dot{P}_{p} \dot{P}_{q}}(\omega)$ is an even function of $\omega$, whereas
$\text{Im}\,
\tilde G^{R}_{\dot{P}_{p} \dot{P}_{q}}(\omega)$ is an odd function of $\omega$ \cite{hartnoll_herzog}.

In a time-reversal-invariant theory, in which momentum is the only
almost-conserved observable, $N=0$, and Eq.~\eqref{mmf} gives us a diagonal component of the electrical conductivity tensor as
\begin{align}
\label{eqcond1}
\sigma_{p p} (\omega, T) &\equiv
\, \frac{ \chi_{J_p P_p}^2
 }
{ M_{P_p P_p } (\omega) - i  \,\omega \,\chi_{ P_p P_p}  } \,.
\end{align}
Here, $ \chi_{ P_p P_q} $ is the momentum-momentum static susceptibility, and $\chi_{J_p P_q}$ is the current-momentum static susceptibility.
This immediately tells us that the corresponding DC electrical conductivity is given by
\begin{align}
\label{eqconddc}
\sigma^{\rm dc}_{pp}   & \equiv \sigma_{p p} (0, T) =
\, \frac{  \chi_{J_p P_p}^2 }
{  M_{P_p P_p} (0)   }\,.
\end{align}
Electric and thermal transport generally couple together in charged quantum matter. Hence, we would want to compute not just the electrical conductivity, but a more general matrix of thermoelectric conductivity tensors.
For this, we need to consider the heat current $\mathbf{J}^Q$, which naturally couples to a temperature gradienat $\nabla_r T$ \cite{luttinger_kubo}. At the linear order, we have the generalized Ohm/Fourier law \cite{Sachdev-MIT} 
\begin{align}
\begin{pmatrix}
 J_p \vspace{0.2 cm} \\
{J}^Q_p 
\end{pmatrix} & = 
 \sum \limits_q
\begin{pmatrix}
\sigma_{pq } &  \alpha_{p q }
 \vspace{0.2 cm}  \\
T \, \alpha_{ p q } &  \bar \kappa_{p q }
\end{pmatrix}
\begin{pmatrix}
E_q 
 \vspace{0.2 cm}  \\
- { \partial_{r^q} T } 
\end{pmatrix},
\end{align}
following the notations of Ref.~\cite{hartnoll_144502}. While the thermoelectric conductivity tensor $\alpha $ determines the Peltier,
Seebeck, and Nernst effects, $\bar \kappa $ is the linear response coefficient between the heat
current and the temperature gradient at vanishing electric field. The latter applies to samples connected to conducting leads, allowing for a stationary current flow. In experiments, one often measures the components of the thermal conductivity tensor $\kappa$, which provide the coefficients between the heat current and temperature gradient at vanishing electric current (i.e., $\mathbf J = 0$), and are given by
\begin{align}
\label{eqkappa}
\kappa = \bar \kappa - T\,\alpha \, \sigma^{-1} \,\alpha \,.
\end{align}
Finally, the Nernst response is defined as the electric field ($\mathbf E_{\rm ind}$) induced by a thermal gradient, in the absence of an electric
current, and is given by the relation
\begin{align}
\mathbf E_{\rm ind} = - \vartheta \,\nabla_r T \,, \text{ where }
\vartheta = -\sigma^{-1} \,\alpha \,.
\end{align}

From our discussions, we find that we need to specify a set of three tensors, viz. $\sigma$, $\alpha $, and $\bar \kappa$, from which we can derive the remaining tensors listed above. In conjunction with Eq.~\eqref{eqcond1}, we have the relations
\begin{align}
\bar \kappa_{pq} (\omega, T) &= \frac{1}{T} 
\, \Upsilon_{  J^Q_p  J^Q_q }(\omega) 
\,, \nn
\alpha_{pq} (\omega, T) &= \frac{1}{T} 
\, \Upsilon_{  J_p J^Q_q }(\omega) \,.
\end{align}
Analogous to the electrical DC conductivity tensor, the diagonal components in the low-frequency limit are given by
\begin{align}
\label{eqcond2}
\bar{\kappa}^{\rm dc}_{pp}   & = \frac{1}{T}\, \frac{ \chi^2_{J^Q_p P_p}
 } { M_{P_p P_p } (0)  } \,,
\nn 
\alpha_{p p}^{\rm dc} ( T) & =  \frac{1}{T}\, \chi_{J_p P_p}
  M_{P_p P_p }^{-1} (0) \, \chi_{ P_p J^Q_p}\,.
\end{align}

\subsubsection{Generalized conductivity tensors in the presence of a weak magnetic field}


We will now rederive the expressions for the conductivity tensors in the presence of a uniform magnetic field $\mathbf B$, with $B$ denoting its magnitude \cite{Hayes2016, Giraldo_Gallo_2018, andy_review, Freire-EPL,Freire-EPL_2018}. We will focus on the regime where $ B$ is perturbatively small, such that the cyclotron frequency $\omega_c$ is a perturbatively-small parameter as well, and follow the route of Ref.~\cite{andy_review}.
In this case, the time-reversal symmetry-breaking matrix $N$ is nonvanishing. 
We note that each component of $M$ is of first-order within a perturbation theory --- hence, it will suffice to consider only the $B$-dependent corrections to $M$. This is because, considering $B$-dependent corrections to the static susceptibility matrices will generate a higher-order correction. However, within the memory matrix formalism, any $B$-dependent correction to a parity-even component like $M_{P_p P_q}$ must be $\order{B^2}$. Consequently, the only matrix which may admit an $\order{B} $ correction is $N$. Therefore, we need compute the consequences of a nonzero $B$ on $N$ only.

For nonzero $ B$, the zero-momentum component of the canonical momentum operator $ \boldsymbol{\mathcal{P}}_B $, which generates translations, is no longer equivalent to the physical momentum operator. Instead, it is given by
\begin{align}
\boldsymbol{\mathcal{P}}^{(B)} =
\mathbf{P} + \int d^d \mathbf{r}\,n_c\, \mathbf{A}_B  \,,
\end{align}
where $ \mathbf{A}_B $ is a vector potential corresponding to the applied magnetic field $ \mathbf B $. Although $\mathbf{P}$ is gauge-invariant, the second term is not, since the form of $  \mathbf{A}_B $ depends on the choice of gauge. The effective Hamiltonian also needs to be modified as $ H^{(0)} \rightarrow   H^{(0)}  + H_B \,,$ where
\begin{align}
H_B = -\int d^d \mathbf{r} \,\mathbf{J} \cdot \mathbf{A}_B  \,.
\end{align}

For demonstration purposes, let us choose to consider a two-dimensional system (i.e., we set $d=2 $). In order to evaluate $\dot{P}_x$, for example, a convenient gauge choice is $\mathbf{A}_B = -B\, y \,\hat{\mathbf{x}} $. This leads to
\begin{align}
\dot{P}_x &= \mathrm{i}
\left [ H^{(0)} + H_B, \,P_x \right ]  
\nn & = 
i \left[  H^{(0)} + H_B, \,\mathcal{P}^{(B)}_x + 
\frac{B} {2}
\int d^2 \mathbf{r}\; n_c \, y\right] 
\nn
& = i \left[ H_B, \mathcal{P}^{(B)}_x \right ] 
 +  B \int {d}^2 \mathbf{r} \; \dot{n}_c \,y + \order{B^2} \nn
&= B \int {d}^2 \mathbf{r} \left( \partial_{x} J_x + \dot{n}_c \right)  y 
\nn &  = 
- B \int {d}^2\mathbf{r} \; \partial_y J_y \,y 
 = B \int {d}^2 \mathbf{r}\;  J_y \,.
\end{align}
A similar argument works for $\dot{P}_y $ with the gauge choice $\mathbf{A}_B = B \, x \,\hat{\mathbf{y}}$.
This leads to the expressions for the nonzero elements of $N$ to be
\begin{align}
N_{P_x P_y} &= -N_{P_yP_x} = \chi_{P_x\dot{P}_y} = -B \,\chi_{J_x P_x} \,.
\end{align}
With these ingredients, we obtain
\begin{align}
& M(\omega) + N-i \,\omega \, \chi \nn
& =
\begin{pmatrix}
M_{P_x P_x} - i\,\omega \,\chi_{P_x P_x} & -B \, \chi_{J_x P_x} \\
B \, \chi_{J_x P_x} & M_{P_y P_y} - i\,\omega \,\chi_{P_y P_y}
\end{pmatrix} + \order{B^2}
\nn & =
\begin{pmatrix}
M_{P_x P_x} - i\,\omega \,\chi_{P_x P_x} & -B \, \chi_{J_x P_x} \\
B \, \chi_{J_x P_x} & M_{P_x P_x} - i\,\omega \,\chi_{P_x P_x}
\end{pmatrix} \,.
\end{align}
We can use this expression to evaluate the forms of the tensors $\sigma$, $\alpha$, and $\bar \kappa $.
In particular, the longitudinal and transverse components of the DC electrical conductivity tensor are given by
\begin{align}
& \sigma_{xx}^{\rm dc}   =
\frac{ \chi_{J_z P_z}^2
\, M_{P_x P_x} } 
{M_{P_x P_x }^2 + B^2 \,\chi_{J_z P_z}^2} \, \text{ and }
\nn  & \sigma_{xy} 
= - \, \sigma_{yx}
=\frac{B \, \chi_{J_x P_x}^3} 
{M_{P_x P_x}^2 + B^2 \, \chi_{J_x P_x}^2} \,,
\end{align}
respectively.

\subsection{Generic scaling arguments}

Hyperscaling is the property that the free energy scales by its naive dimension \cite{hyperscaling, subir_book}. In a quantum system in $d$ spatial dimensions, with dynamical critical exponent $ \mathfrak{z} $, the free energy $F$ has the scaling dimension
\begin{align}
[F] =d+ \mathfrak{z}\,. 
\end{align}
Therefore, if hyperscaling is not violated, we should have the temperature-dependence
\begin{align}
F  \sim T^{d/ \mathfrak{z} +  1}\,.
\label{eqfscale}
\end{align}

The spatial components of the stress-energy tensor $\mathcal T_{ p q}$ have the same scaling dimension as the Lagrangian density and, hence,
\begin{align}
[T_{zx}]=d+ \mathfrak{z}\,.
\end{align}
Using the definition of
\begin{align}
\label{eqkubosig1}
\eta & \sim 
 \tilde  G^R_{ \mathcal T _{p q } \mathcal T _{p q }} (\omega) \,,
\end{align}
where $\eta $ is the shear viscosity, this implies that
\begin{align}
[\eta] &= 2\,[T_{zx}] - \mathfrak{z}- [\text{Volume in energy-momentum space}]
\nn & 
= 2\left( d+z \right)-\mathfrak{z} -d-\mathfrak{z} = d \,.
\label{eqeta}
\end{align}
This shows that $\eta$ has the same scaling dimension as the entropy
density $s$, which, by definition, is the derivative of the free energy with respect to $T$. Hence, the ratio $\eta/s$ turns out to be dimensionless.
If hyperscaling is not violated, Eq.~\eqref{eqeta} also implies the scaling form
\begin{align}
\label{eqet1}
\eta (\omega) \sim \omega^{ d/ \mathfrak{z}} 
\end{align}
for optical viscosity (i.e., for $ \omega \gg T $), and 
\begin{align}
\label{eqet2}
\eta^{\rm dc} (T) \sim T^{ d/\mathfrak{z}} 
\end{align}
for DC viscosity (i.e., for $T\gg \omega $).

In view of Eqs.~\eqref{eqcur1} and \eqref{eqkubosig}, the scaling dimension for the conductivity tensor is
\cite{Sachdev-MIT}
\begin{align}
[\sigma_{pq}] &= 2\,[\mathbf J] - \mathfrak{z} - [\text{Volume in energy-momentum space}]\nonumber\\
&= 2\left( d+ \mathfrak{z} -1\right)- \mathfrak{z}-d- \mathfrak{z} =d-2\,,
\end{align}
which implies the scaling form
\begin{align}
\sigma_{p q} (\omega) \sim \omega^{(d-2)/\mathfrak{z}} 
\end{align}
for the optical conductivity. In scenarios when the hyperscaling is violated (see, for example, Ref.~\cite{ips-subir} for an NFL with a finite Fermi surface), the above scaling form is modified to
\begin{align}
\sigma_{p q} (\omega) \sim \omega^{(d-2-\theta)/\mathfrak{z}} \,,
\end{align}
where $\theta$ represents a hyperscaling-violating exponent \cite{liza_brian}.
In Ref.~\cite{ips_hermann1}, our results for the optical conductivity computed in the LAB phase show hyperscaling violation.

\section{Transport properties of the LAB phase}
\label{secmodel}

\begin{figure}
\includegraphics[width=0.25 \textwidth]{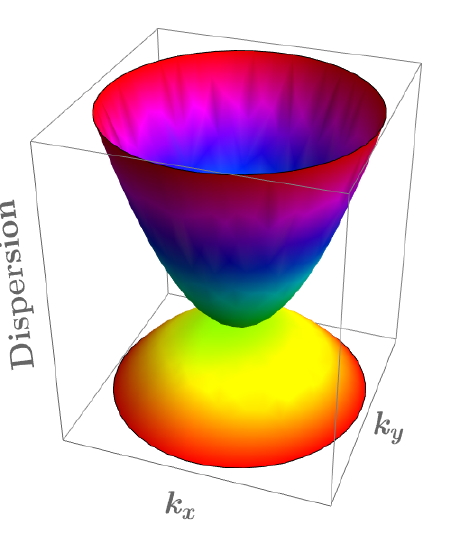}
\caption{Schematics of the low-energy effective dispersion of an isotropic, but band-mass-asymmetric, Luttinger semimetal that harbours a doubly-degenerate quadratic band-touching at the $\Gamma$-point.
\label{fig:bands}}
\end{figure}

As an example of an NFL arising at a Fermi point, we consider the 3d
Luttinger semimetals \cite{Luttinger, moon-xu, rahul-sid, ips-rahul, lukas-herbut, igor16, ips-qbt-sc, ips_qbt_plasmons, ips_qbt_tunnel, Roy_PRB, Herbut-PRB, krempa, polini, Boettcher_2019, ips-sandip, ips_hermann1,ips-hermann2, ips-hermann3, ips-jing}, as discussed in the introduction. In this review, we will focus on isotropic, but band-mass-asymmetric, Luttinger semimetals, and show how various transport properties can be computed in the LAB phase by applying the techniques reviewed in the previous sections.

\subsection{Minimal low-energy effective model}
\label{Eff_model}

The effective low-energy continuum Hamiltonian, in the vicinity of the nodal point of an isotropic band-mass asymmetric Luttinger semimetal, is given by
\begin{align}
\label{bare2}
 \mathcal{H}_0 = \frac{ |\mathbf k |^2} {2 \, m'}
 -\frac{\frac{5}{4}k^2-(\mathbf{k}\cdot \boldsymbol{\mathcal{J}})^2} {2 \,m}\,,
\end{align}
where $\boldsymbol {\mathcal{J}}$ represents the vector angular-momentum operator in the spin-3/2 representation of the SO(3) group. Therefore, the Hamiltonian represents a system of noninteracting pseudospin-$3/2$ quasiparticles. 
The energy eigenvalues evaluate to
\begin{align}
\epsilon_\pm (\mathbf k) = \frac{|\mathbf{k}|^2}{2 \,m'} \pm \frac{|\mathbf{k}|^2}{2 \,m}\,,
\end{align}
where the ``+'' and ``-'' signs refer to the conduction and valence bands, respectively, which are doubly-degenerate.
Fig.~\ref{fig:bands} shows the schematics of the doubly-degenerate dispersion. The symbols $m$ and $m'$ represent the mass parameters of the quadratically dispersing bands. Since the term $\frac{ |\mathbf{k}|^2} {2\,m'}$ multiplies an identity matrix, it causes the effective band masses of the conduction and valence bands to be asymmetric/unequal.

Following Refs. \cite{Murakami, rahul-sid, ips-rahul, ips_qbt_plasmons}, the Hamitonian in Eq.~\eqref{bare2}
can be brought to the form
\begin{align}
\label{eqbare}
 \mathcal{H}_0 = \sum_{a=1}^5 d_a(\mathbf{k}) \,  \Gamma_a   
 + \frac{ |\mathbf k|^2} {2\,m'} \,,
 \end{align} 
where the set of five $\Gamma_a$-matrices forms a rank-four irreducible representation of the Euclidean Clifford algebra. Therefore, they obey the anticommutation relation $\{\Gamma_a,\Gamma_b\} = 2\, \delta_{ab}$. They can always be chosen such that three are real and two are imaginary \cite{herbut_085304}. 

The five $d_a(\mathbf k)$-functions are the real $\ell =2$ spherical harmonics with the following structures~\cite{Murakami,rahul-sid, ips-rahul}:
\begin{align}
\label{ddef}
& d_1(\mathbf{k}) = \frac{\sqrt{3}\, k_y \,k_z}{2m}\,,
\quad  d_2(\mathbf{k}) =  \frac{\sqrt{3}\, k_x\, k_z}{2 \,m}\, ,\quad
 d_3(\mathbf{k}) =  \frac{\sqrt{3} \,k_x\, k_y}{2 \,m}\, \quad\nonumber\\
 &d_4(\mathbf{k}) =\frac{\sqrt{3}  \,  (k_x^2 - k_y^2) }{4 \, m}\,, \quad
 d_5(\mathbf{k}) = \frac{2\, k_z^2 - k_x^2 - k_y^2}{4 \, m} \,.
\end{align}
Henceforth, we will use the notation 
\begin{align}
\mathbf{d}_{\mathbf{k}}
\equiv \left[ d_{1}(\mathbf{k}),
\, d_{2}(\mathbf{k}), \, d_{3}(\mathbf{k}), \, d_{4}(\mathbf{k}), \,d_{5}(\mathbf{k})
\right ]
\end{align}
to refer to the five-component vector consisting of the five $d_a(\mathbf k)$-functions. 
Analogously, $\mathbf \Gamma $ will denote the vector whose components are the five $\Gamma_a$-matrices.

Adding the Coulomb interactions via a non-dynamical scalar boson field $\varphi$, the Euclidean action of the resulting interacting system can be straightforwardly written as
\begin{align}
S_{\rm int} & =  \int d\tau \,d^3{\mathbf r}
\Big[  
\psi^{\dag} (\tau, \mathbf r)
 \left \lbrace  \partial_{\tau} + \mathcal{H}_0 + i \, e 
 \,\varphi \right \rbrace
  \psi (\tau, \mathbf r)
 \nn &
\hspace{ 2 cm } 
+ \frac{c}{2} \left \lbrace  \nabla \varphi  (\tau, \mathbf r) \right \rbrace^2  \Big ] \,.
\end{align}
Here, $\psi$ denotes the fermionic field, which is a four-component spinor, and $c$ is a constant equal to $1/(4 \,\pi)$.

\begin{figure}
\includegraphics[width=0.35 \textwidth]{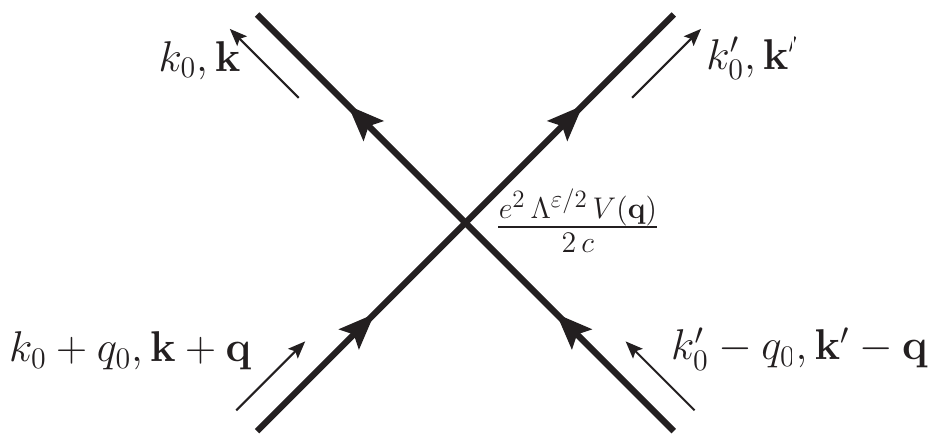}
\caption{The four-fermion vertex arising due to Coulomb interactions.
\label{figvert1}}
\end{figure}

We now extend the theory to a generic number of spatial dimensions $d$, to make it possible to apply dimensional regularization. Finally, we integrate out $\varphi$ to express the Coulomb interaction as an effective four-fermion interaction vertex, resulting in the form 
\begin{widetext}
\begin{align}
S & = 
\int \frac{d\tau \,d^d {\mathbf k} } {(2\,\pi)^d}\, 
\psi^{\dag}(\tau,\mathbf k)
\left(\partial_{\tau} + \mathcal{H}_0 \right)
 \psi  (\tau, \mathbf k) \nonumber\\
& \quad + \frac{e^2\,\Lambda^{\varepsilon}}{2 \,c }
\int \frac{d\tau\,
d^d {\mathbf k}\, d^d {\mathbf k'}\,d^d {\mathbf q} }
{(2 \, \pi)^{3d}}\, 
V(\mathbf q)\,{\psi}^{\dag}(\tau,{\mathbf k}+\mathbf q)
\,\psi (\tau,{\mathbf k})\,
 {\psi}^{\dag} ( \tau ,{\mathbf k}'-\mathbf q)\, 
\psi (\tau,{\mathbf k}')   
\nn & = 
\int \frac{d k_0 \,d^d {\mathbf k} } {(2\,\pi)^{d+1}}\, 
{\tilde \psi}^{\dag}( k_0 ,\mathbf k)
\left( - i  \,k_0 + \mathcal{H}_0 \right)
 {\tilde \psi}  ( k_0, \mathbf k) 
\nn & 
\quad + \frac{e^2\,\Lambda^{\varepsilon/2}}{2 \,c } 
\int \frac{d q_0\, d k_0 \,dk_0'\,
d^d {\mathbf q}\, d^d {\mathbf k}\, d^d {\mathbf k'}}
{(2\pi)^{3(d+1)}}\, 
V(\mathbf q)\,\tilde{\psi}^{\dag} ( k_0 ,\mathbf k)\,
\tilde{\psi}^{\dag}( k_0' ,{\mathbf k}')\, 
\tilde{\psi} (k_0 + q_0 ,{\mathbf k}+\mathbf q) \,
\tilde{\psi} ( k_0'-q_0,{\mathbf k}'-\mathbf q)\,,
\label{fullaction}
\end{align}
\end{widetext}
where $V(\mathbf q) = 1/|\mathbf q|^2$. In the momentum space, the Coulomb interaction vertex is given by
$\frac{e^2\, \Lambda^{\varepsilon}}{2 \,c }V(|\mathbf q|)$, and  ${\tilde \psi}^{\dagger} $ and ${\tilde \psi}$ represent Fourier-transformed fermionic fields. The four-fermion vertex is depicted schematically in Fig.~\ref{figvert1}.
We have also scaled $ {e^2}/{c}$ by using the floating mass scale $\Lambda$ (of the RG flow) so that its engineering dimension vanishes at $d=4-\varepsilon $.
Here, we have determined the tree-level scaling dimensions of the fields and couplings by setting the scaling dimension of $\mathbf k$ as $[\mathbf k] = 1$. Hence, the various engineering dimensions are given by: $[\tau]=- \mathfrak{z} =-2$ (where $ \mathfrak{z} $ is the dynamical critical exponent), $[1/m] = [1/{m'}] = \mathfrak{z} -2 $, and $[e^2] = 2\, \mathfrak{z} -d $ (before using the scaling factor $\Lambda^\varepsilon$).

\begin{figure*}
	\centering
	\subfigure[]{\includegraphics[width=0.25 \textwidth]{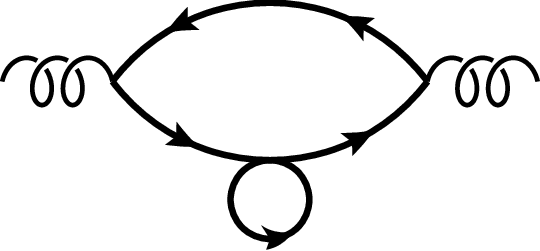} \label{fig2}} \quad
	\subfigure[]{\includegraphics[width=0.25 \textwidth]{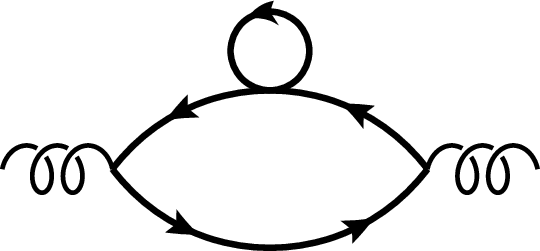} \label{fig3}} \quad
	\subfigure[]{\includegraphics[width=0.33 \textwidth]{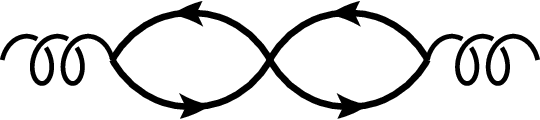} \label{fig4}}
	\caption{\label{fig2loop} Representation of the Feynman diagrams for the current-current correlator
at two-loop order. Subfigures (a) and (b) denote the diagrams with fermionic self-energy insertions, while subfigure (c) represents the diagram with the insertion of a vertex correction.
}.
\end{figure*}

The bare Green's function for each fermionic flavour, obtained from $\mathcal H_0$, is given by
\begin{align}
G_0(k_0, \mathbf{k}) =  
\frac{ \mathrm{i}\, k_0 - \frac{ |\mathbf k|^2}{2\,m'}  
+ \mathbf{d}_{\mathbf{k}} \cdot{\mathbf{\Gamma}}}
{-\left ( \mathrm{i}\, k_0 - \frac{ |\mathbf k| ^2}{2\, m'}  \right )^2 
+ \left |\mathbf{d}_{\mathbf{k}} \right |^2}\,,
\label{baregf}
\end{align}
where $ \left |\mathbf{d}_{\mathbf{k}} \right | ^2 $ evaluates to $ { |\mathbf k|^4} /( 4\,m^2)$.
From the one-loop fermionic self-energy, the upper critical dimension of the interacting system [described by Eq.~\eqref{fullaction}] turns out to be $d_c = 4$ . When we derive the RG flow equations of the model, a stable NFL fixed point in the infrared (which is the LAB fixed point) is found to exist for $d = 4 -\varepsilon$ (analogous to the Wilson-Fisher fixed point of the bosonic $\phi^4$-theory in $4-\varepsilon$ dimensions \cite{Wilson_Fisher}). As a result, we can extract the critical scalings of the system by a controlled approximation using an $\varepsilon $-expansion about $d_c$ \cite{Abrikosov,moon-xu}.
At the LAB fixed point, the coupling constant $e$ takes the value
\begin{align}
{e^*}^2= \frac{60\,\pi^2\,c\,\varepsilon} 
{ 19 \,m} \,.
\end{align}
The fixed-point value of the dynamical critical exponent $ \mathfrak{z} $ is given by $ \mathfrak{z}^*=2-4\,\varepsilon/ 19 $ \cite{moon-xu}. It is to be noted that the results obtained using dimensional regularization can also be obtained by large-$N$ methods.

Employing the Noether's theorem \cite{Peskin}, the current ($\mathbf{J}$) and the momentum ($\mathbf{P}$) operators, associated with the invariance of the action of Eq.~\eqref{fullaction} under the global $U(1)$ symmetry and the continuous spatial translations, are given by \cite{ips_hermann1}
\begin{align}
\label {eqjz}
& \mathbf{J}(q_0,\mathbf q)
\nn &= 
\int\frac{ dk_0 \,d^d\mathbf{k}}
{(2\,\pi)^{d+1}}\,
\tilde{\psi}^{\dagger}(k_0+q_0, \mathbf{k} +\mathbf q)
\left [\nabla_{\mathbf{k}}\mathbf{d(k)}\cdot\mathbf{\Gamma} \right ]
\tilde{\psi} (k_0,\mathbf{k})
\end{align}
and
\begin{align}
& \mathbf{P} (q_0,\mathbf{q}) \nn
&= \int\frac{ dk_0 \,d^d\mathbf{k}}
{(2\,\pi)^{d+1}}\,
\left(  \mathbf{k} +\mathbf{q}/2 \right)
\tilde{\psi} ^{\dagger}(k_0+q_0,\mathbf{k}+\mathbf q)\,
\tilde{\psi} (k_0,\mathbf{k})\,,
\end{align}
respectively.
Analogously, the thermal current operator can be expressed as \cite{ips-hermann2}
\begin{widetext}
\begin{align}
\mathbf{J}^Q{(t,\mathbf{q})} &=\frac{1}{2}
\int\frac{d^d \mathbf k}{(2\pi)^d }
\left[\partial_t{\psi}^{\dagger}(t,\mathbf{k+q})
\left(\nabla \mathbf{d}_\mathbf{k}\cdot\mathbf{\Gamma}+\frac{\mathbf{k}}{m'}\right) \psi(t,\mathbf{k})\right.
+{\psi}^{\dagger}(t,\mathbf{k+q})\left(\nabla \mathbf{d}_\mathbf{k}\cdot\mathbf{\Gamma}
+ \frac{\mathbf{k}}{m'}\right)\partial_t{\psi} (t,\mathbf{k})\bigg]\,,
\end{align}
\end{widetext}
where $\partial_t{\psi} \equiv \mathrm{i}\,[ \mathcal H_0,\psi]$.

\subsection{Optical electrical conductivity}
\label{secoptical2}

Using the Kubo formula derived in Sec.~\ref{seckubo} [cf. Eq.~\eqref{eqkubosig}], each longitudinal component of the isotropic optical conductivity tensor at $ T=0$ is equal to \cite{ips_hermann1}
\begin{align}
\sigma_{zz}(\omega,T) = -\frac{\left \langle J_z \,J_z \right \rangle( k_0)}
{k_0} 
\bigg|_{ i \, k_0 \rightarrow \omega +  i \,0^+}\,,
\label{eq:Kubo}
\end{align}
with $J_z$ given by Eq.~\eqref {eqjz}. Here, the symbol $\left \langle \cdots \right \rangle$ denotes a correlator evaluated using the action in the Matsubara frequency space [in particular, using the second line of Eq.~\eqref{fullaction}].
By calculating the correlator $ \left \langle J_z \,J_z \right \rangle $ in $d=4-\varepsilon$ up to two-loop order (cf. Fig.~\ref{fig2loop}), we obtain \cite{ips_hermann1}
\begin{align}
\sigma_{zz} (\omega)\sim \omega^{1-\frac{\varepsilon }{2}
+ \frac{ 5 \,\varepsilon } {114}  } .
\end{align}
This scaling dependence differs from the form $\omega^{(d-2)/ \mathfrak{z} }$, which indicates hyperscaling violation, similar to the NFLs arising at hot-spots of a finite Fermi surface in a 2d fermion-boson system \cite{lee_prx}.

\subsection{DC electrical conductivity}

To find the DC electrical conductivity, we resort to the memory matrix formalism of Sec.~\ref{secmem}, as we are then considering the limit $T\gg \omega $. 
We first calculate $\chi_{J_z P_z}$  and $M_{P_z P_z}(0)$, appearing in Eq.~\eqref{eqconddc}.
For this calculation, we work directly in $d=3$, as the memory matrix approach inherently takes care of the NFL behaviour of the system, without the need for a controlled expansion (e.g., by using small-$\varepsilon$ or large-$N$).
Since we are required to provide a mechanism of momentum relaxation via adding a weak disorder, it is not valid in the $T\rightarrow 0$ limit. This is because the impurity term in Eq.~\eqref{eqdisorder} is a relevant perturbation \cite{rahul-sid, ips-rahul}, when we add it to the action described in Eq.~\eqref{fullaction}. Consequently, the assumption of weak disorder is valid only in the $T \gg \omega $ limit, where the temperature plays the role of a cutoff in the RG flows (thus preventing the disorder strength from flowing to nonperturbative values), and the perturbative evaluation of the memory matrix up to $\order{W_0^2}$ stands on a firm ground. 

While computing the momentum integrals, we are required to impose an ultraviolet (UV) cutoff $\Lambda_0$ at the outset, which is physical, since there is a natural lattice cutoff for any solid state system.
By taking into account the diagrams with self-energy feedback and vertex corrections (see Fig. \ref{fig2loop}), and evaluting the corresponding integrals numerically, we have obtained
$\sigma_{zz}^{\rm dc} \sim T^{n_\sigma } $, where $2\lesssim  n_\sigma \lesssim 4$ \cite{ips_hermann1}.
We note that this prediction compares well with the experimental data \cite{Pramanik} for (Y$_{1-x}$Pr$_x$)$_2$Ir$_2$O$_7$,
which predicts the same scaling relation, with the value $n_\sigma \approx 2.98$ at zero doping (i.e., when the chemical potential cuts the nodal point).

\subsection{Thermal and thermoelectric response}
\label{thermal}

We apply the memory matrix formalism to calculate the thermal and thermoelectrical response, up to two-loop order, by taking into account Feynman diagrams similar to those shown in Fig. \ref{fig2loop}. The final results are found to be \cite{ips-hermann2}
\begin{align}
{\bar{\kappa}}^{\rm dc} & 
\sim T^{-n_\kappa }, \text{ where } 0\lesssim n_\kappa \lesssim 1\,, \\
\alpha^{\rm dc} & \sim 
T^{n_\alpha }\,, \text{ where } 1/2\lesssim n_\alpha \lesssim 3/2\,.
\label{thermoelectric}
\end{align}
For $T < 1$, the main effect of the Coulomb interactions is to increase(decrease) the value of the exponent $n_\kappa $($n_\sigma$), thereby enhancing both $ {\bar{\kappa}}^{\rm dc} $ and $\alpha^{\rm dc}$.
Since the experimentally relevant thermal conductivity is given by $\kappa$, we compute it using Eq.~\eqref{eqkappa}. The result shows
that $\kappa$ vanishes at leading order \cite{ips-hermann2}. Consequently, in this system,
the thermal conductivity at zero electric current could potentially be dominated by the phonon contribution, even below the Debye temperature.

Regarding the thermoelectric properties of the LAB phase, its efficiency is better captured in terms of either the Seebeck coefficient $ \mathcal S = \left( \sigma^{\rm dc} \right)^{-1}  \alpha^{\rm dc}$, or the thermoelectric figure of merit $  \mathcal S^2\,\sigma^{\rm dc}\, T \left(\kappa^{\rm dc} \right)^{-1} $.
Analyzing their scaling forms, we have found that they can be quite high \cite{ips-hermann2}, which suggests that the Luttinger semimetals might be extremely useful for thermoelectric applications.

\subsection{Raman response}
\label{raman}

\begin{figure}[t]
	\centering
	\subfigure[]{\includegraphics[width=0.15\textwidth]{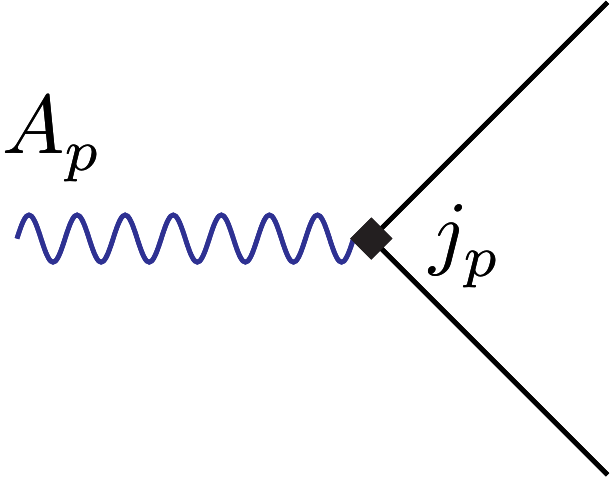} \label{figtree1}}\hspace{4 cm}
\subfigure[]{\includegraphics[width=0.12 \textwidth]{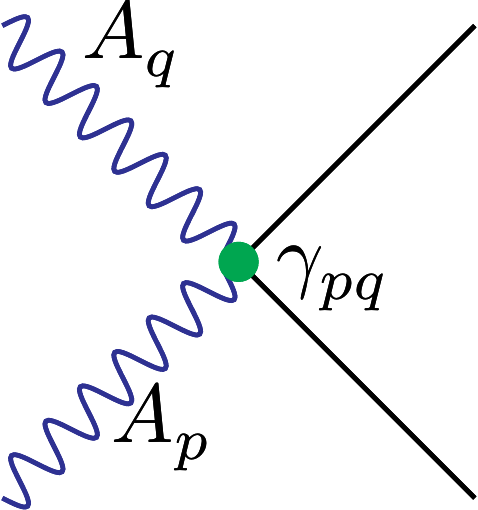} \label{figtree2}}
	\caption{\label{figvert}
Electron-photon coupling vertices for the Raman response. The blue curly lines represent the incident and scattered photons. (a) The first type of vertex represents coupling the electron’s current to a single photon, and is denoted with a black square. (b) The second type of vertex represents coupling the electron’s charge to two photons, and is denoted by a green dot.}
\end{figure}

\begin{figure*}[]
	\centering
\includegraphics[width=0.25\textwidth]{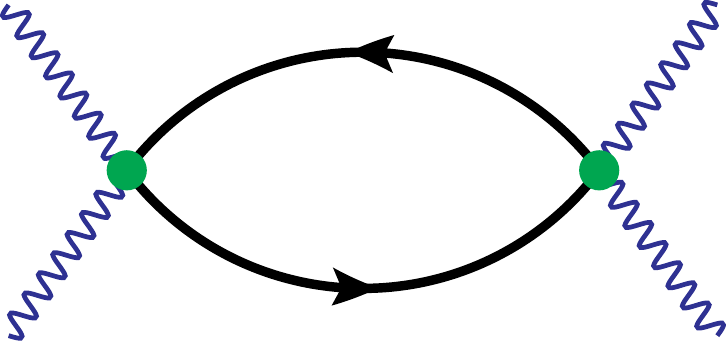}
\caption{Feynman diagram for the contribution to the Raman response at one-loop order.
\label{fig1loopraman}}
\end{figure*}

\begin{figure*}[]
	\centering
	\subfigure[]{\includegraphics[width=0.25\textwidth]{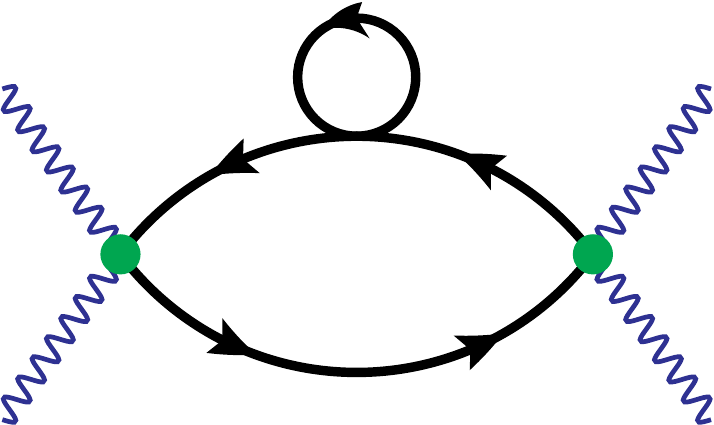} \label{figr2}}\hspace{1 cm}
	\subfigure[]{\includegraphics[width=0.25\textwidth]{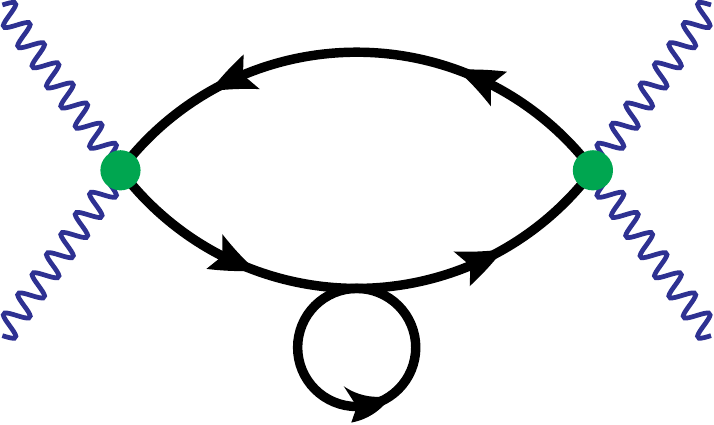} \label{figr3}}\hspace{1 cm}
	\subfigure[]{\includegraphics[width=0.25\textwidth]{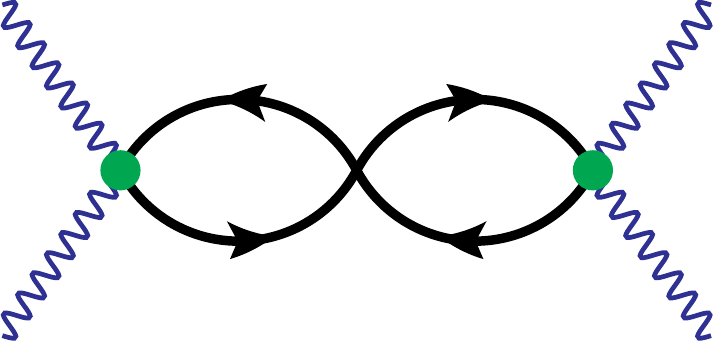} \label{figr4}}
	\caption{\label{fig2loopraman}Feynman diagrams for the contributions to the Raman response at two-loop order, with subfigures (a) and (b) representing the self-energy-induced corrections, and subfigure (c) depicting the vertex-induced correction.}
\end{figure*}

Raman scattering is another fundamental tool that provides valuable information about the dynamics of the system \cite{Bozovic,Cooper_Slakey,Staufer,Slakey}. A comprehensive review about this experimental technique can be found in Ref. \cite{Devereaux_RMP}. Raman experiments involve the coupling of the electrons to an electromagnetic field representing the
incoming and outgoing photons. This is incorporated by adding a gauge coupling via the Peierls substitution
$\bk \rightarrow \bk -\mathfrak{q}\,\bA/c$, where $\bA$ is the vector potential. In the resulting Hamiltonian, the terms depending on the vector potential are given by \cite{ips-hermann3}
\begin{align}
\label{Raman_coupling}
  \mathcal H_{\mathbf{A}} & 
= -\frac{\mathfrak{q}} {c} \,A_p \, j_p(\bk) 
 + \frac{ \mathfrak{q}^2 }{2\, {c^2}}\,A_p \,A_q \, \gamma_{p q}(\bk)\,,
\end{align}
where $ j_p =\frac{k_p} {m'} 
+ \partial_{k^p} d_a(\bk)\,  \Gamma_a $
represents the $p^{\rm th}$-component of the current operator, and
\begin{align}
& \gamma_{xx} =\frac{\Gamma_0}{m'} +\frac{\sqrt{3}\,\Gamma_4-\Gamma_5}{2\,m}\,,\quad
\gamma_{yy} = \frac{\Gamma_0}{m'} - \frac{\sqrt{3}\, \Gamma_4+\Gamma_5}{2\,m}\,, 
\nn &
\gamma_{zz} = \frac{\Gamma_0}{m'} +\frac{\Gamma_5}{m}\,,
\quad 
\gamma_{xy}= \gamma_{yx}= \frac{\sqrt{3}\, \Gamma_3}{m}\,,
\nn & 
\gamma_{yz}= \gamma_{zy}= \frac{\sqrt{3}\, \Gamma_1}{m}\,,\quad
\gamma_{xz}= \gamma_{zx}= \frac{\sqrt{3} \,\Gamma_2}{m} \,.
\end{align}
We note that Eq.~\eqref{Raman_coupling} gives rise to two types of electron-photon coupling vertices.
While calculating the loop diagrams, we find that the contributing vertices can only be either $(\Gamma_0,\Gamma_0)$ or $(\Gamma_a,\Gamma_b)$ for $a,b \in \lbrace 1, \cdots, 5 \rbrace $, since the remaining cross-terms turn out to vanish identically \cite{ips-hermann3}.

Quantizing the vector potential, one obtains
\begin{align}
\mathbf{A}(\mathbf{k})= \sqrt{\frac{c^2}
{\omega_{\mathbf{k}}\,V}}
\left (\hat{\mathbf{e}}_{\mathbf{k}} 
\, a_{-\mathbf{k}}+\hat{\mathbf{e}}^*_{\mathbf{k}} \,a^{\dagger}_{\mathbf{k}} \right ),
\end{align}
where $V$ is the volume. The operators
$a_{\mathbf q}^\dagger$ and $a_{\mathbf q} $ are the creation and annihilation operators, respectively, of the photons with the dispersion relation $\omega_{\mathbf k} = c\,|\mathbf k|$, and having a polarization direction defined by $\hat{\mathbf e}_{\mathbf k}$.

The Raman scattering cross-section within the Born approximation is given by
\begin{align}
    \frac{d^2\sigma}
 {d\Omega \,d\omega_{\rm in} }
  \propto \sum_{F,I} \frac{\exp(-\beta E_I)}{Z}\, 
 | \mathcal M_{FI}|^2\,\delta(E_F+\omega_{\rm fi}-E_I-\omega_{\rm in})\,,
\end{align}
where $I$ and $F$ represent the initial and final states, respectively, of the Luttinger semimetal, and
$Z$ is the (canonical) partition function. Furthermore, $ \mathcal M_{FI} =\langle F \,|\,\mathcal M\,|\,I \rangle $, where $\mathcal M $ is the effective light-scattering operator. The summation over $F$ and $I$ stands for a thermodynamic average over all possible initial and over final states of the system, possessing energies $E_I$ and $E_F$, respectively,
with the momentum vectors inside the solid angle element $d\Omega$. 
Moreover, $\omega =\omega_{\rm in}-\omega_{\rm fi} $ is the frequency and $\mathbf q$ is the momentum transferred by the photons.

We define the operators
${\rho}_0=\psi^{\dagger}\,\psi$ and ${\rho}_a=\psi^{\dagger}\,\Gamma_a\psi$,
since their two-point correlators will contribute to the Raman response $|\mathcal M_{FI}|^2$.
If we consider scattering in the visible range, a good approximation for the relevance of such a quantity is the zero-momentum limit. Hence, we will focus on expressions for the correlators $\left \langle \rho_0\,\rho_0 \right \rangle (k_0)$ and $\left \langle \rho_a\,\rho_b \right \rangle (k_0)$ in the zero-momentum limit, restricting up to the two-loop order.
The Feynman diagrams for computing $ |\mathcal M_{FI}|^2$ involve vertices of two types, as depicted in Fig.~\ref{figvert}. However, only diagrams consisting solely of green vertices involve non-resonant
scatterings, while the others give rise to resonant and mixed scatterings which can be neglected in the low-energy limit \cite{Devereaux_RMP}. Consequently, we consider here only the leading-order Feynman diagrams for $|\mathcal M_{FI}|^2$ contributed by the non-resonant scatterings.

In the $T=0 $ limit, we employ the $\varepsilon $-expansion at the NFL fixed point to evaluate the relevant correlators. 
Figs.~\ref{fig1loopraman} and \ref{fig2loopraman} show the Feynman diagrams at the one-loop and two-loop orders, respectively.
The final results come out to be \cite{ips-hermann3}:
\begin{align}
\langle \rho_a\,\rho_b \rangle( \omega) & =0 \,,\nn
\langle \rho_a\,\rho_b \rangle( \omega)
& \simeq
-\frac{  m^{2-\frac{\varepsilon }{2}}\, 
|\omega|^{1-\frac{\varepsilon }{2}+\frac{ \varepsilon }
{ 38}}
\,\delta_{ ab } }{10\, \pi }
\left(  \frac{m}{\Lambda^2}\right)^{ \frac{ \varepsilon }
{ 38 }}\,.
\end{align}
This shows that the Raman response in the LAB phase should scale as $|\omega|^{1-\varepsilon /2
+ \varepsilon / 38 }
\overset{\varepsilon=1}  =
|\omega|^{1/2 + 1/38 } $, in the regime where $\omega\gg T$.

Next, we review our results illustrating the behavior obtained for the Raman response in the $T\gg \omega$ regime. For this part, analogous to the calculations for the generalized conductivity tensors, we use the memory matrix formalism. 
As explained in Sec.~\ref{secmem}, the existence of a nearly-conserved operator is essential for applying the memory matrix formalism, which we have taken to be the momentum so far. However, the momentum operator turns out to have no influence on the Raman response \cite{wang2020low}, because it has no overlap with the operators $\lbrace \rho_a \rbrace $. For this reason,
we use here $\rho_a$ itself for computing the Raman response, since it is also a nearly-conserved operator in the presence of weak disorder.
Therefore, we set $\mathcal{O} = \rho_a $ in Eq.~\eqref{eqimpO}, and
the time evolution of $ \rho_a $ is given by $ {\dot{\rho}}_a   = 
i \left [ \mathcal H_0 +  H_{\textrm{imp}} ,  \,\rho_a \right ] $.

\begin{figure}[t]
\includegraphics[width=0.45\textwidth]{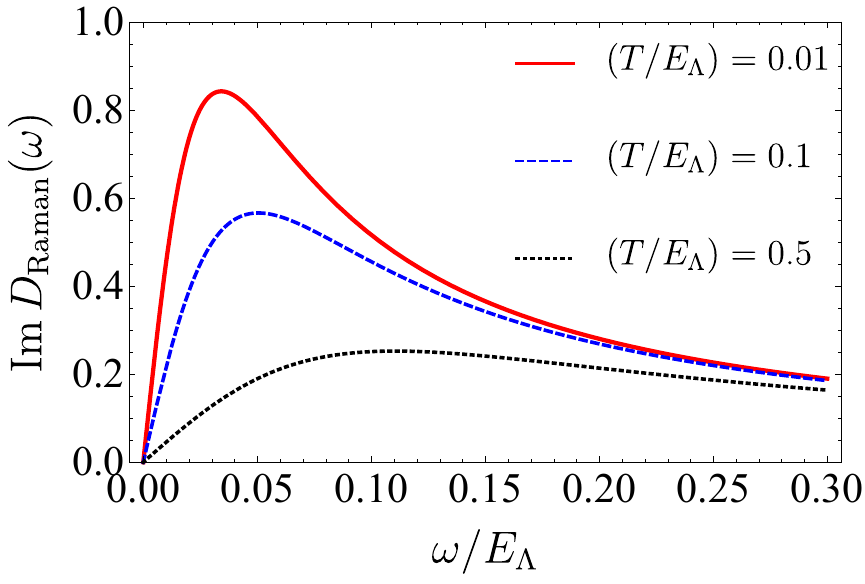}
\caption{\label{raman_fig}
Raman response as a function of frequency $\omega$, for some representative values of the temperature $T$ (as shown in the plotlegends). The chosen parameter values for numerical evaluation are $\Lambda_0=10$, $m=1$, $m'=2$, $e^2/c=1$, and $W_0=1$. The parameters $T$ and $\omega$ are in units of $E_{\Lambda} = \Lambda_0^2/(2\,m)$, which is the ultraviolet energy cutoff used. $E_{\Lambda}$ is of the order of the energy scale, up to which the dispersion of the conduction and valence bands can be taken to be quadratic (valid only in the vicinity of the nodal point of the Lutiinger semimetal).
} 
\end{figure}

The Raman response at $T>0$ is defined as \cite{Berg-PRB, wang2020low}
\begin{align}
D_{\text{Raman}}(\omega,T)
= i\, \omega\, \tilde \Upsilon_{\rho_a \rho_b} (\omega) \,,
\end{align}
which can be approximated as
\begin{align}
\label{D_Raman}
D_{\text{Raman}}(\omega,T)\approx \chi_{\rho_a\rho_b}
\, \frac{ i \,\omega}
{{M_{\rho_b\rho_b}(\omega})- i \,\omega
\,\chi_{\rho_b\rho_b}}
\,  \chi_{\rho_b \rho_a}\,.
\end{align}
In this case, the memory matrix has the components
\begin{align}
 M_{\rho_a\rho_b}(\omega) 
\approx
{\text{Im}\,\tilde G^R_{\dot{\rho}_a\dot{\rho}_b}(\omega,T)}
/{\omega} \,,
\end{align}
for small $\omega $-values. 
For the LAB phase, the above expression reduces to
\begin{align}
M_{\rho_a \rho_b}(\omega) =
\frac{ W_0^2} {\omega } 
\int\frac{d^3 \mathbf{k}}
{(2 \, \pi)^3}\, {\text{Im}\,\widetilde{\Pi}_{ab}^R( \omega, \mathbf{k})}\,,
\end{align}
where
\begin{align}
& \widetilde{\Pi}_{ab}^R( \omega, \mathbf{k})
 =
\widetilde{\Pi}_{ab} (k_0, \mathbf{k}) \vert_{i \,k_0 \rightarrow \omega+i \,0^+}\,,
\text{ and } \nn
& \widetilde{\Pi}_{ab} ( k_0, \mathbf{k})  \nn & = 
-T \sum \limits_{ q_0}
\int\frac{d^3 \mathbf{q}} {(2 \,\pi)^3}
\text{Tr}[\, \Gamma_a\, G_0(q_0 + k_0, \mathbf q +\mathbf k)
\, \Gamma_b \,G_0(q_0, \mathbf q)\,] \,.
\end{align}
At low frequencies and finite temperatures, the real
part can be approximated by its $\omega \rightarrow 0$ value, which vanishes as a power law of $T$ \cite{Berg-PRB}.
Therefore, we focus on the imaginary part, which, using the fact that $\chi_{\rho_a\rho_b} \propto \delta_{ ab }$,
takes the form \cite{ips-hermann3}
\begin{align}
\label{Raman_final}
\text{Im}\,D_{\text{Raman}}(\omega,T)
& = \frac{\omega \,M_{\rho_a \rho_a}
\,\chi^2_{\rho_a\rho_a}}
{\omega^2 \,\chi^2_{\rho_a\rho_a} + M^2_{\rho_a\rho_a}}
\equiv
\frac{\omega\,\widetilde{\Gamma}\,\chi_{\rho_a\rho_a}}
{\omega^2+\widetilde{\Gamma}^2}\,, \nn
\widetilde{\Gamma} &=
M_{\rho_a\rho_a} \, \chi^{-1}_{\rho_a\rho_a}\,.
\end{align}
Fig.~\ref{raman_fig} shows a representative behaviour of $\text{Im}D_{\text{Raman}}(\omega,T)$. From the nature of the curves, we can clearly see that the Raman response in the LAB phase exhibits a quasi-elastic peak at $\omega \approx \omega_{\rm max}
= M_{\rho_a\rho_a} (0) /\chi_{\rho_a\rho_a} $, with a peak-height equal to $\chi_{\rho_a\rho_a} /2$. 
The static susceptibility, to the leading order, can be fitted to the functional form of $\chi_{\rho_a\rho_a} \simeq 
a_1 + a_2 \,T^{1/2}+ a_3/T$, where $a_1$, $a_2$, and $a_3$ correspond to temperature-independent constants. Furthermore, $M_{\rho_a\rho_a}(0)$ is either $T$-independent, from a numerical point of view, or displays an extremely weak $T$-dependence, which is not observed within our numerical accuracy \cite{ips-hermann3}.

\subsection{Shear viscosity and entropy density}
\label{secshear}

In the hydrodynamic regime, the shear viscosity $\eta$ is one of the fundamental physical properties that describe the inherent characteristics of the system. In fluid mechanics, for example, $\eta$ determines whether the hydrodynamic flow
will be laminar or turbulent. In strongly-correlated systems, $\eta$ takes an even more important role --- it expresses the degree of many-body quantum entanglement of the electronic phase. In a landmark paper by Kovtun, Son and Starinets \cite{DTSon-PRL_2005}, the authors have shown (by employing the AdS/CFT correspondence) that in strongly-interacting quantum field theories,
the ratio of $\eta$ with the entropy density $s$ satisfies an inequality given by $(\eta/s)\geq\hbar/(4\pi k_B)$. If the universal lower bound is approximately-saturated, the system is said to exhibit the ``minimal-viscosity scenario''. Important examples of systems, that satisfy the lower bound, include quark-gluon plasma \cite{DTSon-PRL_2005} generated in heavy-ion colliders, ultracold quantum gases trapped in optical lattices tuned to the unitary limit \cite{Cao}, and ultraclean graphene in the vicinity of the charge neutrality point \cite{Fritz-PRB}. 
Here, we review our results for $\eta$ and $s$ for the LAB phase.

Using the Kubo formula \cite{Taylor,zwerger}, the isotropic optical shear viscosity in the $T=0$ limit is given by
\begin{align}
 \eta (\omega) &= \lim_{\mathbf{q}\to 0} 
 \frac{ \langle \mathcal T_{ xy} \, \mathcal T_{ xy}\rangle (\omega, \mathbf q) } 
 {\omega}
\nn & =
\lim_{\mathbf{q}\to 0} 
 \frac{ \langle \mathcal T_{yz} \, \mathcal T_{yz}\rangle (\omega, \mathbf q) } 
 {\omega} 
\nn & =
\lim_{\mathbf{q}\to 0} 
 \frac{ \langle \mathcal T_{zx} \, \mathcal T_{zx}\rangle (\omega, \mathbf q) }
  {\omega} \, ,
\end{align}
where ${\mathcal T}_{\mu \nu}$ is the stress tensor. The variables $\mu$ and $\nu$ span over the Matsubara frequency and the spatial components. For the LAB phase, the stress tensor takes the form 
\begin{align}
{\mathcal T}_{\mu\nu} (q_0,\mathbf q)
&=   \int\frac{dk_0 \, d^3\mathbf{k}}
{(2\,\pi)^4 }\,
\left( k_{\nu} + q_{\nu} /2 \right)
\tilde\psi^{\dagger}( k_0 + q_0, \mathbf{k} +\mathbf q)\nonumber\\
& \qquad
\times\left [ \partial_{k^{\mu}}
(\mathbf{d_{k}} \cdot\mathbf{\Gamma}) \right ]\tilde\psi( k_0, \mathbf{k}).
\end{align}
Evaluating the above expression using the fixed-point action, the optical viscosity is found to scale as \cite{ips-hermann3}
\begin{equation}
\eta(\omega)\sim \omega^{2-\frac{\varepsilon }{2}
-\frac{ 367\, \varepsilon } {2736}}\,.
\end{equation}
Comparing with Eq.~\eqref{eqet1}, we find that there is a small hyperscaling violation proportional to $\varepsilon$.

Using the memory matrix formalism, the temperature-dependent isotropic shear viscosity is given by \cite{ips-hermann3}
\begin{align}
\label{eta}
\eta(\omega,T)= &\chi_{ {\mathcal T}_{ p q}  {\mathcal T}_{ p q}} 
\left[ \frac{1}
{M_{ {\mathcal T}_{ p q}  {\mathcal T}_{ p q}} - i \,\omega
\,\chi_{\mathcal T_{\mu \nu} \mathcal T_{p q}} }\right ] 
\chi_{ \mathcal T_{p q} \mathcal T_{p q}} .
\end{align}
We point out that, here, we have used the fact that the stress tensor turns out to be another nearly-conserved operator for the LAB phase in the limit of weak disorder \cite{ips-hermann3}.

In the DC limit, we obtain
\begin{align}
\eta^{\rm dc} (T)
=\chi^2_{ {\mathcal T}_{zx} \, {\mathcal T}_{zx}} /
M_{ {\mathcal T}_{zx} 
\, {\mathcal T}_{zx}} .
\end{align} 
In Ref. \cite{ips-hermann3}, we have evaluated the right-hand side to get the result
\begin{align}\label{eta }
\eta^{\rm dc} \sim T^{\lambda},
\end{align} 
where $\lambda $ turns out to be a non-universal exponent lying within the range $0<\lambda< 1$.

The next step is to calculate the free energy at $T>0$. Since the bosons have
no dynamics, the contributions to this quantity come only from the free fermions and the perturbative corrections due to the Coulomb interactions \cite{ips-hermann3}. We compute the latter by using the $\varepsilon$-expansion about $d_c = 4$.
The contribution from the free fermions is given by $\Delta F^{(0)}  =F^{(0)}   - F^{(0)} (0)$, where
\begin{align}
\Delta F^{(0)}  
= - 2 \int \frac{ d^{d}\mathbf k} {(2\,\pi)^{d}}
\,\Bigg [ T \sum \limits_{ l=\pm}\ln \left( 1 +  e^{- \frac{ l\,|\mathbf k|^2 } 
{ 2\,m \, T} }\right) 
- \frac{ |\mathbf k|^2}{2\,m} \Bigg ]\,,
\end{align}
where we have subtracted off the contribution from the $T$-independent ground state energy, and have included a factor of 2 to account for the double-degeneracy of the bands.
This final result takes the form \cite{ips-hermann3}
\begin{align}
 \Delta F^{(0)}  
&\simeq - \frac{  \bar \eta(3)
\left(   m \, T \right)^{ 3-\frac{\varepsilon} {2} } } { \pi^2 \,m}
 - \frac{ 3\, \zeta(3)
\left(   m \, T \right)^{ 3-\frac{\varepsilon} {2} } }
 { 4\, \pi^2 \,m} \,,
\end{align}
with $ \bar \eta(u)= \frac{\int_0^\infty dt \,t^{u-2} \,\ln(1+e^{-u})}
{\Gamma(u-1)}$ symbolizing the Dirichlet eta function, and we have set $\varepsilon= 0$ in the numerical prefactors of the scaling arguments. 
The lowest-order correction in the free energy, due to the Coulomb interactions, reads \cite{ips-hermann3}
\begin{align}
& F_{\rm int}
 = 
  \frac{e^2\,\Lambda^{\varepsilon}\,T^2}  {c } 
\nn &
\hspace{ 0.75 cm} \times  \sum_{\Omega_n , \,\omega_{n' } } 
 \int \frac{   d^{ d} \mathbf q \,
   d^{ d} \mathbf k  }   {  (2 \, \pi)^{ 2d } }
\frac{ \mathrm{Tr}\left  [ 
 G ( \omega_{n' } + \Omega_n , \mathbf k + \mathbf  q)  
 \, G( \omega_{n' }, \mathbf  k)\right ]} {|\mathbf q|^2} ,
\end{align}
where $\omega_{n'}$ and $\Omega_n $ represent the fermionic and bosonic Matsubara frequencies, respectively. Setting $d = 4-\varepsilon$, we need to isolate the contributions due to poles in the parameter $\varepsilon$. To the lowest order in $\varepsilon$, these poles are obtained by evaluating one Matsubara frequency sum at $T>0$ exactly, and the other one as an integral in the limit of $T\rightarrow 0$. Employing these steps, we get the final scaling form of the free energy as \cite{ips-hermann3}
\begin{align}
& F  \simeq 
- \frac{ 3\, \zeta(3) 
\,  m ^{ 2-\frac{\varepsilon} {2} }\,
T^{3 -\frac{\varepsilon} {2} -  \frac{ 4\,\varepsilon}{ 19 }}
}
 { 4\, \pi^2 } 
\left( \frac {\Lambda^2 }{m } \right)^{\frac{ 4\,\varepsilon}{ 19}}\,.
\end{align}
Extrapolating our results to the physical scenario of $d=3$, we find that
the free energy scales as 
\begin{equation}
F(T)\sim  T^{3 -\frac{\varepsilon} {2} -  \frac{ 4\,\varepsilon}{ 19 }}
\overset{\epsilon=1}  =
T^{ \frac{5} {2} -  \frac{4}{ 19 }}
\,.
\label{eqEscale}
\end{equation}
This implies that the specific heat scales as
\begin{equation}
C(T)\sim  T^{ 2 -\frac{\varepsilon} {2} -  \frac{ 4\,\varepsilon}{ 19}}
\overset{\epsilon=1}  =
T^{ \frac{3} {2} -  \frac{4}{ 19 }}\,.
\label{eqcv}
\end{equation}
Comparing with Eq.~\eqref{eqfscale}, and taking into account the fixed point value of $z^*=2-\frac{4\,\varepsilon} {19 }$ for the dynamical critical exponent, we find that there is a hyperscaling violation, which is proportional to $\varepsilon$.

Using the fact that the entropy density is the derivative of the free energy with respect to the temperature, we 
get $s  \sim  T^{ 2 -\frac{\varepsilon} {2} -  \frac{ 4\,\varepsilon}{ 19 }}$. Setting $d=3$ (i.e., $\varepsilon=1$), this leads to
\begin{align}
s  \sim  T^{2-\frac{27}{38}}\,.
\label{eqs_scale}
\end{align}
Therefore, 
\begin{align}
\left(\frac{\eta^{\rm dc} } {s}\right) \sim T^{\lambda-\frac{49}{38}},
\end{align}
which indicates that $\eta/s$ always tends to diverge at low temperatures, rather than saturating to a constant universal value. This divergent behavior bears some resemblance to the result found in the finite-Fermi-suface NFL arising at the Ising-nematic critical point \cite{patel2}. In that particular scenario, the divergence emerges due to the violation of the hyperscaling property, whose origin can be related to the presence of a finite sharply-defined Fermi surface. We would like to point out that this divergent behavior of $\eta/s$, as a function of $T$, contrasts with the results obtained in Refs.~\cite{dumitrescu,Herbut-PRB}, which employed a quantum BE method.

\section{Discussion and Outlook}
\label{final}

In this review, we have examined the recent progress in constructing non-quasiparticle transport theories, which can be applied to various NFL phases. In particular, we have identified the Kubo formalism and the memory matrix approach as two useful frameworks to extract the transport properties. As an application of these two methods, we have demonstrated the computation of generalized conductivity tensors, Raman response, free energy, viscosity, and entropy density in the so-called LAB phase of the Luttinger semimetals. The LAB phase emerges when the chemical potential cuts the nodal-point, and the pseudospin-$3/2$ quasiparticles are acted upon by unscreened Coulomb interactions. We have discussed the history and the phenomenology of identifying the materials harbouring the LAB phase.

Overall, the investigation of NFL phases arising in nodal-point semimetals has seen great progress over the last decade. From a more theoretical standpoint, some future directions worthy of investigations are as follows:

\begin{enumerate}

\item The effect of phonons as a relaxation mechanism in the finite-temperature transport properties: As we have seen in our discussions for the LAB phase, phonons are expected to be relevant for some of the response characteristics. We would like to point out that this is in accord with the results reported in recent experimental works \cite{Rosalin_2023}.

\item Finding a way to go beyond the weak disorder limit: In the memory matrix formalism, when we want to slowly relax one of the conserved charges, we add a weak disorder, as discussed in detail. But we may want to compute the effects of strong disorder as well, when the perturbative disorder approximation will not work. One way to do so might be to follow in the footsteps of the SYK-Yukawa models \cite{Esterlis,Wang_Yukawa,Patel_SYK,Aldape}), where the systems inherently have random couplings. Another possibility is to apply the invariant measure approach (IMA) \cite{ips-klaus}.

\item The calculation of other fundamental magnetotransport coefficients like the planar Hall and planar thermal Hall effects \cite{shama, li18_giant, GP_Das, ips-serena, ips_rahul_ph_strain, ips-rahul-jpcm}, Nernst response \cite{Behnia_Nernst, ips-kush, ips-kush-review}, Magnus Hall effect \cite{papaj_magnus, amit-magnus, ips_magnus}, to name a few.

\item Computation of many-body quantum chaos parameters, such as the Lyapunov exponent $\lambda_L$ and the ``butterfly-effect'' velocity $v_B$, by using the out-of-time-order correlators (OTOCs) \cite{Stanford2016}, for example:
Although these quantities are not experimentally measured, it was proposed in Refs. \cite{Blake,Blake2} that they are related to important physical quantities such as the charge and energy diffusivities. The inverse of the Lyapunov exponent represents the time in which the quantum information, associated with a local perturbation, gets ``scrambled'' into non-local 
degrees of freedom. On the other hand, $ v_B$ refers to the speed at which the effects of such a perturbation propagate in the system. Recently, Maldacena \textit{et al.} \cite{Maldacena2016} has shown that the Lyapunov time, $\tau_L \equiv 1/\lambda_L$, satisfies the inequality $\tau_L\geq \hbar/(2\pi k_B T)$. The Lyapunov time saturates to the universal lower bound of the inequality in scenarios like the SYK model \cite{MaldacenaSYK2016}, and if the corresponding quantum field theory is holographically dual to a black hole \cite{Shenker2014}.  

\end{enumerate}

On the experimental front, although some recent explorations~\cite{Kondo_2015,armitage_expt,Rosalin_2023,osti_1435592} have found smoking-gun signatures that might turn out to be precursors to the physics of the LAB state, there is still no unambiguous evidence for the 
NFL phase appearing at the nodal point. More precise experiments are needed in this direction. Apart from the predictions for several transport signatures that we have reviewed here, other complementary characteristics, amenable to experimental measurements, will provide important information about these systems. One important tool might be to apply the angle-resolved photoemission spectroscopy (ARPES) technique \cite{Damascelli}, that measures the single-particle lifetime for the low-energy excitations in the system. In such an experiment, a strong departure from the Fermi liquid behavior should be detected if the NFL phase can be accessed. Another promising experimental technique is the momentum-resolved electron-energy-loss spectroscopy (M-EELS) \cite{Mitrano_2018,Husain,PMID:37558882}, that probes the dynamic charge response resolved in momentum.

\section*{Acknowledgments}
IM's research, leading to these results, has received funding from the European Union's Horizon 2020 research and innovation programme under the Marie Skłodowska-Curie grant agreement number 754340. H.F. acknowledges funding obtained from the Conselho Nacional de Desenvolvimento Cient\'{i}fico e Tecnol\'{o}gico (CNPq) under grant numbers 311428/2021-5 and 404274/2023-4.

\bibliography{biblio_rev}

\begin{thebibliography}{191}%
\makeatletter
\providecommand \@ifxundefined [1]{%
 \@ifx{#1\undefined}
}%
\providecommand \@ifnum [1]{%
 \ifnum #1\expandafter \@firstoftwo
 \else \expandafter \@secondoftwo
 \fi
}%
\providecommand \@ifx [1]{%
 \ifx #1\expandafter \@firstoftwo
 \else \expandafter \@secondoftwo
 \fi
}%
\providecommand \natexlab [1]{#1}%
\providecommand \enquote  [1]{``#1''}%
\providecommand \bibnamefont  [1]{#1}%
\providecommand \bibfnamefont [1]{#1}%
\providecommand \citenamefont [1]{#1}%
\providecommand \href@noop [0]{\@secondoftwo}%
\providecommand \href [0]{\begingroup \@sanitize@url \@href}%
\providecommand \@href[1]{\@@startlink{#1}\@@href}%
\providecommand \@@href[1]{\endgroup#1\@@endlink}%
\providecommand \@sanitize@url [0]{\catcode `\\12\catcode `\$12\catcode
  `\&12\catcode `\#12\catcode `\^12\catcode `\_12\catcode `\%12\relax}%
\providecommand \@@startlink[1]{}%
\providecommand \@@endlink[0]{}%
\providecommand \url  [0]{\begingroup\@sanitize@url \@url }%
\providecommand \@url [1]{\endgroup\@href {#1}{\urlprefix }}%
\providecommand \urlprefix  [0]{URL }%
\providecommand \Eprint [0]{\href }%
\providecommand \doibase [0]{https://doi.org/}%
\providecommand \selectlanguage [0]{\@gobble}%
\providecommand \bibinfo  [0]{\@secondoftwo}%
\providecommand \bibfield  [0]{\@secondoftwo}%
\providecommand \translation [1]{[#1]}%
\providecommand \BibitemOpen [0]{}%
\providecommand \bibitemStop [0]{}%
\providecommand \bibitemNoStop [0]{.\EOS\space}%
\providecommand \EOS [0]{\spacefactor3000\relax}%
\providecommand \BibitemShut  [1]{\csname bibitem#1\endcsname}%
\let\auto@bib@innerbib\@empty
\bibitem [{\citenamefont {{Nayak}}\ and\ \citenamefont
  {{Wilczek}}(1994{\natexlab{a}})}]{nayak}%
  \BibitemOpen
  \bibfield  {author} {\bibinfo {author} {\bibfnamefont {C.}~\bibnamefont
  {{Nayak}}}\ and\ \bibinfo {author} {\bibfnamefont {F.}~\bibnamefont
  {{Wilczek}}},\ }\bibfield  {title} {\bibinfo {title} {{Renormalization group
  approach to low temperature properties of a non-Fermi liquid metal}},\ }\href
  {https://doi.org/10.1016/0550-3213(94)90158-9} {\bibfield  {journal}
  {\bibinfo  {journal} {Nuclear Physics B}\ }\textbf {\bibinfo {volume}
  {430}},\ \bibinfo {pages} {534} (\bibinfo {year}
  {1994}{\natexlab{a}})}\BibitemShut {NoStop}%
\bibitem [{\citenamefont {{Nayak}}\ and\ \citenamefont
  {{Wilczek}}(1994{\natexlab{b}})}]{nayak1}%
  \BibitemOpen
  \bibfield  {author} {\bibinfo {author} {\bibfnamefont {C.}~\bibnamefont
  {{Nayak}}}\ and\ \bibinfo {author} {\bibfnamefont {F.}~\bibnamefont
  {{Wilczek}}},\ }\bibfield  {title} {\bibinfo {title} {{Non-Fermi liquid fixed
  point in 2 + 1 dimensions}},\ }\href
  {https://doi.org/10.1016/0550-3213(94)90477-4} {\bibfield  {journal}
  {\bibinfo  {journal} {Nuclear Physics B}\ }\textbf {\bibinfo {volume}
  {417}},\ \bibinfo {pages} {359} (\bibinfo {year}
  {1994}{\natexlab{b}})}\BibitemShut {NoStop}%
\bibitem [{\citenamefont {{Lawler}}\ \emph {et~al.}(2006)\citenamefont
  {{Lawler}}, \citenamefont {{Barci}}, \citenamefont {{Fern{\'a}ndez}},
  \citenamefont {{Fradkin}},\ and\ \citenamefont {{Oxman}}}]{lawler1}%
  \BibitemOpen
  \bibfield  {author} {\bibinfo {author} {\bibfnamefont {M.~J.}\ \bibnamefont
  {{Lawler}}}, \bibinfo {author} {\bibfnamefont {D.~G.}\ \bibnamefont
  {{Barci}}}, \bibinfo {author} {\bibfnamefont {V.}~\bibnamefont
  {{Fern{\'a}ndez}}}, \bibinfo {author} {\bibfnamefont {E.}~\bibnamefont
  {{Fradkin}}},\ and\ \bibinfo {author} {\bibfnamefont {L.}~\bibnamefont
  {{Oxman}}},\ }\bibfield  {title} {\bibinfo {title} {{Nonperturbative behavior
  of the quantum phase transition to a nematic Fermi fluid}},\ }\href
  {https://doi.org/10.1103/PhysRevB.73.085101} {\bibfield  {journal} {\bibinfo
  {journal} {\prb}\ }\textbf {\bibinfo {volume} {73}},\ \bibinfo {eid} {085101}
  (\bibinfo {year} {2006})}\BibitemShut {NoStop}%
\bibitem [{\citenamefont {Mross}\ \emph {et~al.}(2010)\citenamefont {Mross},
  \citenamefont {McGreevy}, \citenamefont {Liu},\ and\ \citenamefont
  {Senthil}}]{mross}%
  \BibitemOpen
  \bibfield  {author} {\bibinfo {author} {\bibfnamefont {D.~F.}\ \bibnamefont
  {Mross}}, \bibinfo {author} {\bibfnamefont {J.}~\bibnamefont {McGreevy}},
  \bibinfo {author} {\bibfnamefont {H.}~\bibnamefont {Liu}},\ and\ \bibinfo
  {author} {\bibfnamefont {T.}~\bibnamefont {Senthil}},\ }\bibfield  {title}
  {\bibinfo {title} {{Controlled expansion for certain non-Fermi-liquid
  metals}},\ }\href {https://doi.org/10.1103/PhysRevB.82.045121} {\bibfield
  {journal} {\bibinfo  {journal} {Phys. Rev. B}\ }\textbf {\bibinfo {volume}
  {82}},\ \bibinfo {pages} {045121} (\bibinfo {year} {2010})}\BibitemShut
  {NoStop}%
\bibitem [{\citenamefont {{Jiang}}\ \emph {et~al.}(2013)\citenamefont
  {{Jiang}}, \citenamefont {{Block}}, \citenamefont {{Mishmash}}, \citenamefont
  {{Garrison}}, \citenamefont {{Sheng}}, \citenamefont {{Motrunich}},\ and\
  \citenamefont {{Fisher}}}]{Jiang}%
  \BibitemOpen
  \bibfield  {author} {\bibinfo {author} {\bibfnamefont {H.-C.}\ \bibnamefont
  {{Jiang}}}, \bibinfo {author} {\bibfnamefont {M.~S.}\ \bibnamefont
  {{Block}}}, \bibinfo {author} {\bibfnamefont {R.~V.}\ \bibnamefont
  {{Mishmash}}}, \bibinfo {author} {\bibfnamefont {J.~R.}\ \bibnamefont
  {{Garrison}}}, \bibinfo {author} {\bibfnamefont {D.~N.}\ \bibnamefont
  {{Sheng}}}, \bibinfo {author} {\bibfnamefont {O.~I.}\ \bibnamefont
  {{Motrunich}}},\ and\ \bibinfo {author} {\bibfnamefont {M.~P.~A.}\
  \bibnamefont {{Fisher}}},\ }\bibfield  {title} {\bibinfo {title}
  {{Non-Fermi-liquid d-wave metal phase of strongly interacting electrons}},\
  }\href {https://doi.org/10.1038/nature11732} {\bibfield  {journal} {\bibinfo
  {journal} {\nat}\ }\textbf {\bibinfo {volume} {493}},\ \bibinfo {pages} {39}
  (\bibinfo {year} {2013})}\BibitemShut {NoStop}%
\bibitem [{\citenamefont {Metlitski}\ and\ \citenamefont
  {Sachdev}(2010{\natexlab{a}})}]{metlsach1}%
  \BibitemOpen
  \bibfield  {author} {\bibinfo {author} {\bibfnamefont {M.~A.}\ \bibnamefont
  {Metlitski}}\ and\ \bibinfo {author} {\bibfnamefont {S.}~\bibnamefont
  {Sachdev}},\ }\bibfield  {title} {\bibinfo {title} {{Quantum phase
  transitions of metals in two spatial dimensions. I. Ising-nematic order}},\
  }\href {https://doi.org/10.1103/PhysRevB.82.075127} {\bibfield  {journal}
  {\bibinfo  {journal} {Phys. Rev. B}\ }\textbf {\bibinfo {volume} {82}},\
  \bibinfo {pages} {075127} (\bibinfo {year} {2010}{\natexlab{a}})}\BibitemShut
  {NoStop}%
\bibitem [{\citenamefont {Metlitski}\ and\ \citenamefont
  {Sachdev}(2010{\natexlab{b}})}]{metlsach}%
  \BibitemOpen
  \bibfield  {author} {\bibinfo {author} {\bibfnamefont {M.~A.}\ \bibnamefont
  {Metlitski}}\ and\ \bibinfo {author} {\bibfnamefont {S.}~\bibnamefont
  {Sachdev}},\ }\bibfield  {title} {\bibinfo {title} {{Quantum phase
  transitions of metals in two spatial dimensions. II. Spin density wave
  order}},\ }\href {https://doi.org/10.1103/PhysRevB.82.075128} {\bibfield
  {journal} {\bibinfo  {journal} {Phys. Rev. B}\ }\textbf {\bibinfo {volume}
  {82}},\ \bibinfo {pages} {075128} (\bibinfo {year}
  {2010}{\natexlab{b}})}\BibitemShut {NoStop}%
\bibitem [{\citenamefont {Chung}\ \emph {et~al.}(2013)\citenamefont {Chung},
  \citenamefont {Mandal}, \citenamefont {Raghu},\ and\ \citenamefont
  {Chakravarty}}]{ips2}%
  \BibitemOpen
  \bibfield  {author} {\bibinfo {author} {\bibfnamefont {S.~B.}\ \bibnamefont
  {Chung}}, \bibinfo {author} {\bibfnamefont {I.}~\bibnamefont {Mandal}},
  \bibinfo {author} {\bibfnamefont {S.}~\bibnamefont {Raghu}},\ and\ \bibinfo
  {author} {\bibfnamefont {S.}~\bibnamefont {Chakravarty}},\ }\bibfield
  {title} {\bibinfo {title} {Higher angular momentum pairing from transverse
  gauge interactions},\ }\href {https://doi.org/10.1103/PhysRevB.88.045127}
  {\bibfield  {journal} {\bibinfo  {journal} {Phys. Rev. B}\ }\textbf {\bibinfo
  {volume} {88}},\ \bibinfo {pages} {045127} (\bibinfo {year}
  {2013})}\BibitemShut {NoStop}%
\bibitem [{\citenamefont {Wang}\ \emph {et~al.}(2014)\citenamefont {Wang},
  \citenamefont {Mandal}, \citenamefont {Chung},\ and\ \citenamefont
  {Chakravarty}}]{ips3}%
  \BibitemOpen
  \bibfield  {author} {\bibinfo {author} {\bibfnamefont {Z.}~\bibnamefont
  {Wang}}, \bibinfo {author} {\bibfnamefont {I.}~\bibnamefont {Mandal}},
  \bibinfo {author} {\bibfnamefont {S.~B.}\ \bibnamefont {Chung}},\ and\
  \bibinfo {author} {\bibfnamefont {S.}~\bibnamefont {Chakravarty}},\
  }\bibfield  {title} {\bibinfo {title} {{Pairing in half-filled Landau
  level}},\ }\href
  {https://doi.org/http://dx.doi.org/10.1016/j.aop.2014.09.021} {\bibfield
  {journal} {\bibinfo  {journal} {Annals of Physics}\ }\textbf {\bibinfo
  {volume} {351}},\ \bibinfo {pages} {727 } (\bibinfo {year}
  {2014})}\BibitemShut {NoStop}%
\bibitem [{\citenamefont {Sur}\ and\ \citenamefont {Lee}(2014)}]{Shouvik1}%
  \BibitemOpen
  \bibfield  {author} {\bibinfo {author} {\bibfnamefont {S.}~\bibnamefont
  {Sur}}\ and\ \bibinfo {author} {\bibfnamefont {S.-S.}\ \bibnamefont {Lee}},\
  }\bibfield  {title} {\bibinfo {title} {{Chiral non-Fermi liquids}},\ }\href
  {https://doi.org/10.1103/PhysRevB.90.045121} {\bibfield  {journal} {\bibinfo
  {journal} {Phys. Rev. B}\ }\textbf {\bibinfo {volume} {90}},\ \bibinfo
  {pages} {045121} (\bibinfo {year} {2014})}\BibitemShut {NoStop}%
\bibitem [{\citenamefont {Dalidovich}\ and\ \citenamefont
  {Lee}(2013)}]{Lee-Dalid}%
  \BibitemOpen
  \bibfield  {author} {\bibinfo {author} {\bibfnamefont {D.}~\bibnamefont
  {Dalidovich}}\ and\ \bibinfo {author} {\bibfnamefont {S.-S.}\ \bibnamefont
  {Lee}},\ }\bibfield  {title} {\bibinfo {title} {{Perturbative non-Fermi
  liquids from dimensional regularization}},\ }\href
  {https://doi.org/10.1103/PhysRevB.88.245106} {\bibfield  {journal} {\bibinfo
  {journal} {Phys. Rev. B}\ }\textbf {\bibinfo {volume} {88}},\ \bibinfo
  {pages} {245106} (\bibinfo {year} {2013})}\BibitemShut {NoStop}%
\bibitem [{\citenamefont {Sur}\ and\ \citenamefont {Lee}(2015)}]{shouvik2}%
  \BibitemOpen
  \bibfield  {author} {\bibinfo {author} {\bibfnamefont {S.}~\bibnamefont
  {Sur}}\ and\ \bibinfo {author} {\bibfnamefont {S.-S.}\ \bibnamefont {Lee}},\
  }\bibfield  {title} {\bibinfo {title} {Quasilocal strange metal},\ }\href
  {https://doi.org/10.1103/PhysRevB.91.125136} {\bibfield  {journal} {\bibinfo
  {journal} {Phys. Rev. B}\ }\textbf {\bibinfo {volume} {91}},\ \bibinfo
  {pages} {125136} (\bibinfo {year} {2015})}\BibitemShut {NoStop}%
\bibitem [{\citenamefont {Mandal}\ and\ \citenamefont
  {Lee}(2015)}]{ips-uv-ir1}%
  \BibitemOpen
  \bibfield  {author} {\bibinfo {author} {\bibfnamefont {I.}~\bibnamefont
  {Mandal}}\ and\ \bibinfo {author} {\bibfnamefont {S.-S.}\ \bibnamefont
  {Lee}},\ }\bibfield  {title} {\bibinfo {title} {{Ultraviolet/infrared mixing
  in non-Fermi liquids}},\ }\href {https://doi.org/10.1103/PhysRevB.92.035141}
  {\bibfield  {journal} {\bibinfo  {journal} {Phys. Rev. B}\ }\textbf {\bibinfo
  {volume} {92}},\ \bibinfo {pages} {035141} (\bibinfo {year}
  {2015})}\BibitemShut {NoStop}%
\bibitem [{\citenamefont {Mandal}(2016{\natexlab{a}})}]{ips-uv-ir2}%
  \BibitemOpen
  \bibfield  {author} {\bibinfo {author} {\bibfnamefont {I.}~\bibnamefont
  {Mandal}},\ }\bibfield  {title} {\bibinfo {title} {{UV/IR mixing in non-Fermi
  liquids: Higher-loop corrections in different energy ranges}},\ }\href
  {https://doi.org/10.1140/epjb/e2016-70509-4} {\bibfield  {journal} {\bibinfo
  {journal} {Eur. Phys. J. B}\ }\textbf {\bibinfo {volume} {89}},\ \bibinfo
  {pages} {278} (\bibinfo {year} {2016}{\natexlab{a}})}\BibitemShut {NoStop}%
\bibitem [{\citenamefont {de~Carvalho}\ \emph {et~al.}(2015)\citenamefont
  {de~Carvalho}, \citenamefont {Kloss}, \citenamefont {Montiel}, \citenamefont
  {Freire},\ and\ \citenamefont {P\'epin}}]{Freire_Pepin_1}%
  \BibitemOpen
  \bibfield  {author} {\bibinfo {author} {\bibfnamefont {V.~S.}\ \bibnamefont
  {de~Carvalho}}, \bibinfo {author} {\bibfnamefont {T.}~\bibnamefont {Kloss}},
  \bibinfo {author} {\bibfnamefont {X.}~\bibnamefont {Montiel}}, \bibinfo
  {author} {\bibfnamefont {H.}~\bibnamefont {Freire}},\ and\ \bibinfo {author}
  {\bibfnamefont {C.}~\bibnamefont {P\'epin}},\ }\bibfield  {title} {\bibinfo
  {title} {{Strong competition between {\ensuremath{\Theta}}$_{I
  I}$-loop-current order and d -wave charge order along the diagonal direction
  in a two-dimensional hot spot model}},\ }\href
  {https://link.aps.org/doi/10.1103/PhysRevB.92.075123} {\bibfield  {journal}
  {\bibinfo  {journal} {Phys. Rev. B}\ }\textbf {\bibinfo {volume} {92}},\
  \bibinfo {pages} {075123} (\bibinfo {year} {2015})}\BibitemShut {NoStop}%
\bibitem [{\citenamefont {de~Carvalho}\ \emph {et~al.}(2016)\citenamefont
  {de~Carvalho}, \citenamefont {P\'epin},\ and\ \citenamefont
  {Freire}}]{Freire_Pepin_2}%
  \BibitemOpen
  \bibfield  {author} {\bibinfo {author} {\bibfnamefont {V.~S.}\ \bibnamefont
  {de~Carvalho}}, \bibinfo {author} {\bibfnamefont {C.}~\bibnamefont
  {P\'epin}},\ and\ \bibinfo {author} {\bibfnamefont {H.}~\bibnamefont
  {Freire}},\ }\bibfield  {title} {\bibinfo {title} {{Coexistence of
  {\ensuremath{\Theta}}$_{I I}$-loop-current order with checkerboard d -wave
  CDW/PDW order in a hot-spot model for cuprate superconductors}},\ }\href
  {https://link.aps.org/doi/10.1103/PhysRevB.93.115144} {\bibfield  {journal}
  {\bibinfo  {journal} {Phys. Rev. B}\ }\textbf {\bibinfo {volume} {93}},\
  \bibinfo {pages} {115144} (\bibinfo {year} {2016})}\BibitemShut {NoStop}%
\bibitem [{\citenamefont {Eberlein}\ \emph {et~al.}(2016)\citenamefont
  {Eberlein}, \citenamefont {Mandal},\ and\ \citenamefont
  {Sachdev}}]{ips-subir}%
  \BibitemOpen
  \bibfield  {author} {\bibinfo {author} {\bibfnamefont {A.}~\bibnamefont
  {Eberlein}}, \bibinfo {author} {\bibfnamefont {I.}~\bibnamefont {Mandal}},\
  and\ \bibinfo {author} {\bibfnamefont {S.}~\bibnamefont {Sachdev}},\
  }\bibfield  {title} {\bibinfo {title} {{Hyperscaling violation at the
  Ising-nematic quantum critical point in two-dimensional metals}},\ }\href
  {https://doi.org/10.1103/PhysRevB.94.045133} {\bibfield  {journal} {\bibinfo
  {journal} {Phys. Rev. B}\ }\textbf {\bibinfo {volume} {94}},\ \bibinfo
  {pages} {045133} (\bibinfo {year} {2016})}\BibitemShut {NoStop}%
\bibitem [{\citenamefont {Mandal}(2016{\natexlab{b}})}]{ips-sc}%
  \BibitemOpen
  \bibfield  {author} {\bibinfo {author} {\bibfnamefont {I.}~\bibnamefont
  {Mandal}},\ }\bibfield  {title} {\bibinfo {title} {{Superconducting
  instability in non-Fermi liquids}},\ }\href
  {https://doi.org/10.1103/PhysRevB.94.115138} {\bibfield  {journal} {\bibinfo
  {journal} {Phys. Rev. B}\ }\textbf {\bibinfo {volume} {94}},\ \bibinfo
  {pages} {115138} (\bibinfo {year} {2016}{\natexlab{b}})}\BibitemShut
  {NoStop}%
\bibitem [{\citenamefont {Mandal}(2017)}]{ips-c2}%
  \BibitemOpen
  \bibfield  {author} {\bibinfo {author} {\bibfnamefont {I.}~\bibnamefont
  {Mandal}},\ }\bibfield  {title} {\bibinfo {title} {{Scaling behaviour and
  superconducting instability in anisotropic non-Fermi liquids}},\ }\href
  {https://doi.org/https://doi.org/10.1016/j.aop.2016.11.009} {\bibfield
  {journal} {\bibinfo  {journal} {Annals of Physics}\ }\textbf {\bibinfo
  {volume} {376}},\ \bibinfo {pages} {89 } (\bibinfo {year}
  {2017})}\BibitemShut {NoStop}%
\bibitem [{\citenamefont {Lee}(2018)}]{Lee_2018}%
  \BibitemOpen
  \bibfield  {author} {\bibinfo {author} {\bibfnamefont {S.-S.}\ \bibnamefont
  {Lee}},\ }\bibfield  {title} {\bibinfo {title} {Recent developments in
  non-{F}ermi liquid theory},\ }\href
  {https://doi.org/10.1146/annurev-conmatphys-031016-025531} {\bibfield
  {journal} {\bibinfo  {journal} {Annual Review of Condensed Matter Physics}\
  }\textbf {\bibinfo {volume} {9}},\ \bibinfo {pages} {227} (\bibinfo {year}
  {2018})}\BibitemShut {NoStop}%
\bibitem [{\citenamefont {Pimenov}\ \emph {et~al.}(2018)\citenamefont
  {Pimenov}, \citenamefont {Mandal}, \citenamefont {Piazza},\ and\
  \citenamefont {Punk}}]{ips-fflo}%
  \BibitemOpen
  \bibfield  {author} {\bibinfo {author} {\bibfnamefont {D.}~\bibnamefont
  {Pimenov}}, \bibinfo {author} {\bibfnamefont {I.}~\bibnamefont {Mandal}},
  \bibinfo {author} {\bibfnamefont {F.}~\bibnamefont {Piazza}},\ and\ \bibinfo
  {author} {\bibfnamefont {M.}~\bibnamefont {Punk}},\ }\bibfield  {title}
  {\bibinfo {title} {{Non-Fermi liquid at the FFLO quantum critical point}},\
  }\href {https://doi.org/10.1103/PhysRevB.98.024510} {\bibfield  {journal}
  {\bibinfo  {journal} {Phys. Rev. B}\ }\textbf {\bibinfo {volume} {98}},\
  \bibinfo {pages} {024510} (\bibinfo {year} {2018})}\BibitemShut {NoStop}%
\bibitem [{\citenamefont {Mandal}(2020{\natexlab{a}})}]{ips-nfl-u1}%
  \BibitemOpen
  \bibfield  {author} {\bibinfo {author} {\bibfnamefont {I.}~\bibnamefont
  {Mandal}},\ }\bibfield  {title} {\bibinfo {title} {{Critical Fermi surfaces
  in generic dimensions arising from transverse gauge field interactions}},\
  }\href {https://doi.org/10.1103/PhysRevResearch.2.043277} {\bibfield
  {journal} {\bibinfo  {journal} {Phys. Rev. Research}\ }\textbf {\bibinfo
  {volume} {2}},\ \bibinfo {pages} {043277} (\bibinfo {year}
  {2020}{\natexlab{a}})}\BibitemShut {NoStop}%
\bibitem [{\citenamefont {{Mandal}}(2024)}]{ips_2kf}%
  \BibitemOpen
  \bibfield  {author} {\bibinfo {author} {\bibfnamefont {I.}~\bibnamefont
  {{Mandal}}},\ }\bibfield  {title} {\bibinfo {title} {{Stable non-Fermi liquid
  fixed point at the onset of incommensurate $2k_F$ charge density wave
  order}},\ }\href {https://doi.org/10.1016/j.nuclphysb.2024.116586} {\bibfield
   {journal} {\bibinfo  {journal} {Nucl. Phys. B}\ }\textbf {\bibinfo {volume}
  {1005}},\ \bibinfo {pages} {116586} (\bibinfo {year} {2024})}\BibitemShut
  {NoStop}%
\bibitem [{\citenamefont {Chowdhury}\ \emph {et~al.}(2022)\citenamefont
  {Chowdhury}, \citenamefont {Georges}, \citenamefont {Parcollet},\ and\
  \citenamefont {Sachdev}}]{Chowdhury_2022}%
  \BibitemOpen
  \bibfield  {author} {\bibinfo {author} {\bibfnamefont {D.}~\bibnamefont
  {Chowdhury}}, \bibinfo {author} {\bibfnamefont {A.}~\bibnamefont {Georges}},
  \bibinfo {author} {\bibfnamefont {O.}~\bibnamefont {Parcollet}},\ and\
  \bibinfo {author} {\bibfnamefont {S.}~\bibnamefont {Sachdev}},\ }\bibfield
  {title} {\bibinfo {title} {{Sachdev-Ye-Kitaev models and beyond: Window into
  non-Fermi liquids}},\ }\href {http://dx.doi.org/10.1103/RevModPhys.94.035004}
  {\bibfield  {journal} {\bibinfo  {journal} {Reviews of Modern Physics}\
  }\textbf {\bibinfo {volume} {94}} (\bibinfo {year} {2022})}\BibitemShut
  {NoStop}%
\bibitem [{\citenamefont {Ono}\ \emph {et~al.}(2007)\citenamefont {Ono},
  \citenamefont {Komiya},\ and\ \citenamefont {Ando}}]{Ando_2007}%
  \BibitemOpen
  \bibfield  {author} {\bibinfo {author} {\bibfnamefont {S.}~\bibnamefont
  {Ono}}, \bibinfo {author} {\bibfnamefont {S.}~\bibnamefont {Komiya}},\ and\
  \bibinfo {author} {\bibfnamefont {Y.}~\bibnamefont {Ando}},\ }\bibfield
  {title} {\bibinfo {title} {Strong charge fluctuations manifested in the
  high-temperature hall coefficient of high-${T}_{c}$ cuprates},\ }\href
  {https://doi.org/10.1103/PhysRevB.75.024515} {\bibfield  {journal} {\bibinfo
  {journal} {Phys. Rev. B}\ }\textbf {\bibinfo {volume} {75}},\ \bibinfo
  {pages} {024515} (\bibinfo {year} {2007})}\BibitemShut {NoStop}%
\bibitem [{\citenamefont {Legros}\ \emph {et~al.}(2018)\citenamefont {Legros},
  \citenamefont {Benhabib}, \citenamefont {Tabis}, \citenamefont
  {Lalibert{\'{e}}}, \citenamefont {Dion}, \citenamefont {Lizaire},
  \citenamefont {Vignolle}, \citenamefont {Vignolles}, \citenamefont {Raffy},
  \citenamefont {Li}, \citenamefont {Auban-Senzier}, \citenamefont
  {Doiron-Leyraud}, \citenamefont {Fournier}, \citenamefont {Colson},
  \citenamefont {Taillefer},\ and\ \citenamefont {Proust}}]{Legros_2018}%
  \BibitemOpen
  \bibfield  {author} {\bibinfo {author} {\bibfnamefont {A.}~\bibnamefont
  {Legros}}, \bibinfo {author} {\bibfnamefont {S.}~\bibnamefont {Benhabib}},
  \bibinfo {author} {\bibfnamefont {W.}~\bibnamefont {Tabis}}, \bibinfo
  {author} {\bibfnamefont {F.}~\bibnamefont {Lalibert{\'{e}}}}, \bibinfo
  {author} {\bibfnamefont {M.}~\bibnamefont {Dion}}, \bibinfo {author}
  {\bibfnamefont {M.}~\bibnamefont {Lizaire}}, \bibinfo {author} {\bibfnamefont
  {B.}~\bibnamefont {Vignolle}}, \bibinfo {author} {\bibfnamefont
  {D.}~\bibnamefont {Vignolles}}, \bibinfo {author} {\bibfnamefont
  {H.}~\bibnamefont {Raffy}}, \bibinfo {author} {\bibfnamefont {Z.~Z.}\
  \bibnamefont {Li}}, \bibinfo {author} {\bibfnamefont {P.}~\bibnamefont
  {Auban-Senzier}}, \bibinfo {author} {\bibfnamefont {N.}~\bibnamefont
  {Doiron-Leyraud}}, \bibinfo {author} {\bibfnamefont {P.}~\bibnamefont
  {Fournier}}, \bibinfo {author} {\bibfnamefont {D.}~\bibnamefont {Colson}},
  \bibinfo {author} {\bibfnamefont {L.}~\bibnamefont {Taillefer}},\ and\
  \bibinfo {author} {\bibfnamefont {C.}~\bibnamefont {Proust}},\ }\bibfield
  {title} {\bibinfo {title} {{Universal T-linear resistivity and Planckian
  dissipation in overdoped cuprates}},\ }\href
  {https://doi.org/10.1038/s41567-018-0334-2} {\bibfield  {journal} {\bibinfo
  {journal} {Nature Physics}\ }\textbf {\bibinfo {volume} {15}},\ \bibinfo
  {pages} {142} (\bibinfo {year} {2018})}\BibitemShut {NoStop}%
\bibitem [{\citenamefont {Ayres}\ \emph {et~al.}(2021)\citenamefont {Ayres},
  \citenamefont {Berben}, \citenamefont {{\v{C}}ulo}, \citenamefont {Hsu},
  \citenamefont {van Heumen}, \citenamefont {Huang}, \citenamefont {Zaanen},
  \citenamefont {Kondo}, \citenamefont {Takeuchi}, \citenamefont {Cooper},
  \citenamefont {Putzke}, \citenamefont {Friedemann}, \citenamefont
  {Carrington},\ and\ \citenamefont {Hussey}}]{Ayres2021}%
  \BibitemOpen
  \bibfield  {author} {\bibinfo {author} {\bibfnamefont {J.}~\bibnamefont
  {Ayres}}, \bibinfo {author} {\bibfnamefont {M.}~\bibnamefont {Berben}},
  \bibinfo {author} {\bibfnamefont {M.}~\bibnamefont {{\v{C}}ulo}}, \bibinfo
  {author} {\bibfnamefont {Y.-T.}\ \bibnamefont {Hsu}}, \bibinfo {author}
  {\bibfnamefont {E.}~\bibnamefont {van Heumen}}, \bibinfo {author}
  {\bibfnamefont {Y.}~\bibnamefont {Huang}}, \bibinfo {author} {\bibfnamefont
  {J.}~\bibnamefont {Zaanen}}, \bibinfo {author} {\bibfnamefont
  {T.}~\bibnamefont {Kondo}}, \bibinfo {author} {\bibfnamefont
  {T.}~\bibnamefont {Takeuchi}}, \bibinfo {author} {\bibfnamefont {J.~R.}\
  \bibnamefont {Cooper}}, \bibinfo {author} {\bibfnamefont {C.}~\bibnamefont
  {Putzke}}, \bibinfo {author} {\bibfnamefont {S.}~\bibnamefont {Friedemann}},
  \bibinfo {author} {\bibfnamefont {A.}~\bibnamefont {Carrington}},\ and\
  \bibinfo {author} {\bibfnamefont {N.~E.}\ \bibnamefont {Hussey}},\ }\bibfield
   {title} {\bibinfo {title} {Incoherent transport across the strange-metal
  regime of overdoped cuprates},\ }\href
  {https://doi.org/10.1038/s41586-021-03622-z} {\bibfield  {journal} {\bibinfo
  {journal} {Nature}\ }\textbf {\bibinfo {volume} {595}},\ \bibinfo {pages}
  {661} (\bibinfo {year} {2021})}\BibitemShut {NoStop}%
\bibitem [{\citenamefont {Hayes}\ \emph {et~al.}(2016)\citenamefont {Hayes},
  \citenamefont {McDonald}, \citenamefont {Breznay}, \citenamefont {Helm},
  \citenamefont {Moll}, \citenamefont {Wartenbe}, \citenamefont {Shekhter},\
  and\ \citenamefont {Analytis}}]{Hayes2016}%
  \BibitemOpen
  \bibfield  {author} {\bibinfo {author} {\bibfnamefont {I.~M.}\ \bibnamefont
  {Hayes}}, \bibinfo {author} {\bibfnamefont {R.~D.}\ \bibnamefont {McDonald}},
  \bibinfo {author} {\bibfnamefont {N.~P.}\ \bibnamefont {Breznay}}, \bibinfo
  {author} {\bibfnamefont {T.}~\bibnamefont {Helm}}, \bibinfo {author}
  {\bibfnamefont {P.~J.~W.}\ \bibnamefont {Moll}}, \bibinfo {author}
  {\bibfnamefont {M.}~\bibnamefont {Wartenbe}}, \bibinfo {author}
  {\bibfnamefont {A.}~\bibnamefont {Shekhter}},\ and\ \bibinfo {author}
  {\bibfnamefont {J.~G.}\ \bibnamefont {Analytis}},\ }\bibfield  {title}
  {\bibinfo {title} {Scaling between magnetic field and temperature in the
  high-temperature superconductor {B}a{F}e$_2$({A}s$_{1-x}${P}$_{x}$)$_{2}$},\
  }\href {https://doi.org/10.1038/nphys3773} {\bibfield  {journal} {\bibinfo
  {journal} {Nature Physics}\ }\textbf {\bibinfo {volume} {12}},\ \bibinfo
  {pages} {916} (\bibinfo {year} {2016})}\BibitemShut {NoStop}%
\bibitem [{\citenamefont {Nakajima}\ \emph {et~al.}(2006)\citenamefont
  {Nakajima}, \citenamefont {Izawa}, \citenamefont {Matsuda}, \citenamefont
  {Behnia}, \citenamefont {Kontani}, \citenamefont {Hedo}, \citenamefont
  {Uwatoko}, \citenamefont {Matsumoto}, \citenamefont {Shishido}, \citenamefont
  {Settai},\ and\ \citenamefont {Onuki}}]{Nakajima_2006}%
  \BibitemOpen
  \bibfield  {author} {\bibinfo {author} {\bibfnamefont {Y.}~\bibnamefont
  {Nakajima}}, \bibinfo {author} {\bibfnamefont {K.}~\bibnamefont {Izawa}},
  \bibinfo {author} {\bibfnamefont {Y.}~\bibnamefont {Matsuda}}, \bibinfo
  {author} {\bibfnamefont {K.}~\bibnamefont {Behnia}}, \bibinfo {author}
  {\bibfnamefont {H.}~\bibnamefont {Kontani}}, \bibinfo {author} {\bibfnamefont
  {M.}~\bibnamefont {Hedo}}, \bibinfo {author} {\bibfnamefont {Y.}~\bibnamefont
  {Uwatoko}}, \bibinfo {author} {\bibfnamefont {T.}~\bibnamefont {Matsumoto}},
  \bibinfo {author} {\bibfnamefont {H.}~\bibnamefont {Shishido}}, \bibinfo
  {author} {\bibfnamefont {R.}~\bibnamefont {Settai}},\ and\ \bibinfo {author}
  {\bibfnamefont {Y.}~\bibnamefont {Onuki}},\ }\bibfield  {title} {\bibinfo
  {title} {{Evolution of Hall coefficient in two-dimensional heavy fermion
  CeCoIn5}},\ }\href {https://doi.org/10.1143/jpsj.75.023705} {\bibfield
  {journal} {\bibinfo  {journal} {Journal of the Physical Society of Japan}\
  }\textbf {\bibinfo {volume} {75}},\ \bibinfo {pages} {023705} (\bibinfo
  {year} {2006})}\BibitemShut {NoStop}%
\bibitem [{\citenamefont {Cao}\ \emph {et~al.}(2020)\citenamefont {Cao},
  \citenamefont {Chowdhury}, \citenamefont {Rodan-Legrain}, \citenamefont
  {Rubies-Bigorda}, \citenamefont {Watanabe}, \citenamefont {Taniguchi},
  \citenamefont {Senthil},\ and\ \citenamefont {Jarillo-Herrero}}]{Cao_2020}%
  \BibitemOpen
  \bibfield  {author} {\bibinfo {author} {\bibfnamefont {Y.}~\bibnamefont
  {Cao}}, \bibinfo {author} {\bibfnamefont {D.}~\bibnamefont {Chowdhury}},
  \bibinfo {author} {\bibfnamefont {D.}~\bibnamefont {Rodan-Legrain}}, \bibinfo
  {author} {\bibfnamefont {O.}~\bibnamefont {Rubies-Bigorda}}, \bibinfo
  {author} {\bibfnamefont {K.}~\bibnamefont {Watanabe}}, \bibinfo {author}
  {\bibfnamefont {T.}~\bibnamefont {Taniguchi}}, \bibinfo {author}
  {\bibfnamefont {T.}~\bibnamefont {Senthil}},\ and\ \bibinfo {author}
  {\bibfnamefont {P.}~\bibnamefont {Jarillo-Herrero}},\ }\bibfield  {title}
  {\bibinfo {title} {Strange metal in magic-angle graphene with near
  {P}lanckian dissipation},\ }\href
  {http://dx.doi.org/10.1103/PhysRevLett.124.076801} {\bibfield  {journal}
  {\bibinfo  {journal} {Physical Review Letters}\ }\textbf {\bibinfo {volume}
  {124}} (\bibinfo {year} {2020})}\BibitemShut {NoStop}%
\bibitem [{\citenamefont {Mandal}\ and\ \citenamefont
  {Fernandes}(2023)}]{ips-rafael}%
  \BibitemOpen
  \bibfield  {author} {\bibinfo {author} {\bibfnamefont {I.}~\bibnamefont
  {Mandal}}\ and\ \bibinfo {author} {\bibfnamefont {R.~M.}\ \bibnamefont
  {Fernandes}},\ }\bibfield  {title} {\bibinfo {title} {Valley-polarized
  nematic order in twisted moir\'e systems: {I}n-plane orbital magnetism and
  crossover from non-{F}ermi liquid to {F}ermi liquid},\ }\href
  {https://doi.org/10.1103/PhysRevB.107.125142} {\bibfield  {journal} {\bibinfo
   {journal} {Phys. Rev. B}\ }\textbf {\bibinfo {volume} {107}},\ \bibinfo
  {pages} {125142} (\bibinfo {year} {2023})}\BibitemShut {NoStop}%
\bibitem [{\citenamefont {Patel}\ \emph {et~al.}(2015)\citenamefont {Patel},
  \citenamefont {Strack},\ and\ \citenamefont {Sachdev}}]{subir-aavishkar}%
  \BibitemOpen
  \bibfield  {author} {\bibinfo {author} {\bibfnamefont {A.~A.}\ \bibnamefont
  {Patel}}, \bibinfo {author} {\bibfnamefont {P.}~\bibnamefont {Strack}},\ and\
  \bibinfo {author} {\bibfnamefont {S.}~\bibnamefont {Sachdev}},\ }\bibfield
  {title} {\bibinfo {title} {Hyperscaling at the spin density wave quantum
  critical point in two-dimensional metals},\ }\href
  {https://doi.org/10.1103/PhysRevB.92.165105} {\bibfield  {journal} {\bibinfo
  {journal} {Phys. Rev. B}\ }\textbf {\bibinfo {volume} {92}},\ \bibinfo
  {pages} {165105} (\bibinfo {year} {2015})}\BibitemShut {NoStop}%
\bibitem [{\citenamefont {Eberlein}\ \emph {et~al.}(2017)\citenamefont
  {Eberlein}, \citenamefont {Patel},\ and\ \citenamefont {Sachdev}}]{patel2}%
  \BibitemOpen
  \bibfield  {author} {\bibinfo {author} {\bibfnamefont {A.}~\bibnamefont
  {Eberlein}}, \bibinfo {author} {\bibfnamefont {A.~A.}\ \bibnamefont
  {Patel}},\ and\ \bibinfo {author} {\bibfnamefont {S.}~\bibnamefont
  {Sachdev}},\ }\bibfield  {title} {\bibinfo {title} {{Shear viscosity at the
  Ising-nematic quantum critical point in two-dimensional metals}},\ }\href
  {https://doi.org/10.1103/PhysRevB.95.075127} {\bibfield  {journal} {\bibinfo
  {journal} {Phys. Rev. B}\ }\textbf {\bibinfo {volume} {95}},\ \bibinfo
  {pages} {075127} (\bibinfo {year} {2017})}\BibitemShut {NoStop}%
\bibitem [{\citenamefont {Metlitski}\ \emph {et~al.}(2015)\citenamefont
  {Metlitski}, \citenamefont {Mross}, \citenamefont {Sachdev},\ and\
  \citenamefont {Senthil}}]{max_cooper}%
  \BibitemOpen
  \bibfield  {author} {\bibinfo {author} {\bibfnamefont {M.~A.}\ \bibnamefont
  {Metlitski}}, \bibinfo {author} {\bibfnamefont {D.~F.}\ \bibnamefont
  {Mross}}, \bibinfo {author} {\bibfnamefont {S.}~\bibnamefont {Sachdev}},\
  and\ \bibinfo {author} {\bibfnamefont {T.}~\bibnamefont {Senthil}},\
  }\bibfield  {title} {\bibinfo {title} {{Cooper pairing in non-Fermi
  liquids}},\ }\href {https://doi.org/10.1103/PhysRevB.91.115111} {\bibfield
  {journal} {\bibinfo  {journal} {Phys. Rev. B}\ }\textbf {\bibinfo {volume}
  {91}},\ \bibinfo {pages} {115111} (\bibinfo {year} {2015})}\BibitemShut
  {NoStop}%
\bibitem [{\citenamefont {{Abanov}}\ \emph {et~al.}(2003)\citenamefont
  {{Abanov}}, \citenamefont {{Chubukov}},\ and\ \citenamefont
  {{Schmalian}}}]{chubukov3}%
  \BibitemOpen
  \bibfield  {author} {\bibinfo {author} {\bibfnamefont {A.}~\bibnamefont
  {{Abanov}}}, \bibinfo {author} {\bibfnamefont {A.~V.}\ \bibnamefont
  {{Chubukov}}},\ and\ \bibinfo {author} {\bibfnamefont {J.}~\bibnamefont
  {{Schmalian}}},\ }\bibfield  {title} {\bibinfo {title} {{Quantum-critical
  theory of the spin-fermion model and its application to cuprates: Normal
  state analysis}},\ }\href {https://doi.org/10.1080/0001873021000057123}
  {\bibfield  {journal} {\bibinfo  {journal} {Advances in Physics}\ }\textbf
  {\bibinfo {volume} {52}},\ \bibinfo {pages} {119} (\bibinfo {year}
  {2003})}\BibitemShut {NoStop}%
\bibitem [{\citenamefont {Rech}\ \emph {et~al.}(2006)\citenamefont {Rech},
  \citenamefont {P\'epin},\ and\ \citenamefont {Chubukov}}]{rech}%
  \BibitemOpen
  \bibfield  {author} {\bibinfo {author} {\bibfnamefont {J.}~\bibnamefont
  {Rech}}, \bibinfo {author} {\bibfnamefont {C.}~\bibnamefont {P\'epin}},\ and\
  \bibinfo {author} {\bibfnamefont {A.~V.}\ \bibnamefont {Chubukov}},\
  }\bibfield  {title} {\bibinfo {title} {Quantum critical behavior in itinerant
  electron systems: Eliashberg theory and instability of a ferromagnetic
  quantum critical point},\ }\href {https://doi.org/10.1103/PhysRevB.74.195126}
  {\bibfield  {journal} {\bibinfo  {journal} {Phys. Rev. B}\ }\textbf {\bibinfo
  {volume} {74}},\ \bibinfo {pages} {195126} (\bibinfo {year}
  {2006})}\BibitemShut {NoStop}%
\bibitem [{\citenamefont {Dell'Anna}\ and\ \citenamefont
  {Metzner}(2006)}]{delanna}%
  \BibitemOpen
  \bibfield  {author} {\bibinfo {author} {\bibfnamefont {L.}~\bibnamefont
  {Dell'Anna}}\ and\ \bibinfo {author} {\bibfnamefont {W.}~\bibnamefont
  {Metzner}},\ }\bibfield  {title} {\bibinfo {title} {Fermi surface
  fluctuations and single electron excitations near pomeranchuk instability in
  two dimensions},\ }\href {https://doi.org/10.1103/PhysRevB.73.045127}
  {\bibfield  {journal} {\bibinfo  {journal} {Phys. Rev. B}\ }\textbf {\bibinfo
  {volume} {73}},\ \bibinfo {pages} {045127} (\bibinfo {year}
  {2006})}\BibitemShut {NoStop}%
\bibitem [{\citenamefont {Chubukov}\ \emph {et~al.}(2020)\citenamefont
  {Chubukov}, \citenamefont {Abanov}, \citenamefont {Wang},\ and\ \citenamefont
  {Wu}}]{Chubukov2020}%
  \BibitemOpen
  \bibfield  {author} {\bibinfo {author} {\bibfnamefont {A.~V.}\ \bibnamefont
  {Chubukov}}, \bibinfo {author} {\bibfnamefont {A.}~\bibnamefont {Abanov}},
  \bibinfo {author} {\bibfnamefont {Y.}~\bibnamefont {Wang}},\ and\ \bibinfo
  {author} {\bibfnamefont {Y.-M.}\ \bibnamefont {Wu}},\ }\bibfield  {title}
  {\bibinfo {title} {{The interplay between superconductivity and non-Fermi
  liquid at a quantum-critical point in a metal}},\ }\href
  {https://doi.org/https://doi.org/10.1016/j.aop.2020.168142} {\bibfield
  {journal} {\bibinfo  {journal} {Annals of Physics}\ }\textbf {\bibinfo
  {volume} {417}},\ \bibinfo {pages} {168142} (\bibinfo {year} {2020})},\
  \bibinfo {note} {eliashberg theory at 60: Strong-coupling superconductivity
  and beyond}\BibitemShut {NoStop}%
\bibitem [{\citenamefont {Chakravarty}\ \emph {et~al.}(1995)\citenamefont
  {Chakravarty}, \citenamefont {Norton},\ and\ \citenamefont
  {Sylju\aa{}sen}}]{olav}%
  \BibitemOpen
  \bibfield  {author} {\bibinfo {author} {\bibfnamefont {S.}~\bibnamefont
  {Chakravarty}}, \bibinfo {author} {\bibfnamefont {R.~E.}\ \bibnamefont
  {Norton}},\ and\ \bibinfo {author} {\bibfnamefont {O.~F.}\ \bibnamefont
  {Sylju\aa{}sen}},\ }\bibfield  {title} {\bibinfo {title} {Transverse gauge
  interactions and the vanquished {F}ermi liquid},\ }\href
  {https://doi.org/10.1103/PhysRevLett.74.1423} {\bibfield  {journal} {\bibinfo
   {journal} {Phys. Rev. Lett.}\ }\textbf {\bibinfo {volume} {74}},\ \bibinfo
  {pages} {1423} (\bibinfo {year} {1995})}\BibitemShut {NoStop}%
\bibitem [{\citenamefont {Senthil}(2008)}]{senthil_fs}%
  \BibitemOpen
  \bibfield  {author} {\bibinfo {author} {\bibfnamefont {T.}~\bibnamefont
  {Senthil}},\ }\bibfield  {title} {\bibinfo {title} {{Theory of a continuous
  Mott transition in two dimensions}},\ }\href
  {https://doi.org/10.1103/PhysRevB.78.045109} {\bibfield  {journal} {\bibinfo
  {journal} {Phys. Rev. B}\ }\textbf {\bibinfo {volume} {78}},\ \bibinfo
  {pages} {045109} (\bibinfo {year} {2008})}\BibitemShut {NoStop}%
\bibitem [{\citenamefont {Abrikosov}(1974)}]{Abrikosov}%
  \BibitemOpen
  \bibfield  {author} {\bibinfo {author} {\bibfnamefont {A.~A.}\ \bibnamefont
  {Abrikosov}},\ }\bibfield  {title} {\bibinfo {title} {Calculation of critical
  indices for zero-gap semiconductors},\ }\href@noop {} {\bibfield  {journal}
  {\bibinfo  {journal} {Soviet Journal of Experimental and Theoretical
  Physics}\ }\textbf {\bibinfo {volume} {39}},\ \bibinfo {pages} {709}
  (\bibinfo {year} {1974})}\BibitemShut {NoStop}%
\bibitem [{\citenamefont {{Abrikosov}}\ and\ \citenamefont
  {{Beneslavskii}}(1996)}]{Abrikosov_Beneslavskii}%
  \BibitemOpen
  \bibfield  {author} {\bibinfo {author} {\bibfnamefont {A.~A.}\ \bibnamefont
  {{Abrikosov}}}\ and\ \bibinfo {author} {\bibfnamefont {S.~D.}\ \bibnamefont
  {{Beneslavskii}}},\ }\bibfield  {title} {\bibinfo {title} {Possible existence
  of substances intermediate between metals and dielectrics},\ }in\ \href
  {https://doi.org/10.1142/9789814317344_0010} {\emph {\bibinfo {booktitle} {30
  Years of the Landau Institute - Selected Papers}}}\ (\bibinfo {year} {1996})\
  pp.\ \bibinfo {pages} {64--73}\BibitemShut {NoStop}%
\bibitem [{\citenamefont {Moon}\ \emph {et~al.}(2013)\citenamefont {Moon},
  \citenamefont {Xu}, \citenamefont {Kim},\ and\ \citenamefont
  {Balents}}]{moon-xu}%
  \BibitemOpen
  \bibfield  {author} {\bibinfo {author} {\bibfnamefont {E.-G.}\ \bibnamefont
  {Moon}}, \bibinfo {author} {\bibfnamefont {C.}~\bibnamefont {Xu}}, \bibinfo
  {author} {\bibfnamefont {Y.~B.}\ \bibnamefont {Kim}},\ and\ \bibinfo {author}
  {\bibfnamefont {L.}~\bibnamefont {Balents}},\ }\bibfield  {title} {\bibinfo
  {title} {Non-{F}ermi-liquid and topological states with strong spin-orbit
  coupling},\ }\href {https://doi.org/10.1103/PhysRevLett.111.206401}
  {\bibfield  {journal} {\bibinfo  {journal} {Phys. Rev. Lett.}\ }\textbf
  {\bibinfo {volume} {111}},\ \bibinfo {pages} {206401} (\bibinfo {year}
  {2013})}\BibitemShut {NoStop}%
\bibitem [{\citenamefont {Nandkishore}\ and\ \citenamefont
  {Parameswaran}(2017)}]{rahul-sid}%
  \BibitemOpen
  \bibfield  {author} {\bibinfo {author} {\bibfnamefont {R.~M.}\ \bibnamefont
  {Nandkishore}}\ and\ \bibinfo {author} {\bibfnamefont {S.~A.}\ \bibnamefont
  {Parameswaran}},\ }\bibfield  {title} {\bibinfo {title} {{Disorder-driven
  destruction of a non-Fermi liquid semimetal studied by renormalization group
  analysis}},\ }\href {https://doi.org/10.1103/PhysRevB.95.205106} {\bibfield
  {journal} {\bibinfo  {journal} {Phys. Rev. B}\ }\textbf {\bibinfo {volume}
  {95}},\ \bibinfo {pages} {205106} (\bibinfo {year} {2017})}\BibitemShut
  {NoStop}%
\bibitem [{\citenamefont {{Mandal}}\ and\ \citenamefont
  {{Nandkishore}}(2018)}]{ips-rahul}%
  \BibitemOpen
  \bibfield  {author} {\bibinfo {author} {\bibfnamefont {I.}~\bibnamefont
  {{Mandal}}}\ and\ \bibinfo {author} {\bibfnamefont {R.~M.}\ \bibnamefont
  {{Nandkishore}}},\ }\bibfield  {title} {\bibinfo {title} {{Interplay of
  Coulomb interactions and disorder in three-dimensional quadratic band
  crossings without time-reversal symmetry and with unequal masses for
  conduction and valence bands}},\ }\href
  {https://doi.org/10.1103/PhysRevB.97.125121} {\bibfield  {journal} {\bibinfo
  {journal} {\prb}\ }\textbf {\bibinfo {volume} {97}},\ \bibinfo {eid} {125121}
  (\bibinfo {year} {2018})}\BibitemShut {NoStop}%
\bibitem [{\citenamefont {Mandal}(2018)}]{ips-qbt-sc}%
  \BibitemOpen
  \bibfield  {author} {\bibinfo {author} {\bibfnamefont {I.}~\bibnamefont
  {Mandal}},\ }\bibfield  {title} {\bibinfo {title} {{Fate of superconductivity
  in three-dimensional disordered Luttinger semimetals}},\ }\href
  {https://doi.org/https://doi.org/10.1016/j.aop.2018.03.004} {\bibfield
  {journal} {\bibinfo  {journal} {Annals of Physics}\ }\textbf {\bibinfo
  {volume} {392}},\ \bibinfo {pages} {179 } (\bibinfo {year}
  {2018})}\BibitemShut {NoStop}%
\bibitem [{\citenamefont {Roy}\ \emph {et~al.}(2018)\citenamefont {Roy},
  \citenamefont {Kennett}, \citenamefont {Yang},\ and\ \citenamefont
  {Juri\ifmmode \check{c}\else \v{c}\fi{}i\ifmmode~\acute{c}\else
  \'{c}\fi{}}}]{malcolm-bitan}%
  \BibitemOpen
  \bibfield  {author} {\bibinfo {author} {\bibfnamefont {B.}~\bibnamefont
  {Roy}}, \bibinfo {author} {\bibfnamefont {M.~P.}\ \bibnamefont {Kennett}},
  \bibinfo {author} {\bibfnamefont {K.}~\bibnamefont {Yang}},\ and\ \bibinfo
  {author} {\bibfnamefont {V.}~\bibnamefont {Juri\ifmmode \check{c}\else
  \v{c}\fi{}i\ifmmode~\acute{c}\else \'{c}\fi{}}},\ }\bibfield  {title}
  {\bibinfo {title} {From birefringent electrons to a marginal or non-{F}ermi
  liquid of relativistic spin-$1/2$ fermions: {A}n emergent
  superuniversality},\ }\href {https://doi.org/10.1103/PhysRevLett.121.157602}
  {\bibfield  {journal} {\bibinfo  {journal} {Phys. Rev. Lett.}\ }\textbf
  {\bibinfo {volume} {121}},\ \bibinfo {pages} {157602} (\bibinfo {year}
  {2018})}\BibitemShut {NoStop}%
\bibitem [{\citenamefont {{Mandal}}(2021)}]{ips-biref}%
  \BibitemOpen
  \bibfield  {author} {\bibinfo {author} {\bibfnamefont {I.}~\bibnamefont
  {{Mandal}}},\ }\bibfield  {title} {\bibinfo {title} {{Robust marginal Fermi
  liquid in birefringent semimetals}},\ }\href
  {https://doi.org/10.1016/j.physleta.2021.127707} {\bibfield  {journal}
  {\bibinfo  {journal} {Physics Letters A}\ }\textbf {\bibinfo {volume}
  {418}},\ \bibinfo {eid} {127707} (\bibinfo {year} {2021})}\BibitemShut
  {NoStop}%
\bibitem [{\citenamefont {Kubo}(1956)}]{kubo1}%
  \BibitemOpen
  \bibfield  {author} {\bibinfo {author} {\bibfnamefont {R.}~\bibnamefont
  {Kubo}},\ }\bibfield  {title} {\bibinfo {title} {A general expression for the
  conductivity tensor},\ }\href {https://doi.org/10.1139/p56-140} {\bibfield
  {journal} {\bibinfo  {journal} {Canadian Journal of Physics}\ }\textbf
  {\bibinfo {volume} {34}},\ \bibinfo {pages} {1274} (\bibinfo {year}
  {1956})}\BibitemShut {NoStop}%
\bibitem [{\citenamefont {Kubo}(1957)}]{Kubo_1957}%
  \BibitemOpen
  \bibfield  {author} {\bibinfo {author} {\bibfnamefont {R.}~\bibnamefont
  {Kubo}},\ }\bibfield  {title} {\bibinfo {title} {Statistical-mechanical
  theory of irreversible processes. {I}. {G}eneral theory and simple
  applications to magnetic and conduction problems},\ }\href
  {https://doi.org/10.1143/JPSJ.12.570} {\bibfield  {journal} {\bibinfo
  {journal} {Journal of the Physical Society of Japan}\ }\textbf {\bibinfo
  {volume} {12}},\ \bibinfo {pages} {570} (\bibinfo {year} {1957})}\BibitemShut
  {NoStop}%
\bibitem [{\citenamefont {{Mandal}}\ and\ \citenamefont
  {{Saha}}(2024)}]{ips-kush-review}%
  \BibitemOpen
  \bibfield  {author} {\bibinfo {author} {\bibfnamefont {I.}~\bibnamefont
  {{Mandal}}}\ and\ \bibinfo {author} {\bibfnamefont {K.}~\bibnamefont
  {{Saha}}},\ }\bibfield  {title} {\bibinfo {title} {{Thermoelectric response
  in nodal-point semimetals}},\ }\href {https://doi.org/10.1002/andp.202400016}
  {\bibfield  {journal} {\bibinfo  {journal} {Ann. Phys. (Berlin)}\ }\textbf
  {\bibinfo {volume} {(Early View version)}},\ \bibinfo {pages} {202400016}
  (\bibinfo {year} {2024})},\ \Eprint {https://arxiv.org/abs/2309.10763}
  {arXiv:2309.10763 [cond-mat.mes-hall]} \BibitemShut {NoStop}%
\bibitem [{\citenamefont {Prange}\ and\ \citenamefont
  {Kadanoff}(1964)}]{prange}%
  \BibitemOpen
  \bibfield  {author} {\bibinfo {author} {\bibfnamefont {R.~E.}\ \bibnamefont
  {Prange}}\ and\ \bibinfo {author} {\bibfnamefont {L.~P.}\ \bibnamefont
  {Kadanoff}},\ }\bibfield  {title} {\bibinfo {title} {Transport theory for
  electron-phonon interactions in metals},\ }\href
  {https://doi.org/10.1103/PhysRev.134.A566} {\bibfield  {journal} {\bibinfo
  {journal} {Phys. Rev.}\ }\textbf {\bibinfo {volume} {134}},\ \bibinfo {pages}
  {A566} (\bibinfo {year} {1964})}\BibitemShut {NoStop}%
\bibitem [{\citenamefont {Kim}\ \emph {et~al.}(1995)\citenamefont {Kim},
  \citenamefont {Lee},\ and\ \citenamefont {Wen}}]{kim_qbe}%
  \BibitemOpen
  \bibfield  {author} {\bibinfo {author} {\bibfnamefont {Y.~B.}\ \bibnamefont
  {Kim}}, \bibinfo {author} {\bibfnamefont {P.~A.}\ \bibnamefont {Lee}},\ and\
  \bibinfo {author} {\bibfnamefont {X.-G.}\ \bibnamefont {Wen}},\ }\bibfield
  {title} {\bibinfo {title} {Quantum {B}oltzmann equation of composite fermions
  interacting with a gauge field},\ }\href
  {https://doi.org/10.1103/PhysRevB.52.17275} {\bibfield  {journal} {\bibinfo
  {journal} {Phys. Rev. B}\ }\textbf {\bibinfo {volume} {52}},\ \bibinfo
  {pages} {17275} (\bibinfo {year} {1995})}\BibitemShut {NoStop}%
\bibitem [{\citenamefont {Mandal}(2022)}]{ips-zero-mode}%
  \BibitemOpen
  \bibfield  {author} {\bibinfo {author} {\bibfnamefont {I.}~\bibnamefont
  {Mandal}},\ }\bibfield  {title} {\bibinfo {title} {Zero sound and plasmon
  modes for non-{F}ermi liquids},\ }\href
  {https://doi.org/https://doi.org/10.1016/j.physleta.2022.128292} {\bibfield
  {journal} {\bibinfo  {journal} {Physics Letters A}\ }\textbf {\bibinfo
  {volume} {447}},\ \bibinfo {pages} {128292} (\bibinfo {year}
  {2022})}\BibitemShut {NoStop}%
\bibitem [{\citenamefont {{Islam}}\ and\ \citenamefont
  {{Mandal}}(2023)}]{ips-kazi}%
  \BibitemOpen
  \bibfield  {author} {\bibinfo {author} {\bibfnamefont {K.~R.}\ \bibnamefont
  {{Islam}}}\ and\ \bibinfo {author} {\bibfnamefont {I.}~\bibnamefont
  {{Mandal}}},\ }\bibfield  {title} {\bibinfo {title} {{Generic deformation
  channels for critical Fermi surfaces in the collisionless regime}},\ }\href
  {https://doi.org/10.1016/j.aop.2023.169409} {\bibfield  {journal} {\bibinfo
  {journal} {Annals of Physics}\ }\textbf {\bibinfo {volume} {457}},\ \bibinfo
  {eid} {169409} (\bibinfo {year} {2023})}\BibitemShut {NoStop}%
\bibitem [{\citenamefont {Mori}(1965{\natexlab{a}})}]{mori0}%
  \BibitemOpen
  \bibfield  {author} {\bibinfo {author} {\bibfnamefont {H.}~\bibnamefont
  {Mori}},\ }\bibfield  {title} {\bibinfo {title} {Transport, collective
  motion, and {B}rownian motion},\ }\href {https://doi.org/10.1143/PTP.33.423}
  {\bibfield  {journal} {\bibinfo  {journal} {Progress of Theoretical Physics}\
  }\textbf {\bibinfo {volume} {33}},\ \bibinfo {pages} {423} (\bibinfo {year}
  {1965}{\natexlab{a}})}\BibitemShut {NoStop}%
\bibitem [{\citenamefont {Mori}(1965{\natexlab{b}})}]{Mori}%
  \BibitemOpen
  \bibfield  {author} {\bibinfo {author} {\bibfnamefont {H.}~\bibnamefont
  {Mori}},\ }\bibfield  {title} {\bibinfo {title} {A continued-fraction
  representation of the time-correlation functions},\ }\href
  {https://doi.org/10.1143/PTP.34.399} {\bibfield  {journal} {\bibinfo
  {journal} {Progress of Theoretical Physics}\ }\textbf {\bibinfo {volume}
  {34}},\ \bibinfo {pages} {399} (\bibinfo {year}
  {1965}{\natexlab{b}})}\BibitemShut {NoStop}%
\bibitem [{\citenamefont {Zwanzig}(1960)}]{zwanzig0}%
  \BibitemOpen
  \bibfield  {author} {\bibinfo {author} {\bibfnamefont {R.}~\bibnamefont
  {Zwanzig}},\ }\bibfield  {title} {\bibinfo {title} {Ensemble method in the
  theory of irreversibility},\ }\href {https://doi.org/10.1063/1.1731409}
  {\bibfield  {journal} {\bibinfo  {journal} {The Journal of Chemical Physics}\
  }\textbf {\bibinfo {volume} {33}},\ \bibinfo {pages} {1338} (\bibinfo {year}
  {1960})}\BibitemShut {NoStop}%
\bibitem [{\citenamefont {Zwanzig}(1961)}]{Zwanzig}%
  \BibitemOpen
  \bibfield  {author} {\bibinfo {author} {\bibfnamefont {R.}~\bibnamefont
  {Zwanzig}},\ }\bibfield  {title} {\bibinfo {title} {Memory effects in
  irreversible thermodynamics},\ }\href
  {https://doi.org/10.1103/PhysRev.124.983} {\bibfield  {journal} {\bibinfo
  {journal} {Phys. Rev.}\ }\textbf {\bibinfo {volume} {124}},\ \bibinfo {pages}
  {983} (\bibinfo {year} {1961})}\BibitemShut {NoStop}%
\bibitem [{\citenamefont {G\"otze}\ and\ \citenamefont
  {W\"olfle}(1972)}]{Woefle}%
  \BibitemOpen
  \bibfield  {author} {\bibinfo {author} {\bibfnamefont {W.}~\bibnamefont
  {G\"otze}}\ and\ \bibinfo {author} {\bibfnamefont {P.}~\bibnamefont
  {W\"olfle}},\ }\bibfield  {title} {\bibinfo {title} {Homogeneous dynamical
  conductivity of simple metals},\ }\href
  {https://doi.org/10.1103/PhysRevB.6.1226} {\bibfield  {journal} {\bibinfo
  {journal} {Phys. Rev. B}\ }\textbf {\bibinfo {volume} {6}},\ \bibinfo {pages}
  {1226} (\bibinfo {year} {1972})}\BibitemShut {NoStop}%
\bibitem [{\citenamefont {Forster}(1995)}]{forster-book}%
  \BibitemOpen
  \bibfield  {author} {\bibinfo {author} {\bibfnamefont {D.}~\bibnamefont
  {Forster}},\ }\href {https://books.google.de/books?id=1gjtswEACAAJ} {\emph
  {\bibinfo {title} {{Hydrodynamic fluctuations, broken symmetry, and
  correlation functions}}}},\ Advanced Books Classics\ (\bibinfo  {publisher}
  {CRC Press},\ \bibinfo {year} {1995})\BibitemShut {NoStop}%
\bibitem [{\citenamefont {Lucas}\ and\ \citenamefont
  {Sachdev}(2015)}]{andy_review}%
  \BibitemOpen
  \bibfield  {author} {\bibinfo {author} {\bibfnamefont {A.}~\bibnamefont
  {Lucas}}\ and\ \bibinfo {author} {\bibfnamefont {S.}~\bibnamefont
  {Sachdev}},\ }\bibfield  {title} {\bibinfo {title} {Memory matrix theory of
  magnetotransport in strange metals},\ }\href
  {https://doi.org/10.1103/PhysRevB.91.195122} {\bibfield  {journal} {\bibinfo
  {journal} {Phys. Rev. B}\ }\textbf {\bibinfo {volume} {91}},\ \bibinfo
  {pages} {195122} (\bibinfo {year} {2015})}\BibitemShut {NoStop}%
\bibitem [{\citenamefont {Green}(2022)}]{green_thesis}%
  \BibitemOpen
  \bibfield  {author} {\bibinfo {author} {\bibfnamefont {B.~R.}\ \bibnamefont
  {Green}},\ }\emph {\bibinfo {title} {Memory Function Formalism for the
  Electrical Conductivity of Periodic Systems}},\ \href
  {https://etda.libraries.psu.edu/files/final_submissions/25644} {Ph.D.
  thesis},\ \bibinfo  {school} {The Pennsylvania State University} (\bibinfo
  {year} {2022})\BibitemShut {NoStop}%
\bibitem [{\citenamefont {Sachdev}(2012)}]{sachdev_ads}%
  \BibitemOpen
  \bibfield  {author} {\bibinfo {author} {\bibfnamefont {S.}~\bibnamefont
  {Sachdev}},\ }\bibfield  {title} {\bibinfo {title} {What can gauge-gravity
  duality teach us about condensed matter physics?},\ }\href
  {https://doi.org/10.1146/annurev-conmatphys-020911-125141} {\bibfield
  {journal} {\bibinfo  {journal} {Annual Review of Condensed Matter Physics}\
  }\textbf {\bibinfo {volume} {3}},\ \bibinfo {pages} {9–33} (\bibinfo {year}
  {2012})}\BibitemShut {NoStop}%
\bibitem [{\citenamefont {Hartnoll}(2009)}]{hartnoll_lectures}%
  \BibitemOpen
  \bibfield  {author} {\bibinfo {author} {\bibfnamefont {S.~A.}\ \bibnamefont
  {Hartnoll}},\ }\bibfield  {title} {\bibinfo {title} {Lectures on holographic
  methods for condensed matter physics},\ }\href
  {https://doi.org/10.1088/0264-9381/26/22/224002} {\bibfield  {journal}
  {\bibinfo  {journal} {Classical and Quantum Gravity}\ }\textbf {\bibinfo
  {volume} {26}},\ \bibinfo {pages} {224002} (\bibinfo {year}
  {2009})}\BibitemShut {NoStop}%
\bibitem [{\citenamefont {Sachdev}\ and\ \citenamefont {Ye}(1993)}]{SY_1993}%
  \BibitemOpen
  \bibfield  {author} {\bibinfo {author} {\bibfnamefont {S.}~\bibnamefont
  {Sachdev}}\ and\ \bibinfo {author} {\bibfnamefont {J.}~\bibnamefont {Ye}},\
  }\bibfield  {title} {\bibinfo {title} {{Gapless spin-fluid ground state in a
  random quantum Heisenberg magnet}},\ }\href
  {https://doi.org/10.1103/PhysRevLett.70.3339} {\bibfield  {journal} {\bibinfo
   {journal} {Phys. Rev. Lett.}\ }\textbf {\bibinfo {volume} {70}},\ \bibinfo
  {pages} {3339} (\bibinfo {year} {1993})}\BibitemShut {NoStop}%
\bibitem [{\citenamefont {Kitaev}(2015)}]{Kitaev_SYK}%
  \BibitemOpen
  \bibfield  {author} {\bibinfo {author} {\bibfnamefont {A.}~\bibnamefont
  {Kitaev}},\ }\bibfield  {title} {\bibinfo {title} {A simple model of quantum
  holography},\ }\href
  {http://online.kitp.ucsb.edu/online/entangled15/kitaev/,http:
  //online.kitp.ucsb.edu/online/entangled15/kitaev2/} {\bibfield  {journal}
  {\bibinfo  {journal} {Talks at KITP, April 7 and May 27}\ } (\bibinfo {year}
  {2015})}\BibitemShut {NoStop}%
\bibitem [{\citenamefont {Esterlis}\ and\ \citenamefont
  {Schmalian}(2019)}]{Esterlis}%
  \BibitemOpen
  \bibfield  {author} {\bibinfo {author} {\bibfnamefont {I.}~\bibnamefont
  {Esterlis}}\ and\ \bibinfo {author} {\bibfnamefont {J.}~\bibnamefont
  {Schmalian}},\ }\bibfield  {title} {\bibinfo {title} {{Cooper pairing of
  incoherent electrons: An electron-phonon version of the Sachdev-Ye-Kitaev
  model}},\ }\href {https://doi.org/10.1103/PhysRevB.100.115132} {\bibfield
  {journal} {\bibinfo  {journal} {Phys. Rev. B}\ }\textbf {\bibinfo {volume}
  {100}},\ \bibinfo {pages} {115132} (\bibinfo {year} {2019})}\BibitemShut
  {NoStop}%
\bibitem [{\citenamefont {Wang}(2020)}]{Wang_Yukawa}%
  \BibitemOpen
  \bibfield  {author} {\bibinfo {author} {\bibfnamefont {Y.}~\bibnamefont
  {Wang}},\ }\bibfield  {title} {\bibinfo {title} {Solvable strong-coupling
  quantum-dot model with a non-{F}ermi-liquid pairing transition},\ }\href
  {https://doi.org/10.1103/PhysRevLett.124.017002} {\bibfield  {journal}
  {\bibinfo  {journal} {Phys. Rev. Lett.}\ }\textbf {\bibinfo {volume} {124}},\
  \bibinfo {pages} {017002} (\bibinfo {year} {2020})}\BibitemShut {NoStop}%
\bibitem [{\citenamefont {Patel}\ and\ \citenamefont
  {Sachdev}(2018)}]{Patel_SYK}%
  \BibitemOpen
  \bibfield  {author} {\bibinfo {author} {\bibfnamefont {A.~A.}\ \bibnamefont
  {Patel}}\ and\ \bibinfo {author} {\bibfnamefont {S.}~\bibnamefont
  {Sachdev}},\ }\bibfield  {title} {\bibinfo {title} {Critical strange metal
  from fluctuating gauge fields in a solvable random model},\ }\href
  {https://doi.org/10.1103/PhysRevB.98.125134} {\bibfield  {journal} {\bibinfo
  {journal} {Phys. Rev. B}\ }\textbf {\bibinfo {volume} {98}},\ \bibinfo
  {pages} {125134} (\bibinfo {year} {2018})}\BibitemShut {NoStop}%
\bibitem [{\citenamefont {Haldar}\ \emph {et~al.}(2020)\citenamefont {Haldar},
  \citenamefont {Haldar}, \citenamefont {Bera}, \citenamefont {Mandal},\ and\
  \citenamefont {Banerjee}}]{ips_syk}%
  \BibitemOpen
  \bibfield  {author} {\bibinfo {author} {\bibfnamefont {A.}~\bibnamefont
  {Haldar}}, \bibinfo {author} {\bibfnamefont {P.}~\bibnamefont {Haldar}},
  \bibinfo {author} {\bibfnamefont {S.}~\bibnamefont {Bera}}, \bibinfo {author}
  {\bibfnamefont {I.}~\bibnamefont {Mandal}},\ and\ \bibinfo {author}
  {\bibfnamefont {S.}~\bibnamefont {Banerjee}},\ }\bibfield  {title} {\bibinfo
  {title} {{Quench, thermalization, and residual entropy across a non-Fermi
  liquid to Fermi liquid transition}},\ }\href
  {https://doi.org/10.1103/PhysRevResearch.2.013307} {\bibfield  {journal}
  {\bibinfo  {journal} {Phys. Rev. Res.}\ }\textbf {\bibinfo {volume} {2}},\
  \bibinfo {pages} {013307} (\bibinfo {year} {2020})}\BibitemShut {NoStop}%
\bibitem [{\citenamefont {Aldape}\ \emph {et~al.}(2022)\citenamefont {Aldape},
  \citenamefont {Cookmeyer}, \citenamefont {Patel},\ and\ \citenamefont
  {Altman}}]{Aldape}%
  \BibitemOpen
  \bibfield  {author} {\bibinfo {author} {\bibfnamefont {E.~E.}\ \bibnamefont
  {Aldape}}, \bibinfo {author} {\bibfnamefont {T.}~\bibnamefont {Cookmeyer}},
  \bibinfo {author} {\bibfnamefont {A.~A.}\ \bibnamefont {Patel}},\ and\
  \bibinfo {author} {\bibfnamefont {E.}~\bibnamefont {Altman}},\ }\bibfield
  {title} {\bibinfo {title} {{Solvable theory of a strange metal at the
  breakdown of a heavy Fermi liquid}},\ }\href
  {https://doi.org/10.1103/PhysRevB.105.235111} {\bibfield  {journal} {\bibinfo
   {journal} {Phys. Rev. B}\ }\textbf {\bibinfo {volume} {105}},\ \bibinfo
  {pages} {235111} (\bibinfo {year} {2022})}\BibitemShut {NoStop}%
\bibitem [{\citenamefont {Patel}\ \emph {et~al.}(2023)\citenamefont {Patel},
  \citenamefont {Guo}, \citenamefont {Esterlis},\ and\ \citenamefont
  {Sachdev}}]{Patel_Science}%
  \BibitemOpen
  \bibfield  {author} {\bibinfo {author} {\bibfnamefont {A.~A.}\ \bibnamefont
  {Patel}}, \bibinfo {author} {\bibfnamefont {H.}~\bibnamefont {Guo}}, \bibinfo
  {author} {\bibfnamefont {I.}~\bibnamefont {Esterlis}},\ and\ \bibinfo
  {author} {\bibfnamefont {S.}~\bibnamefont {Sachdev}},\ }\bibfield  {title}
  {\bibinfo {title} {Universal theory of strange metals from spatially random
  interactions},\ }\href {https://doi.org/10.1126/science.abq6011} {\bibfield
  {journal} {\bibinfo  {journal} {Science}\ }\textbf {\bibinfo {volume}
  {381}},\ \bibinfo {pages} {790} (\bibinfo {year} {2023})}\BibitemShut
  {NoStop}%
\bibitem [{\citenamefont {Zaanen}\ \emph {et~al.}(2015)\citenamefont {Zaanen},
  \citenamefont {Liu}, \citenamefont {Sun},\ and\ \citenamefont
  {Schalm}}]{Zaanen-CUP}%
  \BibitemOpen
  \bibfield  {author} {\bibinfo {author} {\bibfnamefont {J.}~\bibnamefont
  {Zaanen}}, \bibinfo {author} {\bibfnamefont {Y.}~\bibnamefont {Liu}},
  \bibinfo {author} {\bibfnamefont {Y.-W.}\ \bibnamefont {Sun}},\ and\ \bibinfo
  {author} {\bibfnamefont {K.}~\bibnamefont {Schalm}},\ }\href@noop {} {\emph
  {\bibinfo {title} {Holographic Duality in Condensed Matter Physics}}}\
  (\bibinfo  {publisher} {Cambridge University Press},\ \bibinfo {address}
  {Cambridge},\ \bibinfo {year} {2015})\BibitemShut {NoStop}%
\bibitem [{\citenamefont {{Hartnoll}}\ \emph {et~al.}(2016)\citenamefont
  {{Hartnoll}}, \citenamefont {{Lucas}},\ and\ \citenamefont
  {{Sachdev}}}]{Sachdev-MIT}%
  \BibitemOpen
  \bibfield  {author} {\bibinfo {author} {\bibfnamefont {S.~A.}\ \bibnamefont
  {{Hartnoll}}}, \bibinfo {author} {\bibfnamefont {A.}~\bibnamefont
  {{Lucas}}},\ and\ \bibinfo {author} {\bibfnamefont {S.}~\bibnamefont
  {{Sachdev}}},\ }\bibfield  {title} {\bibinfo {title} {Holographic quantum
  matter},\ }\href {https://doi.org/10.48550/arXiv.1612.07324} {\bibfield
  {journal} {\bibinfo  {journal} {arXiv e-prints}\ ,\ \bibinfo {eid}
  {arXiv:1612.07324}} (\bibinfo {year} {2016})},\ \Eprint
  {https://arxiv.org/abs/1612.07324} {arXiv:1612.07324 [hep-th]} \BibitemShut
  {NoStop}%
\bibitem [{\citenamefont {Phillips}\ \emph {et~al.}(2022)\citenamefont
  {Phillips}, \citenamefont {Hussey},\ and\ \citenamefont
  {Abbamonte}}]{Phillips_review}%
  \BibitemOpen
  \bibfield  {author} {\bibinfo {author} {\bibfnamefont {P.~W.}\ \bibnamefont
  {Phillips}}, \bibinfo {author} {\bibfnamefont {N.~E.}\ \bibnamefont
  {Hussey}},\ and\ \bibinfo {author} {\bibfnamefont {P.}~\bibnamefont
  {Abbamonte}},\ }\bibfield  {title} {\bibinfo {title} {Stranger than metals},\
  }\href {https://doi.org/10.1126/science.abh4273} {\bibfield  {journal}
  {\bibinfo  {journal} {Science}\ }\textbf {\bibinfo {volume} {377}},\ \bibinfo
  {pages} {eabh4273} (\bibinfo {year} {2022})}\BibitemShut {NoStop}%
\bibitem [{\citenamefont {Else}\ \emph {et~al.}(2021)\citenamefont {Else},
  \citenamefont {Thorngren},\ and\ \citenamefont {Senthil}}]{Else_2021}%
  \BibitemOpen
  \bibfield  {author} {\bibinfo {author} {\bibfnamefont {D.~V.}\ \bibnamefont
  {Else}}, \bibinfo {author} {\bibfnamefont {R.}~\bibnamefont {Thorngren}},\
  and\ \bibinfo {author} {\bibfnamefont {T.}~\bibnamefont {Senthil}},\
  }\bibfield  {title} {\bibinfo {title} {Non-{F}ermi liquids as ersatz {F}ermi
  liquids: {G}eneral constraints on compressible metals},\ }\href
  {https://doi.org/10.1103/PhysRevX.11.021005} {\bibfield  {journal} {\bibinfo
  {journal} {Phys. Rev. X}\ }\textbf {\bibinfo {volume} {11}},\ \bibinfo
  {pages} {021005} (\bibinfo {year} {2021})}\BibitemShut {NoStop}%
\bibitem [{\citenamefont {Varma}\ \emph {et~al.}(1989)\citenamefont {Varma},
  \citenamefont {Littlewood}, \citenamefont {Schmitt-Rink}, \citenamefont
  {Abrahams},\ and\ \citenamefont {Ruckenstein}}]{Varma_1989}%
  \BibitemOpen
  \bibfield  {author} {\bibinfo {author} {\bibfnamefont {C.~M.}\ \bibnamefont
  {Varma}}, \bibinfo {author} {\bibfnamefont {P.~B.}\ \bibnamefont
  {Littlewood}}, \bibinfo {author} {\bibfnamefont {S.}~\bibnamefont
  {Schmitt-Rink}}, \bibinfo {author} {\bibfnamefont {E.}~\bibnamefont
  {Abrahams}},\ and\ \bibinfo {author} {\bibfnamefont {A.~E.}\ \bibnamefont
  {Ruckenstein}},\ }\bibfield  {title} {\bibinfo {title} {{Phenomenology of the
  normal state of Cu-O high-temperature superconductors}},\ }\href
  {https://doi.org/10.1103/PhysRevLett.63.1996} {\bibfield  {journal} {\bibinfo
   {journal} {Phys. Rev. Lett.}\ }\textbf {\bibinfo {volume} {63}},\ \bibinfo
  {pages} {1996} (\bibinfo {year} {1989})}\BibitemShut {NoStop}%
\bibitem [{\citenamefont {Mahajan}\ \emph {et~al.}(2013)\citenamefont
  {Mahajan}, \citenamefont {Barkeshli},\ and\ \citenamefont
  {Hartnoll}}]{Hartnoll-PRB_2013}%
  \BibitemOpen
  \bibfield  {author} {\bibinfo {author} {\bibfnamefont {R.}~\bibnamefont
  {Mahajan}}, \bibinfo {author} {\bibfnamefont {M.}~\bibnamefont {Barkeshli}},\
  and\ \bibinfo {author} {\bibfnamefont {S.~A.}\ \bibnamefont {Hartnoll}},\
  }\bibfield  {title} {\bibinfo {title} {{Non-Fermi liquids and the
  Wiedemann-Franz law}},\ }\href {https://doi.org/10.1103/PhysRevB.88.125107}
  {\bibfield  {journal} {\bibinfo  {journal} {Phys. Rev. B}\ }\textbf {\bibinfo
  {volume} {88}},\ \bibinfo {pages} {125107} (\bibinfo {year}
  {2013})}\BibitemShut {NoStop}%
\bibitem [{\citenamefont {Patel}\ and\ \citenamefont
  {Sachdev}(2014)}]{Patel-PRB}%
  \BibitemOpen
  \bibfield  {author} {\bibinfo {author} {\bibfnamefont {A.~A.}\ \bibnamefont
  {Patel}}\ and\ \bibinfo {author} {\bibfnamefont {S.}~\bibnamefont
  {Sachdev}},\ }\bibfield  {title} {\bibinfo {title} {dc resistivity at the
  onset of spin density wave order in two-dimensional metals},\ }\href
  {https://doi.org/10.1103/PhysRevB.90.165146} {\bibfield  {journal} {\bibinfo
  {journal} {Phys. Rev. B}\ }\textbf {\bibinfo {volume} {90}},\ \bibinfo
  {pages} {165146} (\bibinfo {year} {2014})}\BibitemShut {NoStop}%
\bibitem [{\citenamefont {Vieira}\ \emph {et~al.}(2020)\citenamefont {Vieira},
  \citenamefont {{de Carvalho}},\ and\ \citenamefont
  {Freire}}]{Freire-AP_2020}%
  \BibitemOpen
  \bibfield  {author} {\bibinfo {author} {\bibfnamefont {L.~E.}\ \bibnamefont
  {Vieira}}, \bibinfo {author} {\bibfnamefont {V.~S.}\ \bibnamefont {{de
  Carvalho}}},\ and\ \bibinfo {author} {\bibfnamefont {H.}~\bibnamefont
  {Freire}},\ }\bibfield  {title} {\bibinfo {title} {{DC resistivity near a
  nematic quantum critical point: Effects of weak disorder and acoustic
  phonons}},\ }\href
  {https://doi.org/https://doi.org/10.1016/j.aop.2020.168230", url =
  "http://www.sciencedirect.com/science/article/pii/S0003491620301640}
  {\bibfield  {journal} {\bibinfo  {journal} {Annals of Physics}\ }\textbf
  {\bibinfo {volume} {419}},\ \bibinfo {pages} {168230} (\bibinfo {year}
  {2020})}\BibitemShut {NoStop}%
\bibitem [{\citenamefont {Banerjee}\ \emph {et~al.}(2021)\citenamefont
  {Banerjee}, \citenamefont {Grandadam}, \citenamefont {Freire},\ and\
  \citenamefont {P\'epin}}]{Anurag_2021}%
  \BibitemOpen
  \bibfield  {author} {\bibinfo {author} {\bibfnamefont {A.}~\bibnamefont
  {Banerjee}}, \bibinfo {author} {\bibfnamefont {M.}~\bibnamefont {Grandadam}},
  \bibinfo {author} {\bibfnamefont {H.}~\bibnamefont {Freire}},\ and\ \bibinfo
  {author} {\bibfnamefont {C.}~\bibnamefont {P\'epin}},\ }\bibfield  {title}
  {\bibinfo {title} {Strange metal from incoherent bosons},\ }\href
  {https://doi.org/10.1103/PhysRevB.104.054513} {\bibfield  {journal} {\bibinfo
   {journal} {Phys. Rev. B}\ }\textbf {\bibinfo {volume} {104}},\ \bibinfo
  {pages} {054513} (\bibinfo {year} {2021})}\BibitemShut {NoStop}%
\bibitem [{\citenamefont {P\'epin}\ and\ \citenamefont
  {Freire}(2023)}]{PEPIN2023}%
  \BibitemOpen
  \bibfield  {author} {\bibinfo {author} {\bibfnamefont {C.}~\bibnamefont
  {P\'epin}}\ and\ \bibinfo {author} {\bibfnamefont {H.}~\bibnamefont
  {Freire}},\ }\bibfield  {title} {\bibinfo {title} {Charge order and emergent
  symmetries in cuprate superconductors},\ }\href
  {https://doi.org/https://doi.org/10.1016/j.aop.2023.169233} {\bibfield
  {journal} {\bibinfo  {journal} {Annals of Physics}\ }\textbf {\bibinfo
  {volume} {456}},\ \bibinfo {pages} {169233} (\bibinfo {year}
  {2023})}\BibitemShut {NoStop}%
\bibitem [{\citenamefont {Pangburn}\ \emph {et~al.}(2023)\citenamefont
  {Pangburn}, \citenamefont {Banerjee}, \citenamefont {Freire},\ and\
  \citenamefont {P\'epin}}]{Pangburn_2023}%
  \BibitemOpen
  \bibfield  {author} {\bibinfo {author} {\bibfnamefont {E.}~\bibnamefont
  {Pangburn}}, \bibinfo {author} {\bibfnamefont {A.}~\bibnamefont {Banerjee}},
  \bibinfo {author} {\bibfnamefont {H.}~\bibnamefont {Freire}},\ and\ \bibinfo
  {author} {\bibfnamefont {C.}~\bibnamefont {P\'epin}},\ }\bibfield  {title}
  {\bibinfo {title} {Incoherent transport in a model for the strange metal
  phase: {M}emory-matrix formalism},\ }\href
  {https://doi.org/10.1103/PhysRevB.107.245109} {\bibfield  {journal} {\bibinfo
   {journal} {Phys. Rev. B}\ }\textbf {\bibinfo {volume} {107}},\ \bibinfo
  {pages} {245109} (\bibinfo {year} {2023})}\BibitemShut {NoStop}%
\bibitem [{\citenamefont {Berg}\ \emph {et~al.}(2019)\citenamefont {Berg},
  \citenamefont {Lederer}, \citenamefont {Schattner},\ and\ \citenamefont
  {Trebst}}]{Berg_2019}%
  \BibitemOpen
  \bibfield  {author} {\bibinfo {author} {\bibfnamefont {E.}~\bibnamefont
  {Berg}}, \bibinfo {author} {\bibfnamefont {S.}~\bibnamefont {Lederer}},
  \bibinfo {author} {\bibfnamefont {Y.}~\bibnamefont {Schattner}},\ and\
  \bibinfo {author} {\bibfnamefont {S.}~\bibnamefont {Trebst}},\ }\bibfield
  {title} {\bibinfo {title} {Monte carlo studies of quantum critical metals},\
  }\href {https://doi.org/10.1146/annurev-conmatphys-031218-013339} {\bibfield
  {journal} {\bibinfo  {journal} {Annual Review of Condensed Matter Physics}\
  }\textbf {\bibinfo {volume} {10}},\ \bibinfo {pages} {63} (\bibinfo {year}
  {2019})}\BibitemShut {NoStop}%
\bibitem [{\citenamefont {Liu}\ \emph {et~al.}(2019)\citenamefont {Liu},
  \citenamefont {Pan}, \citenamefont {Xu}, \citenamefont {Sun},\ and\
  \citenamefont {Meng}}]{ZYMeng}%
  \BibitemOpen
  \bibfield  {author} {\bibinfo {author} {\bibfnamefont {Z.~H.}\ \bibnamefont
  {Liu}}, \bibinfo {author} {\bibfnamefont {G.}~\bibnamefont {Pan}}, \bibinfo
  {author} {\bibfnamefont {X.~Y.}\ \bibnamefont {Xu}}, \bibinfo {author}
  {\bibfnamefont {K.}~\bibnamefont {Sun}},\ and\ \bibinfo {author}
  {\bibfnamefont {Z.~Y.}\ \bibnamefont {Meng}},\ }\bibfield  {title} {\bibinfo
  {title} {Itinerant quantum critical point with fermion pockets and
  hotspots},\ }\href {https://doi.org/10.1073/pnas.1901751116} {\bibfield
  {journal} {\bibinfo  {journal} {Proceedings of the National Academy of
  Sciences}\ }\textbf {\bibinfo {volume} {116}},\ \bibinfo {pages} {16760}
  (\bibinfo {year} {2019})}\BibitemShut {NoStop}%
\bibitem [{\citenamefont {Teixeira}\ \emph {et~al.}(2023)\citenamefont
  {Teixeira}, \citenamefont {P\'epin},\ and\ \citenamefont
  {Freire}}]{Teixeira_2023}%
  \BibitemOpen
  \bibfield  {author} {\bibinfo {author} {\bibfnamefont {R.~M.~P.}\
  \bibnamefont {Teixeira}}, \bibinfo {author} {\bibfnamefont {C.}~\bibnamefont
  {P\'epin}},\ and\ \bibinfo {author} {\bibfnamefont {H.}~\bibnamefont
  {Freire}},\ }\bibfield  {title} {\bibinfo {title} {{Strange metallicity in an
  antiferromagnetic quantum critical model: A sign-problem-free quantum Monte
  Carlo study}},\ }\href {https://doi.org/10.1103/PhysRevB.108.085131}
  {\bibfield  {journal} {\bibinfo  {journal} {Phys. Rev. B}\ }\textbf {\bibinfo
  {volume} {108}},\ \bibinfo {pages} {085131} (\bibinfo {year}
  {2023})}\BibitemShut {NoStop}%
\bibitem [{\citenamefont {Kondo}\ \emph {et~al.}(2015)\citenamefont {Kondo},
  \citenamefont {Nakayama}, \citenamefont {Chen}, \citenamefont {Ishikawa},
  \citenamefont {Moon}, \citenamefont {Yamamoto}, \citenamefont {Ota},
  \citenamefont {Malaeb}, \citenamefont {Kanai}, \citenamefont {Nakashima},\
  and\ \citenamefont {et~al.}}]{Kondo_2015}%
  \BibitemOpen
  \bibfield  {author} {\bibinfo {author} {\bibfnamefont {T.}~\bibnamefont
  {Kondo}}, \bibinfo {author} {\bibfnamefont {M.}~\bibnamefont {Nakayama}},
  \bibinfo {author} {\bibfnamefont {R.}~\bibnamefont {Chen}}, \bibinfo {author}
  {\bibfnamefont {J.~J.}\ \bibnamefont {Ishikawa}}, \bibinfo {author}
  {\bibfnamefont {E.-G.}\ \bibnamefont {Moon}}, \bibinfo {author}
  {\bibfnamefont {T.}~\bibnamefont {Yamamoto}}, \bibinfo {author}
  {\bibfnamefont {Y.}~\bibnamefont {Ota}}, \bibinfo {author} {\bibfnamefont
  {W.}~\bibnamefont {Malaeb}}, \bibinfo {author} {\bibfnamefont
  {H.}~\bibnamefont {Kanai}}, \bibinfo {author} {\bibfnamefont
  {Y.}~\bibnamefont {Nakashima}},\ and\ \bibinfo {author} {\bibnamefont
  {et~al.}},\ }\bibfield  {title} {\bibinfo {title} {Quadratic {F}ermi node in
  a 3{D} strongly correlated semimetal},\ }\href
  {http://dx.doi.org/10.1038/ncomms10042} {\bibfield  {journal} {\bibinfo
  {journal} {Nature Communications}\ }\textbf {\bibinfo {volume} {6}} (\bibinfo
  {year} {2015})}\BibitemShut {NoStop}%
\bibitem [{\citenamefont {Butch}\ \emph {et~al.}(2011)\citenamefont {Butch},
  \citenamefont {Syers}, \citenamefont {Kirshenbaum}, \citenamefont {Hope},\
  and\ \citenamefont {Paglione}}]{Paglione}%
  \BibitemOpen
  \bibfield  {author} {\bibinfo {author} {\bibfnamefont {N.~P.}\ \bibnamefont
  {Butch}}, \bibinfo {author} {\bibfnamefont {P.}~\bibnamefont {Syers}},
  \bibinfo {author} {\bibfnamefont {K.}~\bibnamefont {Kirshenbaum}}, \bibinfo
  {author} {\bibfnamefont {A.~P.}\ \bibnamefont {Hope}},\ and\ \bibinfo
  {author} {\bibfnamefont {J.}~\bibnamefont {Paglione}},\ }\bibfield  {title}
  {\bibinfo {title} {{Superconductivity in the topological semimetal YPtBi}},\
  }\href {https://doi.org/10.1103/PhysRevB.84.220504} {\bibfield  {journal}
  {\bibinfo  {journal} {Phys. Rev. B}\ }\textbf {\bibinfo {volume} {84}},\
  \bibinfo {pages} {220504} (\bibinfo {year} {2011})}\BibitemShut {NoStop}%
\bibitem [{\citenamefont {Tafti}\ \emph {et~al.}(2013)\citenamefont {Tafti},
  \citenamefont {Fujii}, \citenamefont {Juneau-Fecteau}, \citenamefont
  {Ren\'e~de Cotret}, \citenamefont {Doiron-Leyraud}, \citenamefont
  {Asamitsu},\ and\ \citenamefont {Taillefer}}]{Taillefer_2013}%
  \BibitemOpen
  \bibfield  {author} {\bibinfo {author} {\bibfnamefont {F.~F.}\ \bibnamefont
  {Tafti}}, \bibinfo {author} {\bibfnamefont {T.}~\bibnamefont {Fujii}},
  \bibinfo {author} {\bibfnamefont {A.}~\bibnamefont {Juneau-Fecteau}},
  \bibinfo {author} {\bibfnamefont {S.}~\bibnamefont {Ren\'e~de Cotret}},
  \bibinfo {author} {\bibfnamefont {N.}~\bibnamefont {Doiron-Leyraud}},
  \bibinfo {author} {\bibfnamefont {A.}~\bibnamefont {Asamitsu}},\ and\
  \bibinfo {author} {\bibfnamefont {L.}~\bibnamefont {Taillefer}},\ }\bibfield
  {title} {\bibinfo {title} {{Superconductivity in the noncentrosymmetric
  half-Heusler compound LuPtBi: A candidate for topological
  superconductivity}},\ }\href {https://doi.org/10.1103/PhysRevB.87.184504}
  {\bibfield  {journal} {\bibinfo  {journal} {Phys. Rev. B}\ }\textbf {\bibinfo
  {volume} {87}},\ \bibinfo {pages} {184504} (\bibinfo {year}
  {2013})}\BibitemShut {NoStop}%
\bibitem [{\citenamefont {Groves}\ and\ \citenamefont {Paul}(1963)}]{Groves}%
  \BibitemOpen
  \bibfield  {author} {\bibinfo {author} {\bibfnamefont {S.}~\bibnamefont
  {Groves}}\ and\ \bibinfo {author} {\bibfnamefont {W.}~\bibnamefont {Paul}},\
  }\bibfield  {title} {\bibinfo {title} {Band structure of gray tin},\ }\href
  {https://doi.org/10.1103/PhysRevLett.11.194} {\bibfield  {journal} {\bibinfo
  {journal} {Phys. Rev. Lett.}\ }\textbf {\bibinfo {volume} {11}},\ \bibinfo
  {pages} {194} (\bibinfo {year} {1963})}\BibitemShut {NoStop}%
\bibitem [{\citenamefont {Chamorro}\ \emph {et~al.}(2024)\citenamefont
  {Chamorro}, \citenamefont {Zuo}, \citenamefont {Bassey}, \citenamefont
  {Watkins}, \citenamefont {Zhu}, \citenamefont {Zohar}, \citenamefont
  {Wyckoff}, \citenamefont {Kinnibrugh}, \citenamefont {Lapidus}, \citenamefont
  {Stemmer}, \citenamefont {Clément}, \citenamefont {Wilson},\ and\
  \citenamefont {Seshadri}}]{Chamorro_2024}%
  \BibitemOpen
  \bibfield  {author} {\bibinfo {author} {\bibfnamefont {J.~R.}\ \bibnamefont
  {Chamorro}}, \bibinfo {author} {\bibfnamefont {J.~L.}\ \bibnamefont {Zuo}},
  \bibinfo {author} {\bibfnamefont {E.~N.}\ \bibnamefont {Bassey}}, \bibinfo
  {author} {\bibfnamefont {A.~K.}\ \bibnamefont {Watkins}}, \bibinfo {author}
  {\bibfnamefont {G.}~\bibnamefont {Zhu}}, \bibinfo {author} {\bibfnamefont
  {A.}~\bibnamefont {Zohar}}, \bibinfo {author} {\bibfnamefont {K.~E.}\
  \bibnamefont {Wyckoff}}, \bibinfo {author} {\bibfnamefont {T.~L.}\
  \bibnamefont {Kinnibrugh}}, \bibinfo {author} {\bibfnamefont {S.~H.}\
  \bibnamefont {Lapidus}}, \bibinfo {author} {\bibfnamefont {S.}~\bibnamefont
  {Stemmer}}, \bibinfo {author} {\bibfnamefont {R.~J.}\ \bibnamefont
  {Clément}}, \bibinfo {author} {\bibfnamefont {S.~D.}\ \bibnamefont
  {Wilson}},\ and\ \bibinfo {author} {\bibfnamefont {R.}~\bibnamefont
  {Seshadri}},\ }\bibfield  {title} {\bibinfo {title} {{Soft-chemical
  synthesis, structure evolution, and insulator-to-metal transition in
  pyrochlore-like $\lambda$-RhO$_{2}$}},\ }\href
  {https://doi.org/10.1021/acs.chemmater.3c02814} {\bibfield  {journal}
  {\bibinfo  {journal} {Chemistry of Materials}\ }\textbf {\bibinfo {volume}
  {36}},\ \bibinfo {pages} {1547} (\bibinfo {year} {2024})}\BibitemShut
  {NoStop}%
\bibitem [{\citenamefont {Herbut}\ and\ \citenamefont
  {Janssen}(2014)}]{Janssen_Herbut_1}%
  \BibitemOpen
  \bibfield  {author} {\bibinfo {author} {\bibfnamefont {I.~F.}\ \bibnamefont
  {Herbut}}\ and\ \bibinfo {author} {\bibfnamefont {L.}~\bibnamefont
  {Janssen}},\ }\bibfield  {title} {\bibinfo {title} {Topological {M}ott
  insulator in three-dimensional systems with quadratic band touching},\ }\href
  {https://doi.org/10.1103/PhysRevLett.113.106401} {\bibfield  {journal}
  {\bibinfo  {journal} {Phys. Rev. Lett.}\ }\textbf {\bibinfo {volume} {113}},\
  \bibinfo {pages} {106401} (\bibinfo {year} {2014})}\BibitemShut {NoStop}%
\bibitem [{\citenamefont {Janssen}\ and\ \citenamefont
  {Herbut}(2017)}]{Janssen_Herbut_2}%
  \BibitemOpen
  \bibfield  {author} {\bibinfo {author} {\bibfnamefont {L.}~\bibnamefont
  {Janssen}}\ and\ \bibinfo {author} {\bibfnamefont {I.~F.}\ \bibnamefont
  {Herbut}},\ }\bibfield  {title} {\bibinfo {title} {{Phase diagram of
  electronic systems with quadratic Fermi nodes in $2 <d <4 : 2+\varepsilon$
  expansion, $4 -\varepsilon$ expansion, and functional renormalization
  group}},\ }\href {https://doi.org/10.1103/PhysRevB.95.075101} {\bibfield
  {journal} {\bibinfo  {journal} {Phys. Rev. B}\ }\textbf {\bibinfo {volume}
  {95}},\ \bibinfo {pages} {075101} (\bibinfo {year} {2017})}\BibitemShut
  {NoStop}%
\bibitem [{\citenamefont {Mandal}(2019)}]{ips_qbt_plasmons}%
  \BibitemOpen
  \bibfield  {author} {\bibinfo {author} {\bibfnamefont {I.}~\bibnamefont
  {Mandal}},\ }\bibfield  {title} {\bibinfo {title} {{Search for plasmons in
  isotropic Luttinger semimetals}},\ }\href
  {https://doi.org/10.1016/j.aop.2019.04.002} {\bibfield  {journal} {\bibinfo
  {journal} {Annals of Physics}\ }\textbf {\bibinfo {volume} {406}},\ \bibinfo
  {pages} {173} (\bibinfo {year} {2019})}\BibitemShut {NoStop}%
\bibitem [{\citenamefont {Mandal}(2020{\natexlab{b}})}]{ips_qbt_tunnel}%
  \BibitemOpen
  \bibfield  {author} {\bibinfo {author} {\bibfnamefont {I.}~\bibnamefont
  {Mandal}},\ }\bibfield  {title} {\bibinfo {title} {{Tunneling in Fermi
  systems with quadratic band crossing points}},\ }\href
  {https://doi.org/10.1016/j.aop.2020.168235} {\bibfield  {journal} {\bibinfo
  {journal} {Annals of Physics}\ }\textbf {\bibinfo {volume} {419}},\ \bibinfo
  {pages} {168235} (\bibinfo {year} {2020}{\natexlab{b}})}\BibitemShut
  {NoStop}%
\bibitem [{\citenamefont {Bera}\ and\ \citenamefont
  {Mandal}(2021)}]{ips-sandip}%
  \BibitemOpen
  \bibfield  {author} {\bibinfo {author} {\bibfnamefont {S.}~\bibnamefont
  {Bera}}\ and\ \bibinfo {author} {\bibfnamefont {I.}~\bibnamefont {Mandal}},\
  }\bibfield  {title} {\bibinfo {title} {{Floquet scattering of quadratic
  band-touching semimetals through a time-periodic potential well}},\ }\href
  {https://doi.org/10.1088/1361-648X/ac020a} {\bibfield  {journal} {\bibinfo
  {journal} {J. Phys. Condens. Matter}\ }\textbf {\bibinfo {volume} {33}},\
  \bibinfo {pages} {295502} (\bibinfo {year} {2021})}\BibitemShut {NoStop}%
\bibitem [{\citenamefont {Tchoumakov}\ and\ \citenamefont
  {Witczak-Krempa}(2019)}]{krempa}%
  \BibitemOpen
  \bibfield  {author} {\bibinfo {author} {\bibfnamefont {S.}~\bibnamefont
  {Tchoumakov}}\ and\ \bibinfo {author} {\bibfnamefont {W.}~\bibnamefont
  {Witczak-Krempa}},\ }\bibfield  {title} {\bibinfo {title} {{Dielectric and
  electronic properties of three-dimensional Luttinger semimetals with a
  quadratic band touching}},\ }\href
  {https://doi.org/10.1103/PhysRevB.100.075104} {\bibfield  {journal} {\bibinfo
   {journal} {Phys. Rev. B}\ }\textbf {\bibinfo {volume} {100}},\ \bibinfo
  {pages} {075104} (\bibinfo {year} {2019})}\BibitemShut {NoStop}%
\bibitem [{\citenamefont {Wang}\ and\ \citenamefont {Mandal}(2023)}]{ips-jing}%
  \BibitemOpen
  \bibfield  {author} {\bibinfo {author} {\bibfnamefont {J.}~\bibnamefont
  {Wang}}\ and\ \bibinfo {author} {\bibfnamefont {I.}~\bibnamefont {Mandal}},\
  }\bibfield  {title} {\bibinfo {title} {{Anatomy of plasmons in generic
  Luttinger semimetals}},\ }\href
  {https://doi.org/10.1140/epjb/s10051-023-00596-x} {\bibfield  {journal}
  {\bibinfo  {journal} {Eur. Phys. J. B}\ }\textbf {\bibinfo {volume} {96}},\
  \bibinfo {pages} {132} (\bibinfo {year} {2023})}\BibitemShut {NoStop}%
\bibitem [{\citenamefont {Kennett}\ \emph {et~al.}(2011)\citenamefont
  {Kennett}, \citenamefont {Komeilizadeh}, \citenamefont {Kaveh},\ and\
  \citenamefont {Smith}}]{malcolm}%
  \BibitemOpen
  \bibfield  {author} {\bibinfo {author} {\bibfnamefont {M.~P.}\ \bibnamefont
  {Kennett}}, \bibinfo {author} {\bibfnamefont {N.}~\bibnamefont
  {Komeilizadeh}}, \bibinfo {author} {\bibfnamefont {K.}~\bibnamefont
  {Kaveh}},\ and\ \bibinfo {author} {\bibfnamefont {P.~M.}\ \bibnamefont
  {Smith}},\ }\bibfield  {title} {\bibinfo {title} {{Birefringent breakup of
  Dirac fermions on a square optical lattice}},\ }\href
  {https://doi.org/10.1103/PhysRevA.83.053636} {\bibfield  {journal} {\bibinfo
  {journal} {Phys. Rev. A}\ }\textbf {\bibinfo {volume} {83}},\ \bibinfo
  {pages} {053636} (\bibinfo {year} {2011})}\BibitemShut {NoStop}%
\bibitem [{\citenamefont {Roy}\ \emph {et~al.}(2012)\citenamefont {Roy},
  \citenamefont {Smith},\ and\ \citenamefont {Kennett}}]{prb.85.235119}%
  \BibitemOpen
  \bibfield  {author} {\bibinfo {author} {\bibfnamefont {B.}~\bibnamefont
  {Roy}}, \bibinfo {author} {\bibfnamefont {P.~M.}\ \bibnamefont {Smith}},\
  and\ \bibinfo {author} {\bibfnamefont {M.~P.}\ \bibnamefont {Kennett}},\
  }\bibfield  {title} {\bibinfo {title} {{Asymmetric spatial structure of zero
  modes for birefringent Dirac fermions}},\ }\href
  {https://doi.org/10.1103/PhysRevB.85.235119} {\bibfield  {journal} {\bibinfo
  {journal} {Phys. Rev. B}\ }\textbf {\bibinfo {volume} {85}},\ \bibinfo
  {pages} {235119} (\bibinfo {year} {2012})}\BibitemShut {NoStop}%
\bibitem [{\citenamefont {Komeilizadeh}\ and\ \citenamefont
  {Kennett}(2014)}]{prb.90.045131}%
  \BibitemOpen
  \bibfield  {author} {\bibinfo {author} {\bibfnamefont {N.}~\bibnamefont
  {Komeilizadeh}}\ and\ \bibinfo {author} {\bibfnamefont {M.~P.}\ \bibnamefont
  {Kennett}},\ }\bibfield  {title} {\bibinfo {title} {Instabilities of a
  birefringent semimetal},\ }\href {https://doi.org/10.1103/PhysRevB.90.045131}
  {\bibfield  {journal} {\bibinfo  {journal} {Phys. Rev. B}\ }\textbf {\bibinfo
  {volume} {90}},\ \bibinfo {pages} {045131} (\bibinfo {year}
  {2014})}\BibitemShut {NoStop}%
\bibitem [{\citenamefont {D\'ora}\ \emph {et~al.}(2011)\citenamefont {D\'ora},
  \citenamefont {Kailasvuori},\ and\ \citenamefont {Moessner}}]{prb84.195422}%
  \BibitemOpen
  \bibfield  {author} {\bibinfo {author} {\bibfnamefont {B.}~\bibnamefont
  {D\'ora}}, \bibinfo {author} {\bibfnamefont {J.}~\bibnamefont
  {Kailasvuori}},\ and\ \bibinfo {author} {\bibfnamefont {R.}~\bibnamefont
  {Moessner}},\ }\bibfield  {title} {\bibinfo {title} {Lattice generalization
  of the dirac equation to general spin and the role of the flat band},\ }\href
  {https://doi.org/10.1103/PhysRevB.84.195422} {\bibfield  {journal} {\bibinfo
  {journal} {Phys. Rev. B}\ }\textbf {\bibinfo {volume} {84}},\ \bibinfo
  {pages} {195422} (\bibinfo {year} {2011})}\BibitemShut {NoStop}%
\bibitem [{\citenamefont {Watanabe}\ \emph {et~al.}(2011)\citenamefont
  {Watanabe}, \citenamefont {Hatsugai},\ and\ \citenamefont
  {Aoki}}]{watanabe_2011}%
  \BibitemOpen
  \bibfield  {author} {\bibinfo {author} {\bibfnamefont {H.}~\bibnamefont
  {Watanabe}}, \bibinfo {author} {\bibfnamefont {Y.}~\bibnamefont {Hatsugai}},\
  and\ \bibinfo {author} {\bibfnamefont {H.}~\bibnamefont {Aoki}},\ }\bibfield
  {title} {\bibinfo {title} {{Manipulation of the Dirac cones and the anomaly
  in the graphene related quantum Hall effect}},\ }\href
  {https://doi.org/10.1088/1742-6596/334/1/012044} {\bibfield  {journal}
  {\bibinfo  {journal} {Journal of Physics: Conference Series}\ }\textbf
  {\bibinfo {volume} {334}},\ \bibinfo {pages} {012044} (\bibinfo {year}
  {2011})}\BibitemShut {NoStop}%
\bibitem [{\citenamefont {Lan}\ \emph {et~al.}(2011{\natexlab{a}})\citenamefont
  {Lan}, \citenamefont {Goldman}, \citenamefont {Bermudez}, \citenamefont
  {Lu},\ and\ \citenamefont {\"Ohberg}}]{prb.84.165115}%
  \BibitemOpen
  \bibfield  {author} {\bibinfo {author} {\bibfnamefont {Z.}~\bibnamefont
  {Lan}}, \bibinfo {author} {\bibfnamefont {N.}~\bibnamefont {Goldman}},
  \bibinfo {author} {\bibfnamefont {A.}~\bibnamefont {Bermudez}}, \bibinfo
  {author} {\bibfnamefont {W.}~\bibnamefont {Lu}},\ and\ \bibinfo {author}
  {\bibfnamefont {P.}~\bibnamefont {\"Ohberg}},\ }\bibfield  {title} {\bibinfo
  {title} {{Dirac-Weyl fermions with arbitrary spin in two-dimensional optical
  superlattices}},\ }\href {https://doi.org/10.1103/PhysRevB.84.165115}
  {\bibfield  {journal} {\bibinfo  {journal} {Phys. Rev. B}\ }\textbf {\bibinfo
  {volume} {84}},\ \bibinfo {pages} {165115} (\bibinfo {year}
  {2011}{\natexlab{a}})}\BibitemShut {NoStop}%
\bibitem [{\citenamefont {Lan}\ \emph {et~al.}(2011{\natexlab{b}})\citenamefont
  {Lan}, \citenamefont {Celi}, \citenamefont {Lu}, \citenamefont {\"Ohberg},\
  and\ \citenamefont {Lewenstein}}]{prl.107.253001}%
  \BibitemOpen
  \bibfield  {author} {\bibinfo {author} {\bibfnamefont {Z.}~\bibnamefont
  {Lan}}, \bibinfo {author} {\bibfnamefont {A.}~\bibnamefont {Celi}}, \bibinfo
  {author} {\bibfnamefont {W.}~\bibnamefont {Lu}}, \bibinfo {author}
  {\bibfnamefont {P.}~\bibnamefont {\"Ohberg}},\ and\ \bibinfo {author}
  {\bibfnamefont {M.}~\bibnamefont {Lewenstein}},\ }\bibfield  {title}
  {\bibinfo {title} {Tunable multiple layered {D}irac cones in optical
  lattices},\ }\href {https://doi.org/10.1103/PhysRevLett.107.253001}
  {\bibfield  {journal} {\bibinfo  {journal} {Phys. Rev. Lett.}\ }\textbf
  {\bibinfo {volume} {107}},\ \bibinfo {pages} {253001} (\bibinfo {year}
  {2011}{\natexlab{b}})}\BibitemShut {NoStop}%
\bibitem [{\citenamefont {Bradlyn}\ \emph {et~al.}(2016)\citenamefont
  {Bradlyn}, \citenamefont {Cano}, \citenamefont {Wang}, \citenamefont
  {Vergniory}, \citenamefont {Felser}, \citenamefont {Cava},\ and\
  \citenamefont {Bernevig}}]{bradlyn}%
  \BibitemOpen
  \bibfield  {author} {\bibinfo {author} {\bibfnamefont {B.}~\bibnamefont
  {Bradlyn}}, \bibinfo {author} {\bibfnamefont {J.}~\bibnamefont {Cano}},
  \bibinfo {author} {\bibfnamefont {Z.}~\bibnamefont {Wang}}, \bibinfo {author}
  {\bibfnamefont {M.~G.}\ \bibnamefont {Vergniory}}, \bibinfo {author}
  {\bibfnamefont {C.}~\bibnamefont {Felser}}, \bibinfo {author} {\bibfnamefont
  {R.~J.}\ \bibnamefont {Cava}},\ and\ \bibinfo {author} {\bibfnamefont
  {B.~A.}\ \bibnamefont {Bernevig}},\ }\bibfield  {title} {\bibinfo {title}
  {{Beyond Dirac and Weyl fermions: Unconventional quasiparticles in
  conventional crystals}},\ }\href {https://doi.org/10.1126/science.aaf5037}
  {\bibfield  {journal} {\bibinfo  {journal} {Science}\ }\textbf {\bibinfo
  {volume} {353}},\ \bibinfo {pages} {aaf5037} (\bibinfo {year}
  {2016})}\BibitemShut {NoStop}%
\bibitem [{\citenamefont {Ezawa}(2016)}]{prb.94.195205}%
  \BibitemOpen
  \bibfield  {author} {\bibinfo {author} {\bibfnamefont {M.}~\bibnamefont
  {Ezawa}},\ }\bibfield  {title} {\bibinfo {title} {{Pseudospin-$\frac{3}{2}$
  fermions, type-II Weyl semimetals, and critical Weyl semimetals in tricolor
  cubic lattices}},\ }\href {https://doi.org/10.1103/PhysRevB.94.195205}
  {\bibfield  {journal} {\bibinfo  {journal} {Phys. Rev. B}\ }\textbf {\bibinfo
  {volume} {94}},\ \bibinfo {pages} {195205} (\bibinfo {year}
  {2016})}\BibitemShut {NoStop}%
\bibitem [{\citenamefont {Hsieh}\ \emph {et~al.}(2014)\citenamefont {Hsieh},
  \citenamefont {Liu},\ and\ \citenamefont {Fu}}]{PhysRevB.90.081112}%
  \BibitemOpen
  \bibfield  {author} {\bibinfo {author} {\bibfnamefont {T.~H.}\ \bibnamefont
  {Hsieh}}, \bibinfo {author} {\bibfnamefont {J.}~\bibnamefont {Liu}},\ and\
  \bibinfo {author} {\bibfnamefont {L.}~\bibnamefont {Fu}},\ }\bibfield
  {title} {\bibinfo {title} {{Topological crystalline insulators and Dirac
  octets in antiperovskites}},\ }\href
  {https://doi.org/10.1103/PhysRevB.90.081112} {\bibfield  {journal} {\bibinfo
  {journal} {Phys. Rev. B}\ }\textbf {\bibinfo {volume} {90}},\ \bibinfo
  {pages} {081112} (\bibinfo {year} {2014})}\BibitemShut {NoStop}%
\bibitem [{\citenamefont {Chen}\ \emph {et~al.}(2017)\citenamefont {Chen},
  \citenamefont {Wang}, \citenamefont {Liu}, \citenamefont {Yu}, \citenamefont
  {Sheng}, \citenamefont {Chen},\ and\ \citenamefont
  {Yang}}]{PhysRevMaterials.1.044201}%
  \BibitemOpen
  \bibfield  {author} {\bibinfo {author} {\bibfnamefont {C.}~\bibnamefont
  {Chen}}, \bibinfo {author} {\bibfnamefont {S.-S.}\ \bibnamefont {Wang}},
  \bibinfo {author} {\bibfnamefont {L.}~\bibnamefont {Liu}}, \bibinfo {author}
  {\bibfnamefont {Z.-M.}\ \bibnamefont {Yu}}, \bibinfo {author} {\bibfnamefont
  {X.-L.}\ \bibnamefont {Sheng}}, \bibinfo {author} {\bibfnamefont
  {Z.}~\bibnamefont {Chen}},\ and\ \bibinfo {author} {\bibfnamefont {S.~A.}\
  \bibnamefont {Yang}},\ }\bibfield  {title} {\bibinfo {title} {{Ternary
  wurtzite CaAgBi materials family: A playground for essential and accidental,
  type-I and type-II Dirac fermions}},\ }\href
  {https://doi.org/10.1103/PhysRevMaterials.1.044201} {\bibfield  {journal}
  {\bibinfo  {journal} {Phys. Rev. Materials}\ }\textbf {\bibinfo {volume}
  {1}},\ \bibinfo {pages} {044201} (\bibinfo {year} {2017})}\BibitemShut
  {NoStop}%
\bibitem [{\citenamefont {Isobe}\ \emph {et~al.}(2016)\citenamefont {Isobe},
  \citenamefont {Yang}, \citenamefont {Chubukov}, \citenamefont {Schmalian},\
  and\ \citenamefont {Nagaosa}}]{Isobe_2016}%
  \BibitemOpen
  \bibfield  {author} {\bibinfo {author} {\bibfnamefont {H.}~\bibnamefont
  {Isobe}}, \bibinfo {author} {\bibfnamefont {B.-J.}\ \bibnamefont {Yang}},
  \bibinfo {author} {\bibfnamefont {A.}~\bibnamefont {Chubukov}}, \bibinfo
  {author} {\bibfnamefont {J.}~\bibnamefont {Schmalian}},\ and\ \bibinfo
  {author} {\bibfnamefont {N.}~\bibnamefont {Nagaosa}},\ }\bibfield  {title}
  {\bibinfo {title} {Emergent non-{F}ermi-liquid at the quantum critical point
  of a topological phase transition in two dimensions},\ }\href
  {https://doi.org/10.1103/PhysRevLett.116.076803} {\bibfield  {journal}
  {\bibinfo  {journal} {Phys. Rev. Lett.}\ }\textbf {\bibinfo {volume} {116}},\
  \bibinfo {pages} {076803} (\bibinfo {year} {2016})}\BibitemShut {NoStop}%
\bibitem [{\citenamefont {{Cho}}\ and\ \citenamefont
  {{Moon}}(2016)}]{cho2016novel}%
  \BibitemOpen
  \bibfield  {author} {\bibinfo {author} {\bibfnamefont {G.~Y.}\ \bibnamefont
  {{Cho}}}\ and\ \bibinfo {author} {\bibfnamefont {E.-G.}\ \bibnamefont
  {{Moon}}},\ }\bibfield  {title} {\bibinfo {title} {{Novel Quantum Criticality
  in Two Dimensional Topological Phase transitions}},\ }\href
  {https://doi.org/10.1038/srep19198} {\bibfield  {journal} {\bibinfo
  {journal} {Scientific Reports}\ }\textbf {\bibinfo {volume} {6}},\ \bibinfo
  {eid} {19198} (\bibinfo {year} {2016})}\BibitemShut {NoStop}%
\bibitem [{\citenamefont {Wang}\ \emph {et~al.}(2019)\citenamefont {Wang},
  \citenamefont {Liu},\ and\ \citenamefont {Zhang}}]{Jing-Rong}%
  \BibitemOpen
  \bibfield  {author} {\bibinfo {author} {\bibfnamefont {J.-R.}\ \bibnamefont
  {Wang}}, \bibinfo {author} {\bibfnamefont {G.-Z.}\ \bibnamefont {Liu}},\ and\
  \bibinfo {author} {\bibfnamefont {C.-J.}\ \bibnamefont {Zhang}},\ }\bibfield
  {title} {\bibinfo {title} {{Topological quantum critical point in a
  triple-Weyl semimetal: Non-Fermi-liquid behavior and instabilities}},\ }\href
  {https://doi.org/10.1103/PhysRevB.99.195119} {\bibfield  {journal} {\bibinfo
  {journal} {Phys. Rev. B}\ }\textbf {\bibinfo {volume} {99}},\ \bibinfo
  {pages} {195119} (\bibinfo {year} {2019})}\BibitemShut {NoStop}%
\bibitem [{\citenamefont {Han}\ \emph {et~al.}(2019)\citenamefont {Han},
  \citenamefont {Lee}, \citenamefont {Moon},\ and\ \citenamefont
  {Min}}]{SangEun}%
  \BibitemOpen
  \bibfield  {author} {\bibinfo {author} {\bibfnamefont {S.}~\bibnamefont
  {Han}}, \bibinfo {author} {\bibfnamefont {C.}~\bibnamefont {Lee}}, \bibinfo
  {author} {\bibfnamefont {E.-G.}\ \bibnamefont {Moon}},\ and\ \bibinfo
  {author} {\bibfnamefont {H.}~\bibnamefont {Min}},\ }\bibfield  {title}
  {\bibinfo {title} {Emergent anisotropic non-{F}ermi liquid at a topological
  phase transition in three dimensions},\ }\href
  {https://doi.org/10.1103/PhysRevLett.122.187601} {\bibfield  {journal}
  {\bibinfo  {journal} {Phys. Rev. Lett.}\ }\textbf {\bibinfo {volume} {122}},\
  \bibinfo {pages} {187601} (\bibinfo {year} {2019})}\BibitemShut {NoStop}%
\bibitem [{\citenamefont {Zhang}\ \emph {et~al.}(2021)\citenamefont {Zhang},
  \citenamefont {Jian},\ and\ \citenamefont {Yao}}]{Shi-Xin}%
  \BibitemOpen
  \bibfield  {author} {\bibinfo {author} {\bibfnamefont {S.-X.}\ \bibnamefont
  {Zhang}}, \bibinfo {author} {\bibfnamefont {S.-K.}\ \bibnamefont {Jian}},\
  and\ \bibinfo {author} {\bibfnamefont {H.}~\bibnamefont {Yao}},\ }\bibfield
  {title} {\bibinfo {title} {Quantum criticality preempted by nematicity},\
  }\href {https://doi.org/10.1103/PhysRevB.103.165129} {\bibfield  {journal}
  {\bibinfo  {journal} {Phys. Rev. B}\ }\textbf {\bibinfo {volume} {103}},\
  \bibinfo {pages} {165129} (\bibinfo {year} {2021})}\BibitemShut {NoStop}%
\bibitem [{\citenamefont {Mandal}\ and\ \citenamefont
  {Freire}(2021)}]{ips_hermann1}%
  \BibitemOpen
  \bibfield  {author} {\bibinfo {author} {\bibfnamefont {I.}~\bibnamefont
  {Mandal}}\ and\ \bibinfo {author} {\bibfnamefont {H.}~\bibnamefont
  {Freire}},\ }\bibfield  {title} {\bibinfo {title} {{Transport in the
  non-Fermi liquid phase of isotropic Luttinger semimetals}},\ }\href
  {https://doi.org/10.1103/PhysRevB.103.195116} {\bibfield  {journal} {\bibinfo
   {journal} {Phys. Rev. B}\ }\textbf {\bibinfo {volume} {103}},\ \bibinfo
  {pages} {195116} (\bibinfo {year} {2021})}\BibitemShut {NoStop}%
\bibitem [{\citenamefont {Freire}\ and\ \citenamefont
  {Mandal}(2021)}]{ips-hermann2}%
  \BibitemOpen
  \bibfield  {author} {\bibinfo {author} {\bibfnamefont {H.}~\bibnamefont
  {Freire}}\ and\ \bibinfo {author} {\bibfnamefont {I.}~\bibnamefont
  {Mandal}},\ }\bibfield  {title} {\bibinfo {title} {{Thermoelectric and
  thermal properties of the weakly disordered non-Fermi liquid phase of
  Luttinger semimetals}},\ }\href
  {https://doi.org/10.1016/j.physleta.2021.127470} {\bibfield  {journal}
  {\bibinfo  {journal} {Physics Letters A}\ }\textbf {\bibinfo {volume}
  {407}},\ \bibinfo {pages} {127470} (\bibinfo {year} {2021})}\BibitemShut
  {NoStop}%
\bibitem [{\citenamefont {Mandal}\ and\ \citenamefont
  {Freire}(2022)}]{ips-hermann3}%
  \BibitemOpen
  \bibfield  {author} {\bibinfo {author} {\bibfnamefont {I.}~\bibnamefont
  {Mandal}}\ and\ \bibinfo {author} {\bibfnamefont {H.}~\bibnamefont
  {Freire}},\ }\bibfield  {title} {\bibinfo {title} {{Raman response and shear
  viscosity in the non-Fermi liquid phase of Luttinger semimetals}},\ }\href
  {https://doi.org/10.1088/1361-648x/ac6785} {\bibfield  {journal} {\bibinfo
  {journal} {Journal of Physics: Condensed Matter}\ }\textbf {\bibinfo {volume}
  {34}},\ \bibinfo {pages} {275604} (\bibinfo {year} {2022})}\BibitemShut
  {NoStop}%
\bibitem [{\citenamefont {Dumitrescu}(2016)}]{dumitrescu_thesis}%
  \BibitemOpen
  \bibfield  {author} {\bibinfo {author} {\bibfnamefont {P.}~\bibnamefont
  {Dumitrescu}},\ }\emph {\bibinfo {title} {Strongly Correlated Electron
  Systems Near Criticality: {F}rom Nodal Semimetals to High-Temperature
  Superconductors}},\ \href {https://escholarship.org/uc/item/2pc8r5t5} {Ph.D.
  thesis},\ \bibinfo  {school} {UC Berkeley} (\bibinfo {year}
  {2016})\BibitemShut {NoStop}%
\bibitem [{\citenamefont {Sachdev}(2011)}]{subir_book}%
  \BibitemOpen
  \bibfield  {author} {\bibinfo {author} {\bibfnamefont {S.}~\bibnamefont
  {Sachdev}},\ }\href {https://doi.org/10.1017/CBO9780511973765} {\emph
  {\bibinfo {title} {Quantum Phase Transitions}}},\ \bibinfo {edition} {2nd}\
  ed.\ (\bibinfo  {publisher} {Cambridge University Press},\ \bibinfo {year}
  {2011})\BibitemShut {NoStop}%
\bibitem [{\citenamefont {Luttinger}(1964)}]{luttinger_kubo}%
  \BibitemOpen
  \bibfield  {author} {\bibinfo {author} {\bibfnamefont {J.~M.}\ \bibnamefont
  {Luttinger}},\ }\bibfield  {title} {\bibinfo {title} {Theory of thermal
  transport coefficients},\ }\href {https://doi.org/10.1103/PhysRev.135.A1505}
  {\bibfield  {journal} {\bibinfo  {journal} {Phys. Rev.}\ }\textbf {\bibinfo
  {volume} {135}},\ \bibinfo {pages} {A1505} (\bibinfo {year}
  {1964})}\BibitemShut {NoStop}%
\bibitem [{\citenamefont {Schmalian}(2018)}]{jorg_lectures}%
  \BibitemOpen
  \bibfield  {author} {\bibinfo {author} {\bibfnamefont {J.}~\bibnamefont
  {Schmalian}},\ }\href
  {https://www.tkm.kit.edu/downloads/ss2018_tkm2/TKM2_revised.pdf} {\bibinfo
  {title} {Lecture notes: Theory of condensed matter {II}}} (\bibinfo {year}
  {2018})\BibitemShut {NoStop}%
\bibitem [{\citenamefont {Bruus}\ and\ \citenamefont
  {Flensberg}(2004)}]{bruus}%
  \BibitemOpen
  \bibfield  {author} {\bibinfo {author} {\bibfnamefont {H.}~\bibnamefont
  {Bruus}}\ and\ \bibinfo {author} {\bibfnamefont {K.}~\bibnamefont
  {Flensberg}},\ }\href {https://doi.org/10.1093/oso/9780198566335.001.0001}
  {\emph {\bibinfo {title} {{Many-Body Quantum Theory in Condensed Matter
  Physics: An Introduction}}}},\ Oxford Graduate Texts\ (\bibinfo  {publisher}
  {OUP Oxford},\ \bibinfo {year} {2004})\BibitemShut {NoStop}%
\bibitem [{\citenamefont {Kadanoff}\ and\ \citenamefont
  {Martin}(1963)}]{kadanoff_hydro}%
  \BibitemOpen
  \bibfield  {author} {\bibinfo {author} {\bibfnamefont {L.~P.}\ \bibnamefont
  {Kadanoff}}\ and\ \bibinfo {author} {\bibfnamefont {P.~C.}\ \bibnamefont
  {Martin}},\ }\bibfield  {title} {\bibinfo {title} {Hydrodynamic equations and
  correlation functions},\ }\href
  {https://doi.org/https://doi.org/10.1016/0003-4916(63)90078-2} {\bibfield
  {journal} {\bibinfo  {journal} {Annals of Physics}\ }\textbf {\bibinfo
  {volume} {24}},\ \bibinfo {pages} {419} (\bibinfo {year} {1963})}\BibitemShut
  {NoStop}%
\bibitem [{\citenamefont {Rosch}\ and\ \citenamefont
  {Andrei}(2000)}]{Rosch-PRL}%
  \BibitemOpen
  \bibfield  {author} {\bibinfo {author} {\bibfnamefont {A.}~\bibnamefont
  {Rosch}}\ and\ \bibinfo {author} {\bibfnamefont {N.}~\bibnamefont {Andrei}},\
  }\bibfield  {title} {\bibinfo {title} {Conductivity of a clean
  one-dimensional wire},\ }\href {https://doi.org/10.1103/PhysRevLett.85.1092}
  {\bibfield  {journal} {\bibinfo  {journal} {Phys. Rev. Lett.}\ }\textbf
  {\bibinfo {volume} {85}},\ \bibinfo {pages} {1092} (\bibinfo {year}
  {2000})}\BibitemShut {NoStop}%
\bibitem [{\citenamefont {Shimshoni}\ \emph {et~al.}(2003)\citenamefont
  {Shimshoni}, \citenamefont {Andrei},\ and\ \citenamefont
  {Rosch}}]{Shimshoni}%
  \BibitemOpen
  \bibfield  {author} {\bibinfo {author} {\bibfnamefont {E.}~\bibnamefont
  {Shimshoni}}, \bibinfo {author} {\bibfnamefont {N.}~\bibnamefont {Andrei}},\
  and\ \bibinfo {author} {\bibfnamefont {A.}~\bibnamefont {Rosch}},\ }\bibfield
   {title} {\bibinfo {title} {Thermal conductivity of spin-$\frac{1}{2}$
  chains},\ }\href {https://doi.org/10.1103/PhysRevB.68.104401} {\bibfield
  {journal} {\bibinfo  {journal} {Phys. Rev. B}\ }\textbf {\bibinfo {volume}
  {68}},\ \bibinfo {pages} {104401} (\bibinfo {year} {2003})}\BibitemShut
  {NoStop}%
\bibitem [{\citenamefont {Freire}(2014)}]{Freire-AP_2014}%
  \BibitemOpen
  \bibfield  {author} {\bibinfo {author} {\bibfnamefont {H.}~\bibnamefont
  {Freire}},\ }\bibfield  {title} {\bibinfo {title} {{Controlled calculation of
  the thermal conductivity for a spinon Fermi surface coupled to a U(1) gauge
  field}},\ }\href {https://doi.org/10.1016/j.aop.2014.07.002} {\bibfield
  {journal} {\bibinfo  {journal} {Ann. Phys. (N. Y.)}\ }\textbf {\bibinfo
  {volume} {349}},\ \bibinfo {pages} {357} (\bibinfo {year}
  {2014})}\BibitemShut {NoStop}%
\bibitem [{\citenamefont {Hartnoll}\ \emph {et~al.}(2014)\citenamefont
  {Hartnoll}, \citenamefont {Mahajan}, \citenamefont {Punk},\ and\
  \citenamefont {Sachdev}}]{Hartnoll_PRB_2014}%
  \BibitemOpen
  \bibfield  {author} {\bibinfo {author} {\bibfnamefont {S.~A.}\ \bibnamefont
  {Hartnoll}}, \bibinfo {author} {\bibfnamefont {R.}~\bibnamefont {Mahajan}},
  \bibinfo {author} {\bibfnamefont {M.}~\bibnamefont {Punk}},\ and\ \bibinfo
  {author} {\bibfnamefont {S.}~\bibnamefont {Sachdev}},\ }\bibfield  {title}
  {\bibinfo {title} {{Transport near the Ising-nematic quantum critical point
  of metals in two dimensions}},\ }\href
  {https://doi.org/10.1103/PhysRevB.89.155130} {\bibfield  {journal} {\bibinfo
  {journal} {Phys. Rev. B}\ }\textbf {\bibinfo {volume} {89}},\ \bibinfo
  {pages} {155130} (\bibinfo {year} {2014})}\BibitemShut {NoStop}%
\bibitem [{\citenamefont {Freire}(2017{\natexlab{a}})}]{Freire-AP_2017}%
  \BibitemOpen
  \bibfield  {author} {\bibinfo {author} {\bibfnamefont {H.}~\bibnamefont
  {Freire}},\ }\bibfield  {title} {\bibinfo {title} {Memory matrix theory of
  the dc resistivity of a disordered antiferromagnetic metal with an effective
  composite operator},\ }\href {https://doi.org/10.1016/j.aop.2017.07.001}
  {\bibfield  {journal} {\bibinfo  {journal} {Annals of Physics}\ }\textbf
  {\bibinfo {volume} {384}},\ \bibinfo {pages} {142} (\bibinfo {year}
  {2017}{\natexlab{a}})}\BibitemShut {NoStop}%
\bibitem [{\citenamefont {Freire}(2017{\natexlab{b}})}]{Freire-EPL}%
  \BibitemOpen
  \bibfield  {author} {\bibinfo {author} {\bibfnamefont {H.}~\bibnamefont
  {Freire}},\ }\bibfield  {title} {\bibinfo {title} {Calculation of the
  magnetotransport for a spin-density-wave quantum critical theory in the
  presence of weak disorder},\ }\href
  {https://doi.org/10.1209/0295-5075/118/57003} {\bibfield  {journal} {\bibinfo
   {journal} {Europhysics Letters}\ }\textbf {\bibinfo {volume} {118}},\
  \bibinfo {pages} {57003} (\bibinfo {year} {2017}{\natexlab{b}})}\BibitemShut
  {NoStop}%
\bibitem [{\citenamefont {Freire}(2018)}]{Freire-EPL_2018}%
  \BibitemOpen
  \bibfield  {author} {\bibinfo {author} {\bibfnamefont {H.}~\bibnamefont
  {Freire}},\ }\bibfield  {title} {\bibinfo {title} {{Thermal and
  thermoelectric transport coefficients for a two-dimensional SDW metal with
  weak disorder: A memory matrix calculation}},\ }\href
  {https://doi.org/10.1209/0295-5075/124/27003} {\bibfield  {journal} {\bibinfo
   {journal} {{EPL} (Europhysics Letters)}\ }\textbf {\bibinfo {volume}
  {124}},\ \bibinfo {pages} {27003} (\bibinfo {year} {2018})}\BibitemShut
  {NoStop}%
\bibitem [{\citenamefont {Wang}\ and\ \citenamefont {Berg}(2019)}]{Berg-PRB}%
  \BibitemOpen
  \bibfield  {author} {\bibinfo {author} {\bibfnamefont {X.}~\bibnamefont
  {Wang}}\ and\ \bibinfo {author} {\bibfnamefont {E.}~\bibnamefont {Berg}},\
  }\bibfield  {title} {\bibinfo {title} {{Scattering mechanisms and electrical
  transport near an Ising nematic quantum critical point}},\ }\href
  {https://doi.org/10.1103/PhysRevB.99.235136} {\bibfield  {journal} {\bibinfo
  {journal} {Phys. Rev. B}\ }\textbf {\bibinfo {volume} {99}},\ \bibinfo
  {pages} {235136} (\bibinfo {year} {2019})}\BibitemShut {NoStop}%
\bibitem [{\citenamefont {Wang}\ and\ \citenamefont
  {Berg}(2022)}]{wang2020low}%
  \BibitemOpen
  \bibfield  {author} {\bibinfo {author} {\bibfnamefont {X.}~\bibnamefont
  {Wang}}\ and\ \bibinfo {author} {\bibfnamefont {E.}~\bibnamefont {Berg}},\
  }\bibfield  {title} {\bibinfo {title} {Low-frequency raman response near the
  ising-nematic quantum critical point: A memory-matrix approach},\ }\href
  {https://doi.org/10.1103/PhysRevB.105.045137} {\bibfield  {journal} {\bibinfo
   {journal} {Phys. Rev. B}\ }\textbf {\bibinfo {volume} {105}},\ \bibinfo
  {pages} {045137} (\bibinfo {year} {2022})}\BibitemShut {NoStop}%
\bibitem [{\citenamefont {Kovtun}\ \emph {et~al.}(2005)\citenamefont {Kovtun},
  \citenamefont {Son},\ and\ \citenamefont {Starinets}}]{DTSon-PRL_2005}%
  \BibitemOpen
  \bibfield  {author} {\bibinfo {author} {\bibfnamefont {P.~K.}\ \bibnamefont
  {Kovtun}}, \bibinfo {author} {\bibfnamefont {D.~T.}\ \bibnamefont {Son}},\
  and\ \bibinfo {author} {\bibfnamefont {A.~O.}\ \bibnamefont {Starinets}},\
  }\bibfield  {title} {\bibinfo {title} {Viscosity in strongly interacting
  quantum field theories from black hole physics},\ }\href
  {https://doi.org/10.1103/PhysRevLett.94.111601} {\bibfield  {journal}
  {\bibinfo  {journal} {Phys. Rev. Lett.}\ }\textbf {\bibinfo {volume} {94}},\
  \bibinfo {pages} {111601} (\bibinfo {year} {2005})}\BibitemShut {NoStop}%
\bibitem [{\citenamefont {Hartnoll}\ and\ \citenamefont
  {Herzog}(2008)}]{hartnoll_herzog}%
  \BibitemOpen
  \bibfield  {author} {\bibinfo {author} {\bibfnamefont {S.~A.}\ \bibnamefont
  {Hartnoll}}\ and\ \bibinfo {author} {\bibfnamefont {C.~P.}\ \bibnamefont
  {Herzog}},\ }\bibfield  {title} {\bibinfo {title} {{Impure AdS/CFT
  correspondence}},\ }\href {https://doi.org/10.1103/PhysRevD.77.106009}
  {\bibfield  {journal} {\bibinfo  {journal} {Phys. Rev. D}\ }\textbf {\bibinfo
  {volume} {77}},\ \bibinfo {pages} {106009} (\bibinfo {year}
  {2008})}\BibitemShut {NoStop}%
\bibitem [{\citenamefont {Davison}\ \emph {et~al.}(2019)\citenamefont
  {Davison}, \citenamefont {Gentle},\ and\ \citenamefont
  {Gout\'eraux}}]{Blaise}%
  \BibitemOpen
  \bibfield  {author} {\bibinfo {author} {\bibfnamefont {R.~A.}\ \bibnamefont
  {Davison}}, \bibinfo {author} {\bibfnamefont {S.~A.}\ \bibnamefont
  {Gentle}},\ and\ \bibinfo {author} {\bibfnamefont {B.}~\bibnamefont
  {Gout\'eraux}},\ }\bibfield  {title} {\bibinfo {title} {Slow relaxation and
  diffusion in holographic quantum critical phases},\ }\href
  {https://doi.org/10.1103/PhysRevLett.123.141601} {\bibfield  {journal}
  {\bibinfo  {journal} {Phys. Rev. Lett.}\ }\textbf {\bibinfo {volume} {123}},\
  \bibinfo {pages} {141601} (\bibinfo {year} {2019})}\BibitemShut {NoStop}%
\bibitem [{\citenamefont {Blake}(2016{\natexlab{a}})}]{Blake}%
  \BibitemOpen
  \bibfield  {author} {\bibinfo {author} {\bibfnamefont {M.}~\bibnamefont
  {Blake}},\ }\bibfield  {title} {\bibinfo {title} {Universal charge diffusion
  and the butterfly effect in holographic theories},\ }\href
  {https://doi.org/10.1103/PhysRevLett.117.091601} {\bibfield  {journal}
  {\bibinfo  {journal} {Phys. Rev. Lett.}\ }\textbf {\bibinfo {volume} {117}},\
  \bibinfo {pages} {091601} (\bibinfo {year} {2016}{\natexlab{a}})}\BibitemShut
  {NoStop}%
\bibitem [{\citenamefont {Blake}(2016{\natexlab{b}})}]{Blake2}%
  \BibitemOpen
  \bibfield  {author} {\bibinfo {author} {\bibfnamefont {M.}~\bibnamefont
  {Blake}},\ }\bibfield  {title} {\bibinfo {title} {Universal diffusion in
  incoherent black holes},\ }\href {https://doi.org/10.1103/PhysRevD.94.086014}
  {\bibfield  {journal} {\bibinfo  {journal} {Phys. Rev. D}\ }\textbf {\bibinfo
  {volume} {94}},\ \bibinfo {pages} {086014} (\bibinfo {year}
  {2016}{\natexlab{b}})}\BibitemShut {NoStop}%
\bibitem [{\citenamefont {Rau}\ and\ \citenamefont {Müller}(1996)}]{rau}%
  \BibitemOpen
  \bibfield  {author} {\bibinfo {author} {\bibfnamefont {J.}~\bibnamefont
  {Rau}}\ and\ \bibinfo {author} {\bibfnamefont {B.}~\bibnamefont {Müller}},\
  }\bibfield  {title} {\bibinfo {title} {From reversible quantum microdynamics
  to irreversible quantum transport},\ }\href
  {https://doi.org/https://doi.org/10.1016/0370-1573(95)00077-1} {\bibfield
  {journal} {\bibinfo  {journal} {Physics Reports}\ }\textbf {\bibinfo {volume}
  {272}},\ \bibinfo {pages} {1} (\bibinfo {year} {1996})}\BibitemShut {NoStop}%
\bibitem [{\citenamefont {Hartnoll}\ and\ \citenamefont
  {Hofman}(2012)}]{hartnoll_241601}%
  \BibitemOpen
  \bibfield  {author} {\bibinfo {author} {\bibfnamefont {S.~A.}\ \bibnamefont
  {Hartnoll}}\ and\ \bibinfo {author} {\bibfnamefont {D.~M.}\ \bibnamefont
  {Hofman}},\ }\bibfield  {title} {\bibinfo {title} {Locally critical
  resistivities from {U}mklapp scattering},\ }\href
  {https://doi.org/10.1103/PhysRevLett.108.241601} {\bibfield  {journal}
  {\bibinfo  {journal} {Phys. Rev. Lett.}\ }\textbf {\bibinfo {volume} {108}},\
  \bibinfo {pages} {241601} (\bibinfo {year} {2012})}\BibitemShut {NoStop}%
\bibitem [{\citenamefont {{Davison}}\ \emph {et~al.}(2015)\citenamefont
  {{Davison}}, \citenamefont {{Gout{\'e}raux}},\ and\ \citenamefont
  {{Hartnoll}}}]{Davison_2015}%
  \BibitemOpen
  \bibfield  {author} {\bibinfo {author} {\bibfnamefont {R.~A.}\ \bibnamefont
  {{Davison}}}, \bibinfo {author} {\bibfnamefont {B.}~\bibnamefont
  {{Gout{\'e}raux}}},\ and\ \bibinfo {author} {\bibfnamefont {S.~A.}\
  \bibnamefont {{Hartnoll}}},\ }\bibfield  {title} {\bibinfo {title}
  {{Incoherent transport in clean quantum critical metals}},\ }\href
  {https://doi.org/10.1007/JHEP10(2015)112} {\bibfield  {journal} {\bibinfo
  {journal} {Journal of High Energy Physics}\ }\textbf {\bibinfo {volume}
  {2015}},\ \bibinfo {eid} {112} (\bibinfo {year} {2015})}\BibitemShut
  {NoStop}%
\bibitem [{\citenamefont {Hartnoll}\ \emph {et~al.}(2007)\citenamefont
  {Hartnoll}, \citenamefont {Kovtun}, \citenamefont {M\"uller},\ and\
  \citenamefont {Sachdev}}]{hartnoll_144502}%
  \BibitemOpen
  \bibfield  {author} {\bibinfo {author} {\bibfnamefont {S.~A.}\ \bibnamefont
  {Hartnoll}}, \bibinfo {author} {\bibfnamefont {P.~K.}\ \bibnamefont
  {Kovtun}}, \bibinfo {author} {\bibfnamefont {M.}~\bibnamefont {M\"uller}},\
  and\ \bibinfo {author} {\bibfnamefont {S.}~\bibnamefont {Sachdev}},\
  }\bibfield  {title} {\bibinfo {title} {{Theory of the Nernst effect near
  quantum phase transitions in condensed matter and in dyonic black holes}},\
  }\href {https://doi.org/10.1103/PhysRevB.76.144502} {\bibfield  {journal}
  {\bibinfo  {journal} {Phys. Rev. B}\ }\textbf {\bibinfo {volume} {76}},\
  \bibinfo {pages} {144502} (\bibinfo {year} {2007})}\BibitemShut {NoStop}%
\bibitem [{\citenamefont {Giraldo-Gallo}\ \emph {et~al.}(2018)\citenamefont
  {Giraldo-Gallo}, \citenamefont {Galvis}, \citenamefont {Stegen},
  \citenamefont {Modic}, \citenamefont {Balakirev}, \citenamefont {Betts},
  \citenamefont {Lian}, \citenamefont {Moir}, \citenamefont {Riggs},
  \citenamefont {Wu}, \citenamefont {Bollinger}, \citenamefont {He},
  \citenamefont {Božović}, \citenamefont {Ramshaw}, \citenamefont {McDonald},
  \citenamefont {Boebinger},\ and\ \citenamefont
  {Shekhter}}]{Giraldo_Gallo_2018}%
  \BibitemOpen
  \bibfield  {author} {\bibinfo {author} {\bibfnamefont {P.}~\bibnamefont
  {Giraldo-Gallo}}, \bibinfo {author} {\bibfnamefont {J.~A.}\ \bibnamefont
  {Galvis}}, \bibinfo {author} {\bibfnamefont {Z.}~\bibnamefont {Stegen}},
  \bibinfo {author} {\bibfnamefont {K.~A.}\ \bibnamefont {Modic}}, \bibinfo
  {author} {\bibfnamefont {F.~F.}\ \bibnamefont {Balakirev}}, \bibinfo {author}
  {\bibfnamefont {J.~B.}\ \bibnamefont {Betts}}, \bibinfo {author}
  {\bibfnamefont {X.}~\bibnamefont {Lian}}, \bibinfo {author} {\bibfnamefont
  {C.}~\bibnamefont {Moir}}, \bibinfo {author} {\bibfnamefont {S.~C.}\
  \bibnamefont {Riggs}}, \bibinfo {author} {\bibfnamefont {J.}~\bibnamefont
  {Wu}}, \bibinfo {author} {\bibfnamefont {A.~T.}\ \bibnamefont {Bollinger}},
  \bibinfo {author} {\bibfnamefont {X.}~\bibnamefont {He}}, \bibinfo {author}
  {\bibfnamefont {I.}~\bibnamefont {Božović}}, \bibinfo {author}
  {\bibfnamefont {B.~J.}\ \bibnamefont {Ramshaw}}, \bibinfo {author}
  {\bibfnamefont {R.~D.}\ \bibnamefont {McDonald}}, \bibinfo {author}
  {\bibfnamefont {G.~S.}\ \bibnamefont {Boebinger}},\ and\ \bibinfo {author}
  {\bibfnamefont {A.}~\bibnamefont {Shekhter}},\ }\bibfield  {title} {\bibinfo
  {title} {Scale-invariant magnetoresistance in a cuprate superconductor},\
  }\href {https://doi.org/10.1126/science.aan3178} {\bibfield  {journal}
  {\bibinfo  {journal} {Science}\ }\textbf {\bibinfo {volume} {361}},\ \bibinfo
  {pages} {479} (\bibinfo {year} {2018})}\BibitemShut {NoStop}%
\bibitem [{\citenamefont {Fisher}\ \emph {et~al.}(1989)\citenamefont {Fisher},
  \citenamefont {Weichman}, \citenamefont {Grinstein},\ and\ \citenamefont
  {Fisher}}]{hyperscaling}%
  \BibitemOpen
  \bibfield  {author} {\bibinfo {author} {\bibfnamefont {M.~P.~A.}\
  \bibnamefont {Fisher}}, \bibinfo {author} {\bibfnamefont {P.~B.}\
  \bibnamefont {Weichman}}, \bibinfo {author} {\bibfnamefont {G.}~\bibnamefont
  {Grinstein}},\ and\ \bibinfo {author} {\bibfnamefont {D.~S.}\ \bibnamefont
  {Fisher}},\ }\bibfield  {title} {\bibinfo {title} {Boson localization and the
  superfluid-insulator transition},\ }\href
  {https://doi.org/10.1103/PhysRevB.40.546} {\bibfield  {journal} {\bibinfo
  {journal} {Phys. Rev. B}\ }\textbf {\bibinfo {volume} {40}},\ \bibinfo
  {pages} {546} (\bibinfo {year} {1989})}\BibitemShut {NoStop}%
\bibitem [{\citenamefont {Huijse}\ \emph {et~al.}(2012)\citenamefont {Huijse},
  \citenamefont {Sachdev},\ and\ \citenamefont {Swingle}}]{liza_brian}%
  \BibitemOpen
  \bibfield  {author} {\bibinfo {author} {\bibfnamefont {L.}~\bibnamefont
  {Huijse}}, \bibinfo {author} {\bibfnamefont {S.}~\bibnamefont {Sachdev}},\
  and\ \bibinfo {author} {\bibfnamefont {B.}~\bibnamefont {Swingle}},\
  }\bibfield  {title} {\bibinfo {title} {Hidden fermi surfaces in compressible
  states of gauge-gravity duality},\ }\href
  {https://doi.org/10.1103/PhysRevB.85.035121} {\bibfield  {journal} {\bibinfo
  {journal} {Phys. Rev. B}\ }\textbf {\bibinfo {volume} {85}},\ \bibinfo
  {pages} {035121} (\bibinfo {year} {2012})}\BibitemShut {NoStop}%
\bibitem [{\citenamefont {Luttinger}(1956)}]{Luttinger}%
  \BibitemOpen
  \bibfield  {author} {\bibinfo {author} {\bibfnamefont {J.~M.}\ \bibnamefont
  {Luttinger}},\ }\bibfield  {title} {\bibinfo {title} {Quantum theory of
  cyclotron resonance in semiconductors: General theory},\ }\href
  {https://link.aps.org/doi/10.1103/PhysRev.102.1030} {\bibfield  {journal}
  {\bibinfo  {journal} {Phys. Rev.}\ }\textbf {\bibinfo {volume} {102}},\
  \bibinfo {pages} {1030} (\bibinfo {year} {1956})}\BibitemShut {NoStop}%
\bibitem [{\citenamefont {Janssen}\ and\ \citenamefont
  {Herbut}(2015)}]{lukas-herbut}%
  \BibitemOpen
  \bibfield  {author} {\bibinfo {author} {\bibfnamefont {L.}~\bibnamefont
  {Janssen}}\ and\ \bibinfo {author} {\bibfnamefont {I.~F.}\ \bibnamefont
  {Herbut}},\ }\bibfield  {title} {\bibinfo {title} {Nematic quantum
  criticality in three-dimensional fermi system with quadratic band touching},\
  }\href {https://doi.org/10.1103/PhysRevB.92.045117} {\bibfield  {journal}
  {\bibinfo  {journal} {Phys. Rev. B}\ }\textbf {\bibinfo {volume} {92}},\
  \bibinfo {pages} {045117} (\bibinfo {year} {2015})}\BibitemShut {NoStop}%
\bibitem [{\citenamefont {Boettcher}\ and\ \citenamefont
  {Herbut}(2016)}]{igor16}%
  \BibitemOpen
  \bibfield  {author} {\bibinfo {author} {\bibfnamefont {I.}~\bibnamefont
  {Boettcher}}\ and\ \bibinfo {author} {\bibfnamefont {I.~F.}\ \bibnamefont
  {Herbut}},\ }\bibfield  {title} {\bibinfo {title} {Superconducting quantum
  criticality in three-dimensional {L}uttinger semimetals},\ }\href
  {https://doi.org/10.1103/PhysRevB.93.205138} {\bibfield  {journal} {\bibinfo
  {journal} {Phys. Rev. B}\ }\textbf {\bibinfo {volume} {93}},\ \bibinfo
  {pages} {205138} (\bibinfo {year} {2016})}\BibitemShut {NoStop}%
\bibitem [{\citenamefont {Roy}\ \emph {et~al.}(2019)\citenamefont {Roy},
  \citenamefont {Ghorashi}, \citenamefont {Foster},\ and\ \citenamefont
  {Nevidomskyy}}]{Roy_PRB}%
  \BibitemOpen
  \bibfield  {author} {\bibinfo {author} {\bibfnamefont {B.}~\bibnamefont
  {Roy}}, \bibinfo {author} {\bibfnamefont {S.~A.~A.}\ \bibnamefont
  {Ghorashi}}, \bibinfo {author} {\bibfnamefont {M.~S.}\ \bibnamefont
  {Foster}},\ and\ \bibinfo {author} {\bibfnamefont {A.~H.}\ \bibnamefont
  {Nevidomskyy}},\ }\bibfield  {title} {\bibinfo {title} {Topological
  superconductivity of spin-$3/2$ carriers in a three-dimensional doped
  {L}uttinger semimetal},\ }\href {https://doi.org/10.1103/PhysRevB.99.054505}
  {\bibfield  {journal} {\bibinfo  {journal} {Phys. Rev. B}\ }\textbf {\bibinfo
  {volume} {99}},\ \bibinfo {pages} {054505} (\bibinfo {year}
  {2019})}\BibitemShut {NoStop}%
\bibitem [{\citenamefont {Link}\ and\ \citenamefont
  {Herbut}(2020)}]{Herbut-PRB}%
  \BibitemOpen
  \bibfield  {author} {\bibinfo {author} {\bibfnamefont {J.~M.}\ \bibnamefont
  {Link}}\ and\ \bibinfo {author} {\bibfnamefont {I.~F.}\ \bibnamefont
  {Herbut}},\ }\bibfield  {title} {\bibinfo {title} {{Hydrodynamic transport in
  the Luttinger-Abrikosov-Beneslavskii non-Fermi liquid}},\ }\href
  {https://doi.org/10.1103/PhysRevB.101.125128} {\bibfield  {journal} {\bibinfo
   {journal} {Phys. Rev. B}\ }\textbf {\bibinfo {volume} {101}},\ \bibinfo
  {pages} {125128} (\bibinfo {year} {2020})}\BibitemShut {NoStop}%
\bibitem [{\citenamefont {Mauri}\ and\ \citenamefont {Polini}(2019)}]{polini}%
  \BibitemOpen
  \bibfield  {author} {\bibinfo {author} {\bibfnamefont {A.}~\bibnamefont
  {Mauri}}\ and\ \bibinfo {author} {\bibfnamefont {M.}~\bibnamefont {Polini}},\
  }\bibfield  {title} {\bibinfo {title} {Dielectric function and plasmons of
  doped three-dimensional {L}uttinger semimetals},\ }\href
  {https://doi.org/10.1103/PhysRevB.100.165115} {\bibfield  {journal} {\bibinfo
   {journal} {Phys. Rev. B}\ }\textbf {\bibinfo {volume} {100}},\ \bibinfo
  {pages} {165115} (\bibinfo {year} {2019})}\BibitemShut {NoStop}%
\bibitem [{\citenamefont {Boettcher}(2019)}]{Boettcher_2019}%
  \BibitemOpen
  \bibfield  {author} {\bibinfo {author} {\bibfnamefont {I.}~\bibnamefont
  {Boettcher}},\ }\bibfield  {title} {\bibinfo {title} {Optical response of
  {L}uttinger semimetals in the normal and superconducting states},\ }\href
  {https://doi.org/10.1103/PhysRevB.99.125146} {\bibfield  {journal} {\bibinfo
  {journal} {Phys. Rev. B}\ }\textbf {\bibinfo {volume} {99}},\ \bibinfo
  {pages} {125146} (\bibinfo {year} {2019})}\BibitemShut {NoStop}%
\bibitem [{\citenamefont {Murakami}\ \emph {et~al.}(2004)\citenamefont
  {Murakami}, \citenamefont {Nagosa},\ and\ \citenamefont {Zhang}}]{Murakami}%
  \BibitemOpen
  \bibfield  {author} {\bibinfo {author} {\bibfnamefont {S.}~\bibnamefont
  {Murakami}}, \bibinfo {author} {\bibfnamefont {N.}~\bibnamefont {Nagosa}},\
  and\ \bibinfo {author} {\bibfnamefont {S.-C.}\ \bibnamefont {Zhang}},\
  }\bibfield  {title} {\bibinfo {title} {$\text{SU}(2)$ non-abelian holonomy
  and dissipationless spin current in semiconductors},\ }\href
  {https://link.aps.org/doi/10.1103/PhysRevB.69.235206} {\bibfield  {journal}
  {\bibinfo  {journal} {Phys. Rev. B}\ }\textbf {\bibinfo {volume} {69}},\
  \bibinfo {pages} {235206} (\bibinfo {year} {2004})}\BibitemShut {NoStop}%
\bibitem [{\citenamefont {Herbut}(2012)}]{herbut_085304}%
  \BibitemOpen
  \bibfield  {author} {\bibinfo {author} {\bibfnamefont {I.~F.}\ \bibnamefont
  {Herbut}},\ }\bibfield  {title} {\bibinfo {title} {{Isospin of topological
  defects in Dirac systems}},\ }\href
  {https://doi.org/10.1103/PhysRevB.85.085304} {\bibfield  {journal} {\bibinfo
  {journal} {Phys. Rev. B}\ }\textbf {\bibinfo {volume} {85}},\ \bibinfo
  {pages} {085304} (\bibinfo {year} {2012})}\BibitemShut {NoStop}%
\bibitem [{\citenamefont {Wilson}\ and\ \citenamefont
  {Fisher}(1972)}]{Wilson_Fisher}%
  \BibitemOpen
  \bibfield  {author} {\bibinfo {author} {\bibfnamefont {K.~G.}\ \bibnamefont
  {Wilson}}\ and\ \bibinfo {author} {\bibfnamefont {M.~E.}\ \bibnamefont
  {Fisher}},\ }\bibfield  {title} {\bibinfo {title} {Critical exponents in 3.99
  dimensions},\ }\href {https://doi.org/10.1103/PhysRevLett.28.240} {\bibfield
  {journal} {\bibinfo  {journal} {Phys. Rev. Lett.}\ }\textbf {\bibinfo
  {volume} {28}},\ \bibinfo {pages} {240} (\bibinfo {year} {1972})}\BibitemShut
  {NoStop}%
\bibitem [{\citenamefont {Peskin}\ and\ \citenamefont
  {Schroeder}(1995)}]{Peskin}%
  \BibitemOpen
  \bibfield  {author} {\bibinfo {author} {\bibfnamefont {M.~E.}\ \bibnamefont
  {Peskin}}\ and\ \bibinfo {author} {\bibfnamefont {D.~V.}\ \bibnamefont
  {Schroeder}},\ }\href@noop {} {\emph {\bibinfo {title} {{An Introduction to
  Quantum Field Theory}}}}\ (\bibinfo  {publisher} {Addison-Wesley},\ \bibinfo
  {address} {Reading},\ \bibinfo {year} {1995})\BibitemShut {NoStop}%
\bibitem [{\citenamefont {Schlief}\ \emph {et~al.}(2017)\citenamefont
  {Schlief}, \citenamefont {Lunts},\ and\ \citenamefont {Lee}}]{lee_prx}%
  \BibitemOpen
  \bibfield  {author} {\bibinfo {author} {\bibfnamefont {A.}~\bibnamefont
  {Schlief}}, \bibinfo {author} {\bibfnamefont {P.}~\bibnamefont {Lunts}},\
  and\ \bibinfo {author} {\bibfnamefont {S.-S.}\ \bibnamefont {Lee}},\
  }\bibfield  {title} {\bibinfo {title} {Exact critical exponents for the
  antiferromagnetic quantum critical metal in two dimensions},\ }\href
  {https://doi.org/10.1103/PhysRevX.7.021010} {\bibfield  {journal} {\bibinfo
  {journal} {Phys. Rev. X}\ }\textbf {\bibinfo {volume} {7}},\ \bibinfo {pages}
  {021010} (\bibinfo {year} {2017})}\BibitemShut {NoStop}%
\bibitem [{\citenamefont {Kumar}\ \emph {et~al.}(2020)\citenamefont {Kumar},
  \citenamefont {Kharkwal}, \citenamefont {Kumar}, \citenamefont {Asokan},
  \citenamefont {Banerjee},\ and\ \citenamefont {Pramanik}}]{Pramanik}%
  \BibitemOpen
  \bibfield  {author} {\bibinfo {author} {\bibfnamefont {H.}~\bibnamefont
  {Kumar}}, \bibinfo {author} {\bibfnamefont {K.~C.}\ \bibnamefont {Kharkwal}},
  \bibinfo {author} {\bibfnamefont {K.}~\bibnamefont {Kumar}}, \bibinfo
  {author} {\bibfnamefont {K.}~\bibnamefont {Asokan}}, \bibinfo {author}
  {\bibfnamefont {A.}~\bibnamefont {Banerjee}},\ and\ \bibinfo {author}
  {\bibfnamefont {A.~K.}\ \bibnamefont {Pramanik}},\ }\bibfield  {title}
  {\bibinfo {title} {{Magnetic and transport properties of the pyrochlore
  iridates
  $({\mathrm{Y}}_{1\ensuremath{-}x}{\mathrm{Pr}}_{x}{)}_{2}{\mathrm{Ir}}_{2}{\mathrm{O}}_{7}$:
  Role of $f\text{\ensuremath{-}}d$ exchange interaction and
  $d\text{\ensuremath{-}}p$ orbital hybridization}},\ }\href
  {https://doi.org/10.1103/PhysRevB.101.064405} {\bibfield  {journal} {\bibinfo
   {journal} {Phys. Rev. B}\ }\textbf {\bibinfo {volume} {101}},\ \bibinfo
  {pages} {064405} (\bibinfo {year} {2020})}\BibitemShut {NoStop}%
\bibitem [{\citenamefont {Bozovic}\ \emph {et~al.}(1987)\citenamefont
  {Bozovic}, \citenamefont {Kirillov}, \citenamefont {Kapitulnik},
  \citenamefont {Char}, \citenamefont {Hahn}, \citenamefont {Beasley},
  \citenamefont {Geballe}, \citenamefont {Kim},\ and\ \citenamefont
  {Heeger}}]{Bozovic}%
  \BibitemOpen
  \bibfield  {author} {\bibinfo {author} {\bibfnamefont {I.}~\bibnamefont
  {Bozovic}}, \bibinfo {author} {\bibfnamefont {D.}~\bibnamefont {Kirillov}},
  \bibinfo {author} {\bibfnamefont {A.}~\bibnamefont {Kapitulnik}}, \bibinfo
  {author} {\bibfnamefont {K.}~\bibnamefont {Char}}, \bibinfo {author}
  {\bibfnamefont {M.~R.}\ \bibnamefont {Hahn}}, \bibinfo {author}
  {\bibfnamefont {M.~R.}\ \bibnamefont {Beasley}}, \bibinfo {author}
  {\bibfnamefont {T.~H.}\ \bibnamefont {Geballe}}, \bibinfo {author}
  {\bibfnamefont {Y.~H.}\ \bibnamefont {Kim}},\ and\ \bibinfo {author}
  {\bibfnamefont {A.~J.}\ \bibnamefont {Heeger}},\ }\bibfield  {title}
  {\bibinfo {title} {Optical measurements on oriented thin
  ${\mathrm{yba}}_{2}$${\mathrm{cu}}_{3}$${\mathrm{o}}_{7\mathrm{\ensuremath{-}}\mathrm{\ensuremath{\delta}}}$
  films: Lack of evidence for excitonic superconductivity},\ }\href
  {https://doi.org/10.1103/PhysRevLett.59.2219} {\bibfield  {journal} {\bibinfo
   {journal} {Phys. Rev. Lett.}\ }\textbf {\bibinfo {volume} {59}},\ \bibinfo
  {pages} {2219} (\bibinfo {year} {1987})}\BibitemShut {NoStop}%
\bibitem [{\citenamefont {Cooper}\ \emph {et~al.}(1988)\citenamefont {Cooper},
  \citenamefont {Slakey}, \citenamefont {Klein}, \citenamefont {Rice},
  \citenamefont {Bukowski},\ and\ \citenamefont {Ginsberg}}]{Cooper_Slakey}%
  \BibitemOpen
  \bibfield  {author} {\bibinfo {author} {\bibfnamefont {S.~L.}\ \bibnamefont
  {Cooper}}, \bibinfo {author} {\bibfnamefont {F.}~\bibnamefont {Slakey}},
  \bibinfo {author} {\bibfnamefont {M.~V.}\ \bibnamefont {Klein}}, \bibinfo
  {author} {\bibfnamefont {J.~P.}\ \bibnamefont {Rice}}, \bibinfo {author}
  {\bibfnamefont {E.~D.}\ \bibnamefont {Bukowski}},\ and\ \bibinfo {author}
  {\bibfnamefont {D.~M.}\ \bibnamefont {Ginsberg}},\ }\bibfield  {title}
  {\bibinfo {title} {{Gap anisotropy and phonon self-energy effects in
  single-crystal YBa$_2$Cu$_3$O$_{7-\delta}$}},\ }\href
  {https://doi.org/10.1103/PhysRevB.38.11934} {\bibfield  {journal} {\bibinfo
  {journal} {Phys. Rev. B}\ }\textbf {\bibinfo {volume} {38}},\ \bibinfo
  {pages} {11934} (\bibinfo {year} {1988})}\BibitemShut {NoStop}%
\bibitem [{\citenamefont {Staufer}\ \emph {et~al.}(1991)\citenamefont
  {Staufer}, \citenamefont {Hackl},\ and\ \citenamefont {Müller}}]{Staufer}%
  \BibitemOpen
  \bibfield  {author} {\bibinfo {author} {\bibfnamefont {T.}~\bibnamefont
  {Staufer}}, \bibinfo {author} {\bibfnamefont {R.}~\bibnamefont {Hackl}},\
  and\ \bibinfo {author} {\bibfnamefont {P.}~\bibnamefont {Müller}},\
  }\bibfield  {title} {\bibinfo {title} {{Some new aspects of the electronic
  response in Cu-O superconductors}},\ }\href
  {https://doi.org/https://doi.org/10.1016/0038-1098(91)90494-G} {\bibfield
  {journal} {\bibinfo  {journal} {Solid State Communications}\ }\textbf
  {\bibinfo {volume} {79}},\ \bibinfo {pages} {409} (\bibinfo {year}
  {1991})}\BibitemShut {NoStop}%
\bibitem [{\citenamefont {Slakey}\ \emph {et~al.}(1991)\citenamefont {Slakey},
  \citenamefont {Klein}, \citenamefont {Rice},\ and\ \citenamefont
  {Ginsberg}}]{Slakey}%
  \BibitemOpen
  \bibfield  {author} {\bibinfo {author} {\bibfnamefont {F.}~\bibnamefont
  {Slakey}}, \bibinfo {author} {\bibfnamefont {M.~V.}\ \bibnamefont {Klein}},
  \bibinfo {author} {\bibfnamefont {J.~P.}\ \bibnamefont {Rice}},\ and\
  \bibinfo {author} {\bibfnamefont {D.~M.}\ \bibnamefont {Ginsberg}},\
  }\bibfield  {title} {\bibinfo {title} {{Raman investigation of the
  YBa$_2$Cu$_3$O$_7$ imaginary response function}},\ }\href
  {https://doi.org/10.1103/PhysRevB.43.3764} {\bibfield  {journal} {\bibinfo
  {journal} {Phys. Rev. B}\ }\textbf {\bibinfo {volume} {43}},\ \bibinfo
  {pages} {3764} (\bibinfo {year} {1991})}\BibitemShut {NoStop}%
\bibitem [{\citenamefont {Devereaux}\ and\ \citenamefont
  {Hackl}(2007)}]{Devereaux_RMP}%
  \BibitemOpen
  \bibfield  {author} {\bibinfo {author} {\bibfnamefont {T.~P.}\ \bibnamefont
  {Devereaux}}\ and\ \bibinfo {author} {\bibfnamefont {R.}~\bibnamefont
  {Hackl}},\ }\bibfield  {title} {\bibinfo {title} {Inelastic light scattering
  from correlated electrons},\ }\href
  {https://doi.org/10.1103/revmodphys.79.175} {\bibfield  {journal} {\bibinfo
  {journal} {Reviews of Modern Physics}\ }\textbf {\bibinfo {volume} {79}},\
  \bibinfo {pages} {175} (\bibinfo {year} {2007})}\BibitemShut {NoStop}%
\bibitem [{\citenamefont {Cao}\ \emph {et~al.}(2011)\citenamefont {Cao},
  \citenamefont {Elliott}, \citenamefont {Joseph}, \citenamefont {Wu},
  \citenamefont {Petricka}, \citenamefont {Sch{\"a}fer},\ and\ \citenamefont
  {Thomas}}]{Cao}%
  \BibitemOpen
  \bibfield  {author} {\bibinfo {author} {\bibfnamefont {C.}~\bibnamefont
  {Cao}}, \bibinfo {author} {\bibfnamefont {E.}~\bibnamefont {Elliott}},
  \bibinfo {author} {\bibfnamefont {J.}~\bibnamefont {Joseph}}, \bibinfo
  {author} {\bibfnamefont {H.}~\bibnamefont {Wu}}, \bibinfo {author}
  {\bibfnamefont {J.}~\bibnamefont {Petricka}}, \bibinfo {author}
  {\bibfnamefont {T.}~\bibnamefont {Sch{\"a}fer}},\ and\ \bibinfo {author}
  {\bibfnamefont {J.~E.}\ \bibnamefont {Thomas}},\ }\bibfield  {title}
  {\bibinfo {title} {Universal quantum viscosity in a unitary {F}ermi gas},\
  }\href {https://doi.org/10.1126/science.1195219} {\bibfield  {journal}
  {\bibinfo  {journal} {Science}\ }\textbf {\bibinfo {volume} {331}},\ \bibinfo
  {pages} {58} (\bibinfo {year} {2011})}\BibitemShut {NoStop}%
\bibitem [{\citenamefont {Fritz}\ \emph {et~al.}(2008)\citenamefont {Fritz},
  \citenamefont {Schmalian}, \citenamefont {M\"uller},\ and\ \citenamefont
  {Sachdev}}]{Fritz-PRB}%
  \BibitemOpen
  \bibfield  {author} {\bibinfo {author} {\bibfnamefont {L.}~\bibnamefont
  {Fritz}}, \bibinfo {author} {\bibfnamefont {J.}~\bibnamefont {Schmalian}},
  \bibinfo {author} {\bibfnamefont {M.}~\bibnamefont {M\"uller}},\ and\
  \bibinfo {author} {\bibfnamefont {S.}~\bibnamefont {Sachdev}},\ }\bibfield
  {title} {\bibinfo {title} {Quantum critical transport in clean graphene},\
  }\href {https://doi.org/10.1103/PhysRevB.78.085416} {\bibfield  {journal}
  {\bibinfo  {journal} {Phys. Rev. B}\ }\textbf {\bibinfo {volume} {78}},\
  \bibinfo {pages} {085416} (\bibinfo {year} {2008})}\BibitemShut {NoStop}%
\bibitem [{\citenamefont {Taylor}\ and\ \citenamefont
  {Randeria}(2010)}]{Taylor}%
  \BibitemOpen
  \bibfield  {author} {\bibinfo {author} {\bibfnamefont {E.}~\bibnamefont
  {Taylor}}\ and\ \bibinfo {author} {\bibfnamefont {M.}~\bibnamefont
  {Randeria}},\ }\bibfield  {title} {\bibinfo {title} {Viscosity of strongly
  interacting quantum fluids: Spectral functions and sum rules},\ }\href
  {https://doi.org/10.1103/PhysRevA.81.053610} {\bibfield  {journal} {\bibinfo
  {journal} {Phys. Rev. A}\ }\textbf {\bibinfo {volume} {81}},\ \bibinfo
  {pages} {053610} (\bibinfo {year} {2010})}\BibitemShut {NoStop}%
\bibitem [{\citenamefont {Enss}\ \emph {et~al.}(2011)\citenamefont {Enss}, ,
  \citenamefont {Haussmann},\ and\ \citenamefont {Zwerger}}]{zwerger}%
  \BibitemOpen
  \bibfield  {author} {\bibinfo {author} {\bibfnamefont {T.}~\bibnamefont
  {Enss}}, , \bibinfo {author} {\bibfnamefont {R.}~\bibnamefont {Haussmann}},\
  and\ \bibinfo {author} {\bibfnamefont {W.}~\bibnamefont {Zwerger}},\
  }\bibfield  {title} {\bibinfo {title} {Viscosity and scale invariance in the
  unitary fermi gas},\ }\href {https://doi.org/10.1016/j.aop.2010.10.002}
  {\bibfield  {journal} {\bibinfo  {journal} {Annals of Physics}\ }\textbf
  {\bibinfo {volume} {326}},\ \bibinfo {pages} {770} (\bibinfo {year}
  {2011})}\BibitemShut {NoStop}%
\bibitem [{\citenamefont {Dumitrescu}(2015)}]{dumitrescu}%
  \BibitemOpen
  \bibfield  {author} {\bibinfo {author} {\bibfnamefont {P.~T.}\ \bibnamefont
  {Dumitrescu}},\ }\bibfield  {title} {\bibinfo {title} {Shear viscosity in a
  non-{F}ermi-liquid phase of a quadratic semimetal},\ }\href
  {https://doi.org/10.1103/PhysRevB.92.121102} {\bibfield  {journal} {\bibinfo
  {journal} {Phys. Rev. B}\ }\textbf {\bibinfo {volume} {92}},\ \bibinfo
  {pages} {121102} (\bibinfo {year} {2015})}\BibitemShut {NoStop}%
\bibitem [{\citenamefont {Rosalin}\ \emph {et~al.}(2023)\citenamefont
  {Rosalin}, \citenamefont {Telang}, \citenamefont {Singh}, \citenamefont
  {Muthu},\ and\ \citenamefont {Sood}}]{Rosalin_2023}%
  \BibitemOpen
  \bibfield  {author} {\bibinfo {author} {\bibfnamefont {M.}~\bibnamefont
  {Rosalin}}, \bibinfo {author} {\bibfnamefont {P.}~\bibnamefont {Telang}},
  \bibinfo {author} {\bibfnamefont {S.}~\bibnamefont {Singh}}, \bibinfo
  {author} {\bibfnamefont {D.~V.~S.}\ \bibnamefont {Muthu}},\ and\ \bibinfo
  {author} {\bibfnamefont {A.~K.}\ \bibnamefont {Sood}},\ }\bibfield  {title}
  {\bibinfo {title} {{Non-Fermi-liquid signatures of quadratic band touching
  and phonon anomalies in metallic Pr$_2$Ir$_2$O$_7$}},\ }\href
  {https://doi.org/10.1103/PhysRevB.108.195144} {\bibfield  {journal} {\bibinfo
   {journal} {Phys. Rev. B}\ }\textbf {\bibinfo {volume} {108}},\ \bibinfo
  {pages} {195144} (\bibinfo {year} {2023})}\BibitemShut {NoStop}%
\bibitem [{\citenamefont {{Mandal}}\ and\ \citenamefont
  {{Ziegler}}(2021)}]{ips-klaus}%
  \BibitemOpen
  \bibfield  {author} {\bibinfo {author} {\bibfnamefont {I.}~\bibnamefont
  {{Mandal}}}\ and\ \bibinfo {author} {\bibfnamefont {K.}~\bibnamefont
  {{Ziegler}}},\ }\bibfield  {title} {\bibinfo {title} {{Robust quantum
  transport at particle-hole symmetry}},\ }\href
  {https://doi.org/10.1209/0295-5075/ac1a25} {\bibfield  {journal} {\bibinfo
  {journal} {EPL (EuroPhys. Lett.)}\ }\textbf {\bibinfo {volume} {135}},\
  \bibinfo {eid} {17001} (\bibinfo {year} {2021})}\BibitemShut {NoStop}%
\bibitem [{\citenamefont {Shama}\ \emph {et~al.}(2020)\citenamefont {Shama},
  \citenamefont {Gopal},\ and\ \citenamefont {Singh}}]{shama}%
  \BibitemOpen
  \bibfield  {author} {\bibinfo {author} {\bibnamefont {Shama}}, \bibinfo
  {author} {\bibfnamefont {R.}~\bibnamefont {Gopal}},\ and\ \bibinfo {author}
  {\bibfnamefont {Y.}~\bibnamefont {Singh}},\ }\bibfield  {title} {\bibinfo
  {title} {{Observation of planar Hall effect in the ferromagnetic Weyl
  semimetal Co$_3$Sn$_2$S$_2$}},\ }\href
  {https://doi.org/https://doi.org/10.1016/j.jmmm.2020.166547} {\bibfield
  {journal} {\bibinfo  {journal} {Journal of Magnetism and Magnetic Materials}\
  }\textbf {\bibinfo {volume} {502}},\ \bibinfo {pages} {166547} (\bibinfo
  {year} {2020})}\BibitemShut {NoStop}%
\bibitem [{\citenamefont {Li}\ \emph {et~al.}(2018)\citenamefont {Li},
  \citenamefont {Zhang}, \citenamefont {Zhang}, \citenamefont {Wen},\ and\
  \citenamefont {Zhang}}]{li18_giant}%
  \BibitemOpen
  \bibfield  {author} {\bibinfo {author} {\bibfnamefont {P.}~\bibnamefont
  {Li}}, \bibinfo {author} {\bibfnamefont {C.~H.}\ \bibnamefont {Zhang}},
  \bibinfo {author} {\bibfnamefont {J.~W.}\ \bibnamefont {Zhang}}, \bibinfo
  {author} {\bibfnamefont {Y.}~\bibnamefont {Wen}},\ and\ \bibinfo {author}
  {\bibfnamefont {X.~X.}\ \bibnamefont {Zhang}},\ }\bibfield  {title} {\bibinfo
  {title} {{Giant planar Hall effect in the Dirac semimetal {Z}r{T}e$_{5}$}},\
  }\href {https://doi.org/10.1103/PhysRevB.98.121108} {\bibfield  {journal}
  {\bibinfo  {journal} {Phys. Rev. B}\ }\textbf {\bibinfo {volume} {98}},\
  \bibinfo {pages} {121108} (\bibinfo {year} {2018})}\BibitemShut {NoStop}%
\bibitem [{\citenamefont {Roy~Karmakar}\ \emph {et~al.}(2022)\citenamefont
  {Roy~Karmakar}, \citenamefont {Nandy}, \citenamefont {Taraphder},\ and\
  \citenamefont {Das}}]{GP_Das}%
  \BibitemOpen
  \bibfield  {author} {\bibinfo {author} {\bibfnamefont {A.}~\bibnamefont
  {Roy~Karmakar}}, \bibinfo {author} {\bibfnamefont {S.}~\bibnamefont {Nandy}},
  \bibinfo {author} {\bibfnamefont {A.}~\bibnamefont {Taraphder}},\ and\
  \bibinfo {author} {\bibfnamefont {G.~P.}\ \bibnamefont {Das}},\ }\bibfield
  {title} {\bibinfo {title} {{Giant anomalous thermal Hall effect in tilted
  type-I magnetic Weyl semimetal
  ${\text{Co}}_{3}{\text{Sn}}_{2}{\text{S}}_{2}$}},\ }\href
  {https://doi.org/10.1103/PhysRevB.106.245133} {\bibfield  {journal} {\bibinfo
   {journal} {Phys. Rev. B}\ }\textbf {\bibinfo {volume} {106}},\ \bibinfo
  {pages} {245133} (\bibinfo {year} {2022})}\BibitemShut {NoStop}%
\bibitem [{\citenamefont {{Yadav}}\ \emph {et~al.}(2022)\citenamefont
  {{Yadav}}, \citenamefont {{Fazzini}},\ and\ \citenamefont
  {{Mandal}}}]{ips-serena}%
  \BibitemOpen
  \bibfield  {author} {\bibinfo {author} {\bibfnamefont {S.}~\bibnamefont
  {{Yadav}}}, \bibinfo {author} {\bibfnamefont {S.}~\bibnamefont {{Fazzini}}},\
  and\ \bibinfo {author} {\bibfnamefont {I.}~\bibnamefont {{Mandal}}},\
  }\bibfield  {title} {\bibinfo {title} {{Magneto-transport signatures in
  periodically-driven {W}eyl and multi-{W}eyl semimetals}},\ }\href
  {https://doi.org/10.1016/j.physe.2022.115444} {\bibfield  {journal} {\bibinfo
   {journal} {Physica E Low-Dimensional Systems and Nanostructures}\ }\textbf
  {\bibinfo {volume} {144}},\ \bibinfo {eid} {115444} (\bibinfo {year}
  {2022})}\BibitemShut {NoStop}%
\bibitem [{\citenamefont {Ghosh}\ and\ \citenamefont
  {Mandal}(2024{\natexlab{a}})}]{ips_rahul_ph_strain}%
  \BibitemOpen
  \bibfield  {author} {\bibinfo {author} {\bibfnamefont {R.}~\bibnamefont
  {Ghosh}}\ and\ \bibinfo {author} {\bibfnamefont {I.}~\bibnamefont {Mandal}},\
  }\bibfield  {title} {\bibinfo {title} {{Electric and thermoelectric response
  for Weyl and multi-Weyl semimetals in planar Hall configurations including
  the effects of strain}},\ }\href
  {https://doi.org/https://doi.org/10.1016/j.physe.2024.115914} {\bibfield
  {journal} {\bibinfo  {journal} {Physica E: Low-dimensional Systems and
  Nanostructures}\ }\textbf {\bibinfo {volume} {159}},\ \bibinfo {pages}
  {115914} (\bibinfo {year} {2024}{\natexlab{a}})}\BibitemShut {NoStop}%
\bibitem [{\citenamefont {Ghosh}\ and\ \citenamefont
  {Mandal}(2024{\natexlab{b}})}]{ips-rahul-jpcm}%
  \BibitemOpen
  \bibfield  {author} {\bibinfo {author} {\bibfnamefont {R.}~\bibnamefont
  {Ghosh}}\ and\ \bibinfo {author} {\bibfnamefont {I.}~\bibnamefont {Mandal}},\
  }\bibfield  {title} {\bibinfo {title} {{Direction-dependent conductivity in
  planar Hall set-ups with tilted Weyl/multi-Weyl semimetals}},\ }\href
  {https://doi.org/10.1088/1361-648X/ad38fa} {\bibfield  {journal} {\bibinfo
  {journal} {Journal of Physics Condensed Matter}\ }\textbf {\bibinfo {volume}
  {36}},\ \bibinfo {pages} {275501} (\bibinfo {year}
  {2024}{\natexlab{b}})}\BibitemShut {NoStop}%
\bibitem [{\citenamefont {Behnia}\ \emph {et~al.}(2007)\citenamefont {Behnia},
  \citenamefont {M\'easson},\ and\ \citenamefont {Kopelevich}}]{Behnia_Nernst}%
  \BibitemOpen
  \bibfield  {author} {\bibinfo {author} {\bibfnamefont {K.}~\bibnamefont
  {Behnia}}, \bibinfo {author} {\bibfnamefont {M.-A.}\ \bibnamefont
  {M\'easson}},\ and\ \bibinfo {author} {\bibfnamefont {Y.}~\bibnamefont
  {Kopelevich}},\ }\bibfield  {title} {\bibinfo {title} {Nernst effect in
  semimetals: The effective mass and the figure of merit},\ }\href
  {https://doi.org/10.1103/PhysRevLett.98.076603} {\bibfield  {journal}
  {\bibinfo  {journal} {Phys. Rev. Lett.}\ }\textbf {\bibinfo {volume} {98}},\
  \bibinfo {pages} {076603} (\bibinfo {year} {2007})}\BibitemShut {NoStop}%
\bibitem [{\citenamefont {Mandal}\ and\ \citenamefont {Saha}(2020)}]{ips-kush}%
  \BibitemOpen
  \bibfield  {author} {\bibinfo {author} {\bibfnamefont {I.}~\bibnamefont
  {Mandal}}\ and\ \bibinfo {author} {\bibfnamefont {K.}~\bibnamefont {Saha}},\
  }\bibfield  {title} {\bibinfo {title} {{Thermopower in an anisotropic
  two-dimensional Weyl semimetal}},\ }\href
  {http://dx.doi.org/10.1103/PhysRevB.101.045101} {\bibfield  {journal}
  {\bibinfo  {journal} {Physical Review B}\ }\textbf {\bibinfo {volume} {101}}
  (\bibinfo {year} {2020})}\BibitemShut {NoStop}%
\bibitem [{\citenamefont {Papaj}\ and\ \citenamefont
  {Fu}(2019)}]{papaj_magnus}%
  \BibitemOpen
  \bibfield  {author} {\bibinfo {author} {\bibfnamefont {M.}~\bibnamefont
  {Papaj}}\ and\ \bibinfo {author} {\bibfnamefont {L.}~\bibnamefont {Fu}},\
  }\bibfield  {title} {\bibinfo {title} {Magnus {H}all effect},\ }\href
  {https://doi.org/10.1103/PhysRevLett.123.216802} {\bibfield  {journal}
  {\bibinfo  {journal} {Phys. Rev. Lett.}\ }\textbf {\bibinfo {volume} {123}},\
  \bibinfo {pages} {216802} (\bibinfo {year} {2019})}\BibitemShut {NoStop}%
\bibitem [{\citenamefont {Mandal}\ \emph {et~al.}(2020)\citenamefont {Mandal},
  \citenamefont {Das},\ and\ \citenamefont {Agarwal}}]{amit-magnus}%
  \BibitemOpen
  \bibfield  {author} {\bibinfo {author} {\bibfnamefont {D.}~\bibnamefont
  {Mandal}}, \bibinfo {author} {\bibfnamefont {K.}~\bibnamefont {Das}},\ and\
  \bibinfo {author} {\bibfnamefont {A.}~\bibnamefont {Agarwal}},\ }\bibfield
  {title} {\bibinfo {title} {Magnus {N}ernst and thermal {H}all effect},\
  }\href {https://doi.org/10.1103/PhysRevB.102.205414} {\bibfield  {journal}
  {\bibinfo  {journal} {Phys. Rev. B}\ }\textbf {\bibinfo {volume} {102}},\
  \bibinfo {pages} {205414} (\bibinfo {year} {2020})}\BibitemShut {NoStop}%
\bibitem [{\citenamefont {{Sekh, Sajid}}\ and\ \citenamefont {{Mandal,
  Ipsita}}(2022)}]{ips_magnus}%
  \BibitemOpen
  \bibfield  {author} {\bibinfo {author} {\bibnamefont {{Sekh, Sajid}}}\ and\
  \bibinfo {author} {\bibnamefont {{Mandal, Ipsita}}},\ }\bibfield  {title}
  {\bibinfo {title} {Magnus {H}all effect in three-dimensional topological
  semimetals},\ }\href {https://doi.org/10.1140/epjp/s13360-022-02840-2}
  {\bibfield  {journal} {\bibinfo  {journal} {Eur. Phys. J. Plus}\ }\textbf
  {\bibinfo {volume} {137}},\ \bibinfo {pages} {736} (\bibinfo {year}
  {2022})}\BibitemShut {NoStop}%
\bibitem [{\citenamefont {Stanford}(2016)}]{Stanford2016}%
  \BibitemOpen
  \bibfield  {author} {\bibinfo {author} {\bibfnamefont {D.}~\bibnamefont
  {Stanford}},\ }\bibfield  {title} {\bibinfo {title} {Many-body chaos at weak
  coupling},\ }\href {http://dx.doi.org/10.1007/JHEP10(2016)009} {\bibfield
  {journal} {\bibinfo  {journal} {Journal of High Energy Physics}\ }\textbf
  {\bibinfo {volume} {2016}} (\bibinfo {year} {2016})}\BibitemShut {NoStop}%
\bibitem [{\citenamefont {Maldacena}\ \emph {et~al.}(2016)\citenamefont
  {Maldacena}, \citenamefont {Shenker},\ and\ \citenamefont
  {Stanford}}]{Maldacena2016}%
  \BibitemOpen
  \bibfield  {author} {\bibinfo {author} {\bibfnamefont {J.}~\bibnamefont
  {Maldacena}}, \bibinfo {author} {\bibfnamefont {S.~H.}\ \bibnamefont
  {Shenker}},\ and\ \bibinfo {author} {\bibfnamefont {D.}~\bibnamefont
  {Stanford}},\ }\bibfield  {title} {\bibinfo {title} {A bound on chaos},\
  }\href {http://dx.doi.org/10.1007/JHEP08(2016)106} {\bibfield  {journal}
  {\bibinfo  {journal} {Journal of High Energy Physics}\ }\textbf {\bibinfo
  {volume} {2016}} (\bibinfo {year} {2016})}\BibitemShut {NoStop}%
\bibitem [{\citenamefont {Maldacena}\ and\ \citenamefont
  {Stanford}(2016)}]{MaldacenaSYK2016}%
  \BibitemOpen
  \bibfield  {author} {\bibinfo {author} {\bibfnamefont {J.}~\bibnamefont
  {Maldacena}}\ and\ \bibinfo {author} {\bibfnamefont {D.}~\bibnamefont
  {Stanford}},\ }\bibfield  {title} {\bibinfo {title} {{Remarks on the
  Sachdev-Ye-Kitaev model}},\ }\href
  {http://dx.doi.org/10.1103/PhysRevD.94.106002} {\bibfield  {journal}
  {\bibinfo  {journal} {Physical Review D}\ }\textbf {\bibinfo {volume} {94}}
  (\bibinfo {year} {2016})}\BibitemShut {NoStop}%
\bibitem [{\citenamefont {Shenker}\ and\ \citenamefont
  {Stanford}(2014)}]{Shenker2014}%
  \BibitemOpen
  \bibfield  {author} {\bibinfo {author} {\bibfnamefont {S.~H.}\ \bibnamefont
  {Shenker}}\ and\ \bibinfo {author} {\bibfnamefont {D.}~\bibnamefont
  {Stanford}},\ }\bibfield  {title} {\bibinfo {title} {Black holes and the
  butterfly effect},\ }\href {http://dx.doi.org/10.1007/JHEP03(2014)067}
  {\bibfield  {journal} {\bibinfo  {journal} {Journal of High Energy Physics}\
  }\textbf {\bibinfo {volume} {2014}} (\bibinfo {year} {2014})}\BibitemShut
  {NoStop}%
\bibitem [{\citenamefont {Cheng}\ \emph {et~al.}(2017)\citenamefont {Cheng},
  \citenamefont {Ohtsuki}, \citenamefont {Chaudhuri}, \citenamefont
  {Nakatsuji}, \citenamefont {Lippmaa},\ and\ \citenamefont
  {Armitage}}]{armitage_expt}%
  \BibitemOpen
  \bibfield  {author} {\bibinfo {author} {\bibfnamefont {B.}~\bibnamefont
  {Cheng}}, \bibinfo {author} {\bibfnamefont {T.}~\bibnamefont {Ohtsuki}},
  \bibinfo {author} {\bibfnamefont {D.}~\bibnamefont {Chaudhuri}}, \bibinfo
  {author} {\bibfnamefont {S.}~\bibnamefont {Nakatsuji}}, \bibinfo {author}
  {\bibfnamefont {M.}~\bibnamefont {Lippmaa}},\ and\ \bibinfo {author}
  {\bibfnamefont {N.~P.}\ \bibnamefont {Armitage}},\ }\bibfield  {title}
  {\bibinfo {title} {{Dielectric anomalies and interactions in the
  three-dimensional quadratic band touching Luttinger semimetal
  Pr$_2$Ir$_2$O$_7$}},\ }\href {https://doi.org/10.1038/s41467-017-02121-y}
  {\bibfield  {journal} {\bibinfo  {journal} {Nature Communications}\ }\textbf
  {\bibinfo {volume} {8}},\ \bibinfo {pages} {2097} (\bibinfo {year}
  {2017})}\BibitemShut {NoStop}%
\bibitem [{\citenamefont {Tian}\ \emph {et~al.}(2016)\citenamefont {Tian},
  \citenamefont {Kohama}, \citenamefont {Tomita}, \citenamefont {Ishizuka},
  \citenamefont {Hsieh}, \citenamefont {Ishikawa}, \citenamefont {Kindo},
  \citenamefont {Balents},\ and\ \citenamefont {Nakatsuji}}]{osti_1435592}%
  \BibitemOpen
  \bibfield  {author} {\bibinfo {author} {\bibfnamefont {Z.}~\bibnamefont
  {Tian}}, \bibinfo {author} {\bibfnamefont {Y.}~\bibnamefont {Kohama}},
  \bibinfo {author} {\bibfnamefont {T.}~\bibnamefont {Tomita}}, \bibinfo
  {author} {\bibfnamefont {H.}~\bibnamefont {Ishizuka}}, \bibinfo {author}
  {\bibfnamefont {T.~H.}\ \bibnamefont {Hsieh}}, \bibinfo {author}
  {\bibfnamefont {J.~J.}\ \bibnamefont {Ishikawa}}, \bibinfo {author}
  {\bibfnamefont {K.}~\bibnamefont {Kindo}}, \bibinfo {author} {\bibfnamefont
  {L.}~\bibnamefont {Balents}},\ and\ \bibinfo {author} {\bibfnamefont
  {S.}~\bibnamefont {Nakatsuji}},\ }\bibfield  {title} {\bibinfo {title}
  {{Field-induced quantum metal-insulator transition in the pyrochlore iridate
  Nd$_2$Ir$_2$O$_7$}},\ }\href {https://www.osti.gov/biblio/1435592} {\bibfield
   {journal} {\bibinfo  {journal} {Nature Physics}\ }\textbf {\bibinfo {volume}
  {12}},\ \bibinfo {pages} {134} (\bibinfo {year} {2016})}\BibitemShut
  {NoStop}%
\bibitem [{\citenamefont {Damascelli}\ \emph {et~al.}(2003)\citenamefont
  {Damascelli}, \citenamefont {Hussain},\ and\ \citenamefont
  {Shen}}]{Damascelli}%
  \BibitemOpen
  \bibfield  {author} {\bibinfo {author} {\bibfnamefont {A.}~\bibnamefont
  {Damascelli}}, \bibinfo {author} {\bibfnamefont {Z.}~\bibnamefont
  {Hussain}},\ and\ \bibinfo {author} {\bibfnamefont {Z.-X.}\ \bibnamefont
  {Shen}},\ }\bibfield  {title} {\bibinfo {title} {Angle-resolved photoemission
  studies of the cuprate superconductors},\ }\href
  {https://doi.org/10.1103/RevModPhys.75.473} {\bibfield  {journal} {\bibinfo
  {journal} {Rev. Mod. Phys.}\ }\textbf {\bibinfo {volume} {75}},\ \bibinfo
  {pages} {473} (\bibinfo {year} {2003})}\BibitemShut {NoStop}%
\bibitem [{\citenamefont {Mitrano}\ \emph {et~al.}(2018)\citenamefont
  {Mitrano}, \citenamefont {Husain}, \citenamefont {Vig}, \citenamefont
  {Kogar}, \citenamefont {Rak}, \citenamefont {Rubeck}, \citenamefont
  {Schmalian}, \citenamefont {Uchoa}, \citenamefont {Schneeloch}, \citenamefont
  {Zhong}, \citenamefont {Gu},\ and\ \citenamefont {Abbamonte}}]{Mitrano_2018}%
  \BibitemOpen
  \bibfield  {author} {\bibinfo {author} {\bibfnamefont {M.}~\bibnamefont
  {Mitrano}}, \bibinfo {author} {\bibfnamefont {A.~A.}\ \bibnamefont {Husain}},
  \bibinfo {author} {\bibfnamefont {S.}~\bibnamefont {Vig}}, \bibinfo {author}
  {\bibfnamefont {A.}~\bibnamefont {Kogar}}, \bibinfo {author} {\bibfnamefont
  {M.~S.}\ \bibnamefont {Rak}}, \bibinfo {author} {\bibfnamefont {S.~I.}\
  \bibnamefont {Rubeck}}, \bibinfo {author} {\bibfnamefont {J.}~\bibnamefont
  {Schmalian}}, \bibinfo {author} {\bibfnamefont {B.}~\bibnamefont {Uchoa}},
  \bibinfo {author} {\bibfnamefont {J.}~\bibnamefont {Schneeloch}}, \bibinfo
  {author} {\bibfnamefont {R.}~\bibnamefont {Zhong}}, \bibinfo {author}
  {\bibfnamefont {G.~D.}\ \bibnamefont {Gu}},\ and\ \bibinfo {author}
  {\bibfnamefont {P.}~\bibnamefont {Abbamonte}},\ }\bibfield  {title} {\bibinfo
  {title} {Anomalous density fluctuations in a strange metal},\ }\href
  {https://doi.org/10.1073/pnas.1721495115} {\bibfield  {journal} {\bibinfo
  {journal} {Proceedings of the National Academy of Sciences}\ }\textbf
  {\bibinfo {volume} {115}},\ \bibinfo {pages} {5392} (\bibinfo {year}
  {2018})}\BibitemShut {NoStop}%
\bibitem [{\citenamefont {Husain}\ \emph {et~al.}(2019)\citenamefont {Husain},
  \citenamefont {Mitrano}, \citenamefont {Rak}, \citenamefont {Rubeck},
  \citenamefont {Uchoa}, \citenamefont {March}, \citenamefont {Dwyer},
  \citenamefont {Schneeloch}, \citenamefont {Zhong}, \citenamefont {Gu},\ and\
  \citenamefont {Abbamonte}}]{Husain}%
  \BibitemOpen
  \bibfield  {author} {\bibinfo {author} {\bibfnamefont {A.~A.}\ \bibnamefont
  {Husain}}, \bibinfo {author} {\bibfnamefont {M.}~\bibnamefont {Mitrano}},
  \bibinfo {author} {\bibfnamefont {M.~S.}\ \bibnamefont {Rak}}, \bibinfo
  {author} {\bibfnamefont {S.}~\bibnamefont {Rubeck}}, \bibinfo {author}
  {\bibfnamefont {B.}~\bibnamefont {Uchoa}}, \bibinfo {author} {\bibfnamefont
  {K.}~\bibnamefont {March}}, \bibinfo {author} {\bibfnamefont
  {C.}~\bibnamefont {Dwyer}}, \bibinfo {author} {\bibfnamefont
  {J.}~\bibnamefont {Schneeloch}}, \bibinfo {author} {\bibfnamefont
  {R.}~\bibnamefont {Zhong}}, \bibinfo {author} {\bibfnamefont {G.~D.}\
  \bibnamefont {Gu}},\ and\ \bibinfo {author} {\bibfnamefont {P.}~\bibnamefont
  {Abbamonte}},\ }\bibfield  {title} {\bibinfo {title} {Crossover of charge
  fluctuations across the strange metal phase diagram},\ }\href
  {https://doi.org/10.1103/PhysRevX.9.041062} {\bibfield  {journal} {\bibinfo
  {journal} {Phys. Rev. X}\ }\textbf {\bibinfo {volume} {9}},\ \bibinfo {pages}
  {041062} (\bibinfo {year} {2019})}\BibitemShut {NoStop}%
\bibitem [{\citenamefont {Husain}\ \emph {et~al.}(2023)\citenamefont {Husain},
  \citenamefont {Huang}, \citenamefont {Mitrano}, \citenamefont {Rak},
  \citenamefont {Rubeck}, \citenamefont {Guo}, \citenamefont {Yang},
  \citenamefont {Sow}, \citenamefont {Maeno}, \citenamefont {Uchoa},
  \citenamefont {Chiang}, \citenamefont {Batson}, \citenamefont {Phillips},\
  and\ \citenamefont {Abbamonte}}]{PMID:37558882}%
  \BibitemOpen
  \bibfield  {author} {\bibinfo {author} {\bibfnamefont {A.~A.}\ \bibnamefont
  {Husain}}, \bibinfo {author} {\bibfnamefont {E.~W.}\ \bibnamefont {Huang}},
  \bibinfo {author} {\bibfnamefont {M.}~\bibnamefont {Mitrano}}, \bibinfo
  {author} {\bibfnamefont {M.~S.}\ \bibnamefont {Rak}}, \bibinfo {author}
  {\bibfnamefont {S.~I.}\ \bibnamefont {Rubeck}}, \bibinfo {author}
  {\bibfnamefont {X.}~\bibnamefont {Guo}}, \bibinfo {author} {\bibfnamefont
  {H.}~\bibnamefont {Yang}}, \bibinfo {author} {\bibfnamefont {C.}~\bibnamefont
  {Sow}}, \bibinfo {author} {\bibfnamefont {Y.}~\bibnamefont {Maeno}}, \bibinfo
  {author} {\bibfnamefont {B.}~\bibnamefont {Uchoa}}, \bibinfo {author}
  {\bibfnamefont {T.~C.}\ \bibnamefont {Chiang}}, \bibinfo {author}
  {\bibfnamefont {P.~E.}\ \bibnamefont {Batson}}, \bibinfo {author}
  {\bibfnamefont {P.~W.}\ \bibnamefont {Phillips}},\ and\ \bibinfo {author}
  {\bibfnamefont {P.}~\bibnamefont {Abbamonte}},\ }\bibfield  {title} {\bibinfo
  {title} {{Pines' demon observed as a 3D acoustic plasmon in Sr$_2$RuO$_4$}},\
  }\href {https://doi.org/10.1038/s41586-023-06318-8} {\bibfield  {journal}
  {\bibinfo  {journal} {Nature}\ }\textbf {\bibinfo {volume} {621}},\ \bibinfo
  {pages} {66} (\bibinfo {year} {2023})}\BibitemShut {NoStop}%
\end{thebibliography}%

\appendix

\end{document}